\documentclass[reprint,amsmath,amssymb,aps, prd]{revtex4-2}
\usepackage[colorlinks=true,linkcolor=blue,citecolor=blue,urlcolor=blue]{hyperref}
\usepackage{newtxtext,newtxmath}
\usepackage[T1]{fontenc}
\usepackage{ae,aecompl}
\usepackage[dvipsnames]{xcolor}
\usepackage[left]{lineno}
\usepackage{array, makecell}
\newcolumntype{C}[1]{>{\centering\arraybackslash}p{#1}}
\usepackage{caption}
\usepackage{subcaption}
\usepackage{graphicx}	
\usepackage{amsmath}	
\usepackage[justification=raggedright,singlelinecheck=false]{caption}
\usepackage{tabularray}
\usepackage{booktabs}
\usepackage{hyperref}

\def\bfx{{\bf x}}
\def\bfk{{\bf k}}

\def\calP{{\mathcal P}}

\def\calO{{\mathcal O}}

\def\rme{{\rm e}}

\def\rmk{{\rm k}}
\def\bfk{{\bf k}}

\def\hk{{\hat{k}}}

\def\bfr{{\bf r}}
\def\hr{{\hat{r}}}

\def\calC{{\cal C}}

\def\calN{{\cal N}}
\def\calO{{\cal O}}

\def\calP{{\cal P}}

\def\bfx{{\bf x}}

\newcommand{\av}[1]{\left\langle{#1}\right\rangle} 

\def\ezmock{{\textsc{EZmock}}}
\def\abacus{{\textsc{Abacus}}}
\def\gpch{{h^{-1}\rm{Gpc}}}
\def\mpch{{h^{-1}\rm{Mpc}}}

\newcommand{\six}[6]{\left(\begin{array}{ccc}
									{#1}& {#2}& {#3}\\
									{#4}& {#5}& {#6} \\
\end{array}\right)}

\begin{document}
\label{firstpage}

\title{Parity-odd Four-Point Correlation Function from DESI Data Release 1 Luminous Red Galaxy Sample} 
\author{J. Hou$^{1,2,3,4}$\,$^{\ast}$, R. N. Cahn$^{5}$, J.~Aguilar$^{5}$, S.~Ahlen$^{6}$, D.~Bianchi$^{7,8}$, D.~Brooks$^{9}$, T.~Claybaugh$^{5}$, P.~Doel$^{9}$, J.~E.~Forero-Romero$^{11,12}$, E.~Gaztañaga$^{13,14,15}$, L.~Le~Guillou$^{23}$, G.~Gutierrez$^{16}$, C.~Howlett$^{17}$, D.~Huterer$^{18,19}$, M.~Ishak$^{20}$, R.~Joyce$^{21}$, A.~Kremin$^{5}$, O.~Lahav$^{9}$, C.~Lamman$^{22}$, M.~Landriau$^{5}$, A.~de la Macorra$^{10}$, R.~Miquel$^{24,25}$, S.~Nadathur$^{14}$, G.~Niz$^{26,27}$, W.~J.~Percival$^{28,29,30}$, F.~Prada$^{31}$, I.~P\'erez-R\`afols$^{32}$, G.~Rossi$^{33}$, E.~Sanchez$^{34}$, D.~Schlegel$^{5}$, M.~Schubnell$^{18,19}$, H.~Seo$^{35}$, J.~Silber$^{5}$, D.~Sprayberry$^{21}$, G.~Tarl\'{e}$^{19}$, B.~A.~Weaver$^{21}$, H.~Zou$^{36}$}

\affiliation{$^1$Faculty of Physics, Ludwig-Maximilians-Universit{\"a}t, Scheinerstr. 1, 81679 Munich, Germany}
\affiliation{$^2$Institute of Astronomy, University of Cambridge, Madingley Rd, Cambridge CB3 0HA, UK} 
\affiliation{$^3$Kavli Institute for Cosmology Cambridge, Madingley Road, Cambridge CB3 0HA, UK} 
\affiliation{$^4$Max-Planck-Institut f{\"u}r Extraterrestrische Physik, Postfach 1312, Giessenbachstrasse, 85748 Garching, Germany}
\affiliation{$^5$Lawrence Berkeley National Laboratory, 1 Cyclotron Road, Berkeley, CA 94720, USA}
\affiliation{$^{6}$Department of Physics, Boston University, 590 Commonwealth Avenue, Boston, MA 02215 USA}
\affiliation{$^{7}$Dipartimento di Fisica ``Aldo Pontremoli'', Universit\`a degli Studi di Milano, Via Celoria 16, I-20133 Milano, Italy}
\affiliation{$^{8}$INAF-Osservatorio Astronomico di Brera, Via Brera 28, 20122 Milano, Italy}
\affiliation{$^{9}$Department of Physics \& Astronomy, University College London, Gower Street, London, WC1E 6BT, UK} 
\affiliation{$^{10}$Instituto de F\'{\i}sica, Universidad Nacional Aut\'{o}noma de M\'{e}xico,  Circuito de la Investigaci\'{o}n Cient\'{\i}fica, Ciudad Universitaria, Cd. de M\'{e}xico  C.~P.~04510,  M\'{e}xico}
\affiliation{$^{11}$Departamento de F\'isica, Universidad de los Andes, Cra. 1 No. 18A-10, Edificio Ip, CP 111711, Bogot\'a, Colombia}
\affiliation{$^{12}$Observatorio Astron\'omico, Universidad de los Andes, Cra. 1 No. 18A-10, Edificio H, CP 111711 Bogot\'a, Colombia}
\affiliation{$^{13}$Institut d'Estudis Espacials de Catalunya (IEEC), c/ Esteve Terradas 1, Edifici RDIT, Campus PMT-UPC, 08860 Castelldefels, Spain}
\affiliation{$^{14}$Institute of Cosmology and Gravitation, University of Portsmouth, Dennis Sciama Building, Portsmouth, PO1 3FX, UK} 
\affiliation{$^{15}$Institute of Space Sciences, ICE-CSIC, Campus UAB, Carrer de Can Magrans s/n, 08913 Bellaterra, Barcelona, Spain}
\affiliation{$^{16}$Fermi National Accelerator Laboratory, PO Box 500, Batavia, IL 60510, USA}
\affiliation{$^{17}$School of Mathematics and Physics, University of Queensland, Brisbane, QLD 4072, Australia}
\affiliation{$^{18}$Department of Physics, University of Michigan, 450 Church Street, Ann Arbor, MI 48109, USA}
\affiliation{$^{19}$University of Michigan, 500 S. State Street, Ann Arbor, MI 48109, USA}
\affiliation{$^{20}$Department of Physics, The University of Texas at Dallas, 800 W. Campbell Rd., Richardson, TX 75080, USA}
\affiliation{$^{21}$NSF NOIRLab, 950 N. Cherry Ave., Tucson, AZ 85719, USA}
\affiliation{$^{22}$The Ohio State University, Columbus, 43210 OH, USA}
\affiliation{$^{23}$Sorbonne Universit\'{e}, CNRS/IN2P3, Laboratoire de Physique Nucl\'{e}aire et de Hautes Energies (LPNHE), FR-75005 Paris, France}
\affiliation{$^{24}$Instituci\'{o} Catalana de Recerca i Estudis Avan\c{c}ats, Passeig de Llu\'{\i}s Companys, 23, 08010 Barcelona, Spain}
\affiliation{$^{25}$Institut de F\'{i}sica d’Altes Energies (IFAE), The Barcelona Institute of Science and Technology, Edifici Cn, Campus UAB, 08193, Bellaterra (Barcelona), Spain}
\affiliation{$^{26}$Departamento de F\'{\i}sica, DCI-Campus Le\'{o}n, Universidad de Guanajuato, Loma del Bosque 103, Le\'{o}n, Guanajuato C.~P.~37150, M\'{e}xico}
\affiliation{$^{27}$Instituto Avanzado de Cosmolog\'{\i}a A.~C., San Marcos 11 - Atenas 202. Magdalena Contreras. Ciudad de M\'{e}xico C.~P.~10720, M\'{e}xico}
\affiliation{$^{28}$Department of Physics and Astronomy, University of Waterloo, 200 University Ave W, Waterloo, ON N2L 3G1, Canada}
\affiliation{$^{29}$Perimeter Institute for Theoretical Physics, 31 Caroline St. North, Waterloo, ON N2L 2Y5, Canada}
\affiliation{$^{30}$Waterloo Centre for Astrophysics, University of Waterloo, 200 University Ave W, Waterloo, ON N2L 3G1, Canada}
\affiliation{$^{31}$Instituto de Astrof\'{i}sica de Andaluc\'{i}a (CSIC), Glorieta de la Astronom\'{i}a, s/n, E-18008 Granada, Spain}
\affiliation{$^{32}$Departament de F\'isica, EEBE, Universitat Polit\`ecnica de Catalunya, c/Eduard Maristany 10, 08930 Barcelona, Spain}
\affiliation{$^{33}$Department of Physics and Astronomy, Sejong University, 209 Neungdong-ro, Gwangjin-gu, Seoul 05006, Republic of Korea}
\affiliation{$^{34}$CIEMAT, Avenida Complutense 40, E-28040 Madrid, Spain}
\affiliation{$^{35}$Department of Physics \& Astronomy, Ohio University, 139 University Terrace, Athens, OH 45701, USA}
\affiliation{$^{36}$National Astronomical Observatories, Chinese Academy of Sciences, A20 Datun Road, Chaoyang District, Beijing, 100101, P.~R.~China}

\thanks{Email: \texttt{jiamin.hou@ast.cam.ac.uk}}

\begin{abstract}
The parity-odd four-point function  provides a unique probe of fundamental symmetries and potential new physics in the large-scale structure of the Universe. We present measurements of the parity-odd four-point function using the DESI DR1 LRG sample and assess its detection significance. Our analysis considers both auto- and cross-correlations, using two complementary approaches to the covariance: (i) the full analytic covariance matrix applied to the uncompressed data vector, and (ii) a compressed data vector combined with a hybrid covariance matrix constructed from simulations and analytic estimates. When using the full analytic covariance matrix without corrections, we observe apparent auto-correlation signals with significance up to $4\sigma$. However, this excess is also consistent with a mismatch between the statistical fluctuations estimated from the simulations and those present in the real data. Our findings therefore suggest that the parity-odd signal in the current DESI DR1 LRG sample is consistent with zero. We note, however, that the low completeness of this sample may have a non-negligible impact on the detection sensitivity. Future data releases with improved completeness will be crucial for further investigation.
\end{abstract}


\maketitle

\section{Introduction}
\label{sec:intro}
Parity violation on cosmological scales offers a unique window into fundamental physics, with the potential to reveal new aspects of inflationary dynamics, the properties of dark energy and dark matter, and the mechanisms underlying the matter-antimatter asymmetry in the universe. In recent years, various observational tests have been applied to cosmological data to search for signatures of parity violation~\cite{Minami2020:Birefringence, hou2022:parity, philcox_parity, Philcox2023:CMB_temperature, Philcox2024:CMB_polarization, Ng2023:AmplBirefrigenceGWTC3, Jaraba2025:AstrometryParity}.  These efforts motivate developments of theoretical models, data analysis methods, and cosmological observables. 

For scalar fields, the lowest-order correlation function that is sensitive to parity violation is the four-point correlation function (4PCF) in position space, or its Fourier-space counterpart, the trispectrum. Under the assumption of statistical isotropy, two-point correlation functions depend only on the separation between points and are therefore necessarily parity-even. For three-point functions, while it is possible to construct triple products of vectors, homogeneity and translational symmetry require the wave vector to be coplanar, which causes such terms to vanish. Despite this, it is possible to construct lower-order parity-sensitive statistics by forming composite fields from scalar quantities~\cite{Jeong2012:fossil, Jamieson2024:POP, Kurita2025:parity, Gao2025:parity}, or by considering correlators of cosmological observables that are intrinsically higher-rank tensors in nature~\cite{Inomata2024:ORF, Jenks2023:parametrizeParity}. From this perspective, different cosmological probes and parity-sensitive statistics offer complementary insights, as they can be sensitive to distinct parity-violating mechanisms and probe different physical regimes. 

Ref.\cite{Cahn2021:parity} proposed using the four-point correlation function (4PCF) of galaxy clustering as a probe of parity violation on cosmological scales. Subsequent analyses~\cite{hou2022:parity,philcox_parity} reported intriguing hints of potential parity-violating signals. However, as discussed in~\cite{hou2022:parity}, a key challenge remains the accurate estimation of the underlying statistical fluctuations, which can manifest as discrepancies between the observed data and mock catalogs~\cite{Krolewski2024:parity}. This challenge is a combination of the likely low-signal-noise of the parity violation signal, the high dimensionality of the data vector, and the choice of model-independent search. 

Efforts to address these challenges have taken several directions, including reducing the dimensionality of the data vector with new estimators~\cite{Jamieson2024:POP, Gao2025:parity}, incorporating theoretical insights from specific parity-violating models or templates~\cite{Reinhard2024:axion, BAO2025:anatomy, Cho2025:Axion}, and developing simulations that enable forward modeling and enable simulation-based studies~\cite{Zhu2025:tensor_fossil, Shiim2025:vector_fossil, Jamieson2025:AxionLattice}.

In this work, we take a data-driven approach and focus on the Luminous Red Galaxy (LRG) sample from the Dark Energy Spectroscopic Instrument (DESI) survey~\cite{DESI2013:Snowmass,DESI2016:Science,DESI2016:InstrumentDesign,DESI2022:InstrumentOverview,DESI_DR1_2025,DESI_DR1_cosmology_2024,DESI_DR2_cosmology_2025}. In our companion paper~\cite{Hou2025:DESIeven4PCF}, we analyzed the parity-even 4PCF using these data and here we will present the measurement for the parity-odd component. The DESI DR1 sample, with an average completeness of only about 50\%, is notably more susceptible to observational systematics, particularly the effects of fiber assignment, the procedure of assigning target galaxies to the observation sequence, which inevitably leads to incompleteness, especially during the early stage of the experiment. That the signal-to-noise ratio associated with the parity-odd 4PCF further is likely to be very small complicates the analysis. Additionally, the limited volume of currently available $N$-body simulations introduces challenges in estimating the covariance matrix and increases the impact of sample variance.

Here, we build upon the methods introduced in~\cite{hou2022:parity} by implementing further calibration and correction for both the mocks and the data. We conduct a range of tests to quantify the detection significance and consistently find results consistent with zero. Unlike~\cite{Slepian2025:DESIodd4PCF}, we do not observe any significant discrepancies among the different methods used to assess the detection significance.  We find that neglecting the relevant calibrations and corrections can lead to biased results. 

The paper is structured as follows: in \S\ref{sec:desi_dr1} we provide an overview of the DESI DR1, the LRG sample, the simulations, and the evaluation of systematics, particularly those arising from sample incompleteness in the first-year data of the five-year survey program. 
In \S\ref{sec:methodology} we discuss the methodology used in this paper. In \S\ref{sec:Result} we discuss the detection significance for the DR1 result.
Finally, we conclude in \S\ref{sec:summary}. In the appendix, we present a series of robustness tests, as well as normalization definitions, and the correlation in the mocks.

Our Fourier transform  convention is $\tilde f(\bfk)=\int d^3x\, e^{-i\bfk\cdot \bfx} f(\bfx)$ and $f(\bfx) =\int_{\bfk} e^{i\bfk\cdot \bfx} \tilde{f}(\bfk)$, where $\int_{\bfk}\equiv \int d^3\bfk/(2\pi)^3$.

\section{DESI DR1 LRG Sample and the 4PCF Measurement}
\label{sec:desi_dr1}

\subsection{Brief Overview of the LRG Sample}
The DESI LRG sample is selected from two steps, including an imaging stage for target selection and a spectroscopic stage for redshift measurement~\cite{,DESI_Survey_Operation_2023, DESI_spectroscopic_pipeline_2023, DESI_Fiber_System_2024, DESI_Corrector_2024}. For the imaging selection, most details can be found in~\cite{DESI2024:SampleDefinition} and in our companion paper~\cite{Hou2025:DESIeven4PCF}.
In the spectroscopic stage, each target is assigned a fiber that directs its light to one of the DESI spectrographs. The \textsc{fiberassign} algorithm allocates fibers to targets listed in the merged-target-ledger (MTL) file~\cite{Lasker2025:DESIFiberAssign}. Fiber assignment introduces two limitations: each fiber covers only a small patrol area, so dense regions may contain more targets than available fibers, and closely spaced targets cannot both be observed due to ``fiber collisions''.  Because of the high density of targets, only about 20\% objects in the focal plane can be obtained in each exposure. These effects cause incompleteness early in the survey, which is reduced as multiple passes over the surveyed region accumulate over the five-year program. 

\begin{figure*}
    \centering
    \includegraphics[width=0.85\linewidth]{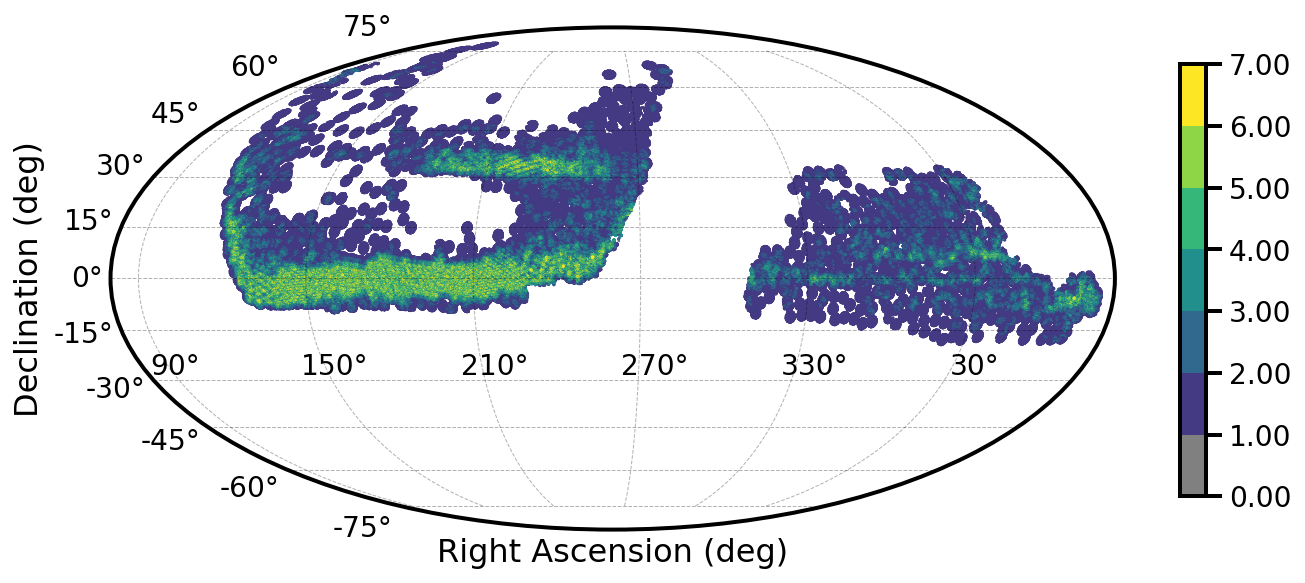}
    \caption{Footprint of DESI-DR1 LRG. The color map shows the number of targets sharing the same unique identifier, which combines the tile ID and fiber location. This quantity is closely correlated with the sample completeness: higher values of \textsc{NTILE} indicate higher completeness. For this paper, we specifically studied two regions, the ${\rm NGC}_1$ and ${\rm NGC}_2$, which have the highest numbers of targets sharing the same identifier (also see Table~\ref{tab:statistic}).}
    \label{fig:footprint}
\end{figure*}


The completeness, the distribution of the number density $n(z)$, and the fiber assignment effect motivate the sample selection in this paper.
For the analysis in this paper, we use the default clustering catalog covering the redshift range $0.4 < z < 1.1$. To account for variations in completeness and redshift evolution, we further divide the LRG sample into three subregions. In the NGC, we focus on areas with higher completeness and define two subregions: NGC-1, with angular cuts in right ascension (RA) and declination (DEC) of $110<\text{RA}<260$, $-10<\text{DEC}<8$, and a redshift range of $0.4<z<0.8$; and NGC-2, defined by $180<\text{RA}<260$, $30<\text{DEC}<40$, with the same redshift limits. For the SGC, we define a single subregion, SGC-3, applying only the redshift cut $0.4<z<0.8$, as completeness is relatively uniform in comparison to the NGC, and this choice preserves statistical constraining power. Fig.~\ref{fig:footprint} shows the DESI survey footprint, with colors indicating the number of tiles as a proxy of relative completeness. The geometric selections, galaxy counts, and corresponding volumes for all regions are listed in Table~\ref{tab:statistic}~\footnote{The fiducial volume $V_{\rm fid}$ and the effective number density $\bar{n}_g$ are free parameters of the analytic covariance~\cite{Hou2022:AnalytCov}. They are obtained by calibrating the analytic covariance to the simulation covariance. Compared to the parity-even 4PCF analysis~\cite{Hou2025:DESIeven4PCF}, the fitted number density here is lower, which is compensated by the fiducial volumes are higher. The results are overall consistent.}. 

\begin{table*}[t]
\centering
\caption{Statistics for the full DR1 LRG sample and subsamples  under different geometric selections. Here we include the number of galaxies $N_{\rm g}$, the number density $\bar{n}_{\rm g}$, and fiducial volume $V_{\rm fid}$. The number density and fiducial volume are obtained from fitting the analytic covariance matrix to the \ezmock s. The effective volume $V_{\rm eff}$ is computed from the trace of the product of the inverse of the analytic covariance corresponding to the full NGC and the covariance for the sky patch of interest ({\it c.f.}~\S\ref{sec:methodology}). The difference between $V_{\rm fid}$ and $V_{\rm eff}$ arises from variations in number density. The effective volume for \abacus\, $V_{\rm eff}^{\rm abcs}$, in the presence of the volume replication effect, is derived from the ratio between the analytic covariance and the \abacus-FFA.}
\begin{tabular}{@{}p{2cm}C{3cm}C{2cm}C{2.4cm}C{2.4cm}C{2.4cm}C{2.4cm}@{}}
\toprule
Region & Selection & $N_{\rm gal}$ & $\bar{n}_{\rm g}~[\mpch]^{-3}$ & $V_{\rm fid}~[\gpch]^3$ & $V_{\rm eff}~[\gpch]^3$ & $V_{\rm eff}^{\rm abcs}~[\gpch]^3$ \\ \midrule
Full-NGC & $0.4<z<1.1$ & 1,476,135 & $2.0\times 10^{-4}$ & 3.24 & 3.24 & 2.76 \\ [3ex]
$\quad$ NGC-1 & \begin{tabular}[c]{@{}l@{}}$110 <\text{RA} < 260$, \\ $-10 <\text{DEC} < 8$, \\ $0.4<z<0.8$ \end{tabular} & 434,054 & $3.1\times 10^{-4}$ & 0.64 & 1.13 & 0.97\\[5ex]
$\quad$ NGC-2 & \begin{tabular}[c]{@{}l@{}}$180 <\text{RA} < 260$, \\ $30 <\text{DEC} < 40$, \\ $0.4<z<0.8$\end{tabular} & 119,034 & $2.7\times 10^{-4}$ & 0.14 & 0.22 & 0.20\\ \midrule
Full-SGC & $0.4<z<1.1$ & 662,492 & $1.3\times 10^{-4}$ & 1.91 & 0.96 & 0.88 \\ [3ex]
$\quad$ SGC-3 & $0.4<z<0.8$ & 398,089 & $1.6\times 10^{-4}$ & 0.82 & 0.59 & 0.54 \\ \bottomrule
\end{tabular}
\label{tab:statistic}
\end{table*}

\subsection{Measurement of the Parity-Odd 4-Point Correlation Functions}
The parity-odd scalar observables incorporate a pseudoscalar, which is essential to isolating the parity information. 
Spherical harmonics are the natural choice when analyzing the angular dependence of the four-point function, as they provide a particularly suitable basis for such decompositions. Given the cosmological principle, which asserts the statistical isotropy of the universe, it is advantageous to decompose physical quantities into a basis that has rotational symmetry.
Each set of spherical harmonics spans a $(2\ell + 1)$-dimensional vector space, and the tensor product of multiple such spaces can be decomposed into a direct sum of irreducible representations characterized by total angular momentum $L$, with $|\ell_i - \ell_j| \leq L \leq \ell_i + \ell_j$. Of particular interest is the singlet state ($L=0$), which is rotationally invariant. Moreover, each spherical harmonic transforms under parity operation as
\begin{eqnarray}
 {\mathbb{\hat P}} [Y_{\ell m}] \rightarrow (-1)^{\ell} Y_{\ell m},
\end{eqnarray}
which naturally separates the parity-even and odd information.

In practice, we are interested in the coefficient of the four-point correlation function, which has six degrees of freedom with three angular momenta and three radial variables.

We define the isotropic functions $\mathcal{P}_{\ell_1 \ell_2 \ell_3}(\hat{r}_1, \hat{r}_2, \hat{r}_3)$ of three arguments by~\cite{Cahn202010}
\begin{align}
\label{eqn:Plll}
\mathcal{P}_{\ell_1 \ell_2 \ell_3}(\hat{r}_1, \hat{r}_2, \hat{r}_3) &= \sum_{m_1 m_2 m_3} 
(-1)^{\ell_1+\ell_2+\ell_3}\six{\ell_1}{\ell_2}{\ell_3}{m_1}{m_2}{m_3}\nonumber\\
&\quad \times Y_{\ell_1 m_1}(\hat{r}_1)
Y_{\ell_2 m_2}(\hat{r}_2)
Y_{\ell_3 m_3}(\hat{r}_3),
\end{align}
where $\hat{r}_1,\hat{r}_2$, and $\hat{r}_3$ are unit vectors in the directions of $\bfr_1,\bfr_2,$ and $\bfr_3$. The isotropic functions are complete and orthonormal: $\av{\calP_{\Lambda'}|\calP_{\Lambda}}=\delta^{\rm k}_{\Lambda'\Lambda}$, with $\delta^{\rm k}_{\Lambda'\Lambda}$ being the Kronecker delta and the collection of angular momenta $\Lambda\equiv\{\ell_1, \ell_2, \ell_3\}$. Here the angular differential is normalized so $\int\, d\hr=4\pi$.  The four-point correlation function can be factorized into a radially dependent part and an angle-dependent part:

\begin{align}
\zeta(\bfr_1,\bfr_2,\bfr_3)=\sum_{\ell_1\ell_2\ell_3}\zeta_{\ell_1\ell_2\ell_3}(r_1,r_2,r_3) \,\mathcal{P}_{\ell_1\ell_2\ell_3}(\hat{r}_1,\hat{r}_2,\hat{r}_3).
\end{align}
Since the angular basis functions are constructed from products of spherical harmonics and possess interchange symmetry among their arguments, the basis is overcomplete if all permutations are included. To avoid this overcompleteness, we impose an ordering convention and restrict to configurations satisfying $r_1<r_2<r_3$. The ordering convention uniquely defines the correspondence between each angular momentum index $\ell_i$ and and its associated radial bin $r_i$.

Thus $\zeta_{\ell_1\ell_2\ell_3}(r_1,r_2,r_3)$ is defined only for $r_1<r_2<r_3$.  It follows that
\begin{align}
    &\zeta_{\ell_1\ell_2\ell_3}(r_1,r_2,r_3) = \\
    &\int d\hr_1\,d\hr_2\,d\hr_3\, {\zeta}(\bfr_1,\bfr_2,\bfr_3) \,\mathcal{P}^*_{\ell_1 \ell_2 \ell_3}(\hat{r}_1, \hat{r}_2, \hat{r}_3),\nonumber
\end{align}
where $\ell_i$, for $i=1,2,3$ are the ``angular momenta'' associated with the three direction vectors $\bfr_i$, and star denotes a complex conjugate.  The sum $\ell_1+\ell_2+\ell_3$ determines the parity of the component: if the sum is even, its parity is even, and similarly for the odd sum. The parity-even isotropic functions are purely real and the parity-odd ones are purely imaginary. 
\

Since the real survey has non-trivial geometry, we further need to correct for the geometry using a random catalog. To do so, we use a generalized extended Landy-Szalay estimator~\cite{SE_3pt, Philcox:encore}. The survey geometry also leads to mixing among different angular modes of $\zeta_{\ell_1\ell_2\ell_3}(r_1,r_2,r_3)$. 
Further details about the survey geometry correction can be found in~\cite{Hou2025:DESIeven4PCF}. 

\begin{figure*}
    \centering
    \includegraphics[width=.96\linewidth]{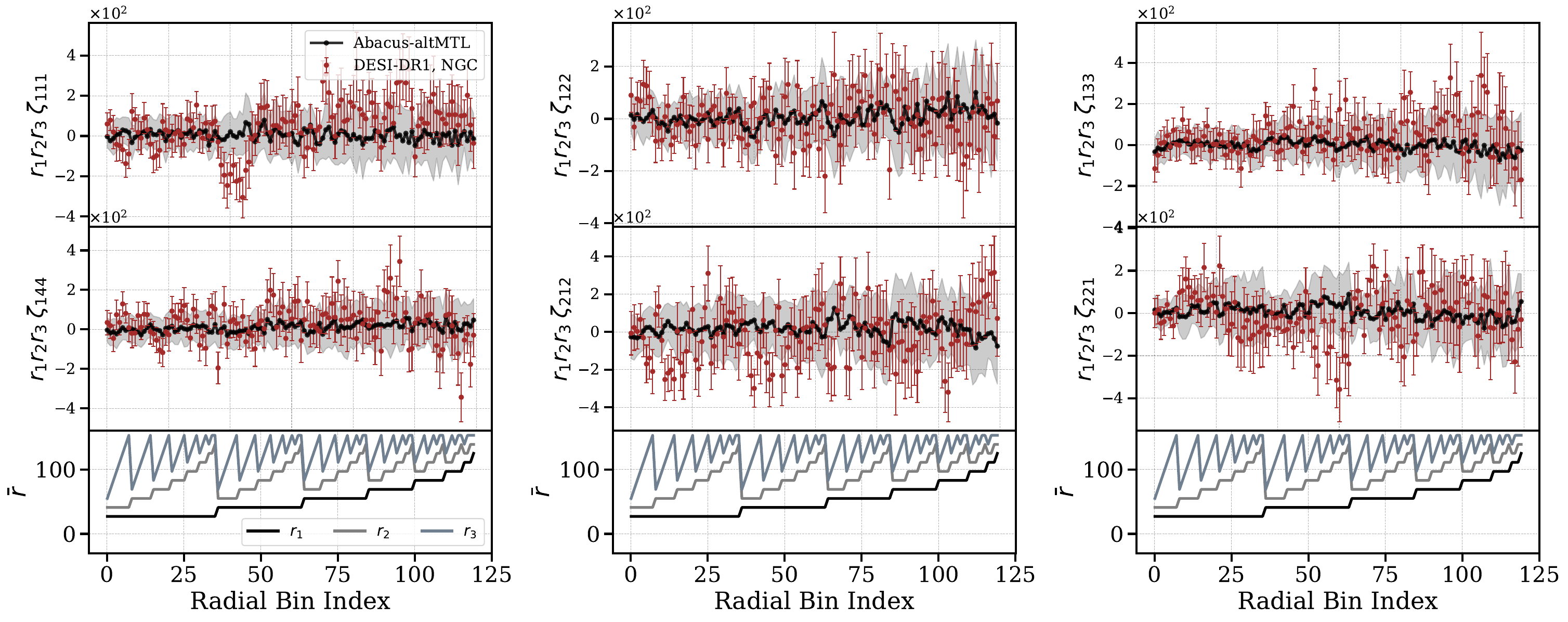}
    \caption{Measurement of the parity-odd 4PCFs using the DESI-DR1 LRG sample in NGC. The top two panels show a subset of the angular channels ($\ell_1,\ell_2,\ell_3$) for the coefficients of the connected 4PCF, weighted by $r_1r_2r_3$. Points with error bars (red) are the measurement from the data; the black curves with shaded gray regions indicate the $1\sigma$ standard deviation from the Abacus simulations with altMTL fiber assignment scheme. The bottom panel shows the arrangement of the three radial bins.}
    \label{fig:4PCF_Iron_AbacusAltmtl_NGC_6ells_odd}
\end{figure*}

\begin{figure*}
    \centering
    \includegraphics[width=.96\linewidth]{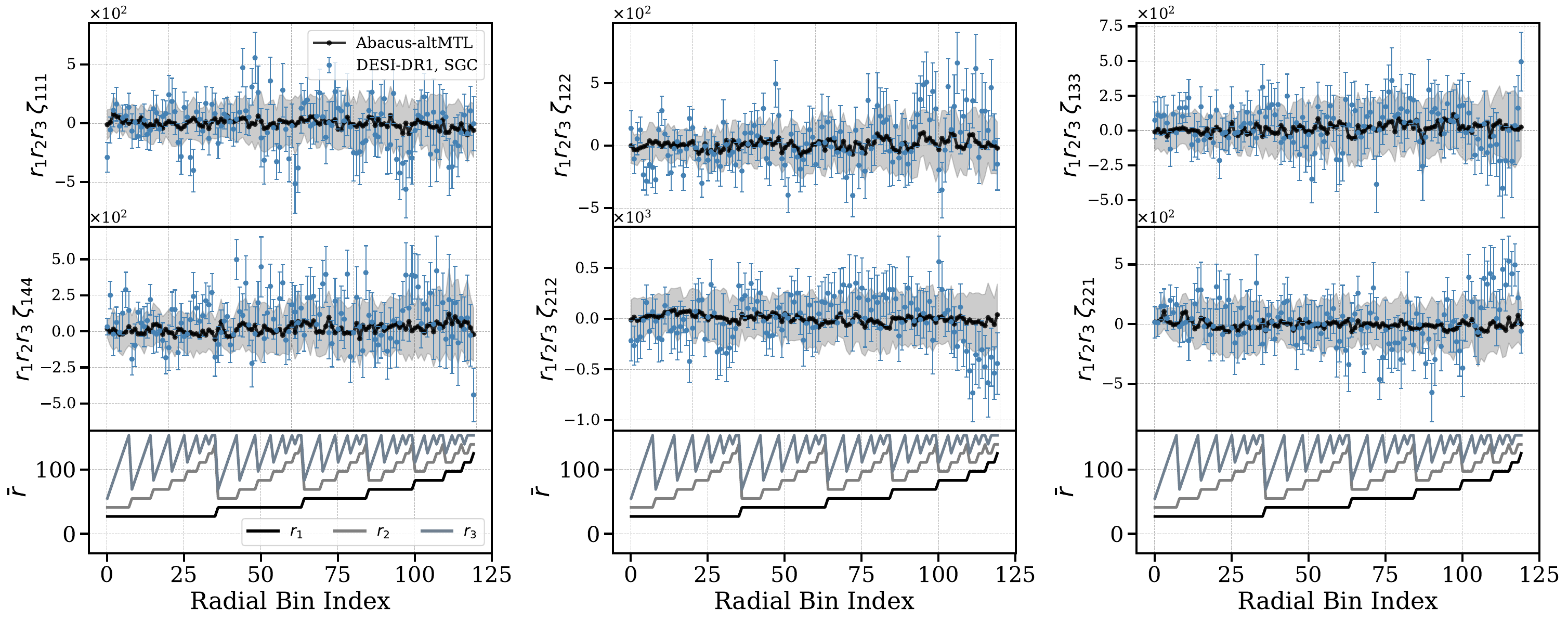}
    \caption{Simlar to Fig.~\ref{fig:4PCF_Iron_AbacusAltmtl_NGC_6ells_odd}, this figure shows the measurement of the parity-odd 4PCFs using the DESI-DR1 LRG sample in SGC. The top two panels show a subset of the angular channels ($\ell_1,\ell_2,\ell_3$) for the coefficients of the connected 4PCF, weighted by $r_1r_2r_3$. Points with error bars (blue) are the measurement from the data; the black curves with shaded gray regions indicate the $1\sigma$ standard deviation from the Abacus simulations with altMTL fiber assignment scheme. The bottom panel shows the arrangement of the three radial bins.}
    \label{fig:4PCF_Iron_AbacusAltmtl_SGC_6ells_odd}
\end{figure*}

We apply this harmonic-based estimator to measure the 4PCF from the DESI DR1 LRG sample~\footnote{We use the GPU version of the public code~\url{https://github.com/oliverphilcox/encore} with extensions including $\ell_{\rm max}$ for the 4PCF measurement}. For each galaxy, we apply default weights following~\cite{Hou2025:DESIeven4PCF, DESI2024:SampleDefinition}, which include the completeness weight, systematic weights, and redshift failure weights. {For fiber-assignment incompleteness, we use the fiducial DESI DR1 completeness scheme, namely $w_{\rm comp}=1/f_{\rm TLID}$ for the data and an $f_{\rm tile}$ correction applied to the random catalog.}
Fig.~\ref{fig:4PCF_Iron_AbacusAltmtl_NGC_6ells_odd} and Fig.~\ref{fig:4PCF_Iron_AbacusAltmtl_SGC_6ells_odd} show the measurement of the parity-odd 4PCF coefficients $\zeta_{\ell_1\ell_2\ell_3}(r_1,r_2,r_3)$ using DESI-DR1 LRG sample in the NGC and SGC, respectively. We use radial scales between $20\,\mpch < r < 160\,\mpch$, divided into ten bins. The top two panels show the 4PCF coefficients $\zeta_{\ell_1\ell_2\ell_3}(r_1,r_2,r_3)$, weighted by the product of $r_1r_2r_3$, as a function of tetrahedron index. Points with error bars are the measurement from the data; the black curves with shaded gray regions indicate the $1\sigma$ standard deviation from the Abacus simulations with altMTL fiber assignment scheme~(see \S\ref{sec:simulations}). The simulations and fiber assignment schemes are described below. The bottom panel shows the arrangement of the three radial bins. {The radial bin index can be expressed as $i = i_1(n-i_2+1) + (i_2-1)$, with $i_1$ and $i_2$ being the first and second bin indices, and $n$ being the total number of radial bins.}

\subsection{Simulations for DR1}
\label{sec:simulations}
Two types of simulations are used for DESI DR1: a suite of 1000 fast simulations, \ezmock, which employs approximate methods for nonlinear gravitational evolution and galaxy bias~\cite{Chuang2015:EZmock,Zhao2021:EZmock}; and a suite of 25 $N$-body simulations, \abacus~\cite{Garrison2019:Abacus,Maksimova2021:Abacus}, in which galaxies are populated via a halo occupation distribution model that describes the galaxy–halo connection~\cite{Yuan2024:DESIAbacusHODLRG}.
The \ezmock\ simulations have a box size of $L_{\rm box} = 6\,\gpch$, which is large enough to be directly post-processed to match the DESI DR1 geometry. In contrast, the \abacus\ mocks have a smaller box size of $L_{\rm box} = 2\,\gpch$, which is insufficient to cover the entire DESI DR1 LRG volume. To compensate, the \abacus\ boxes are tiled to fill the survey footprint. However, this tiling does not introduce new independent modes, resulting in an overestimation of the cosmic variance. We later refer this to as the {\it volume effect}.

Both \ezmock\ and \abacus\ simulations are post-processed into three variants that differ in their fiber-assignment implementation. The ``complete'' mocks represent the parent sample, {\it i.e.} all galaxies that could have been targeted~\footnote{For the \abacus-Complete mocks, we additionally downsample it to match the number density of the altMTL mocks, so that their statistical fluctuations are comparable as those in the real data}. The ``alternative MTL'' (altMTL) mocks are the closest to the survey procedure, since they are generated by running the DESI \textsc{fiberassign} code~\cite{Lasker2025:DESIFiberAssign} on tiles in the same ordering and cadence as the data, including the feedback loop to the target list. The ``fast fiber assignment'' (FFA) mocks emulate the fiber assignment process, which necessarily results in incompleteness,  by repeatedly sampling from the average targeting probability of galaxies. This probability is learned from the data as a function of the number of overlapping tiles and the local angular clustering. We later refer this to as the {\it fiber assignment effect}. 

\begin{table}[t]
\centering
\caption{A brief overview of the available simulations. For \abacus, both fiber assignment schemes are available, while \ezmock\ is only implemented with FFA. In addition, \abacus\ is subject to the volume replication effect and requires additional volume correction, whereas \ezmock\ uses a sufficiently large simulation box and does not require such a correction.}
\label{tab:mock}
\begin{tabular}{@{}C{1cm}C{1.5cm}C{2.5cm}C{2.5cm}C{2cm}@{}}
\toprule
Mock & $N_{\rm sim}$ & \makecell[c]{Fiber\\Assign} & \makecell[c]{Volume\\Replication} \\ \midrule
\abacus & 25 & FFA \& altMTL & Yes \\ [3ex]
\ezmock & 1000 & FFA & No \\
\bottomrule
\end{tabular}
\label{tab:mock}
\end{table}

Both the volume effect and the fiber assignment effect are crucial for obtaining a reliable detection significance, and require additional calibration when computing both the auto-correlation and cross-correlation statistics. We will return to the details of these calibrations when presenting our results in \S\ref{sec:Result}. In Table~\ref{tab:mock} we provide a brief summary of the mocks.

\section{Analysis Methodology}
\label{sec:methodology}

\noindent{\it \textbf{Motivation} \,\,}
To quantify the parity-odd signal, we apply the null hypothesis by assuming the signal is absent from the data, then measure the test statistics from the data to determine to what extent we can reject this null hypothesis.

As previously discussed in~\cite{hou2022:parity}, this test is highly sensitive to the covariance estimation for two related reasons. For a $\chi^2$ statistic, the contributions are quadratic and that both covariance underestimation and coherent systematic biases can add to the total $\chi^2$. Their impact can accumulate across a high-dimensional data vector. Moreover, since we make no assumptions about the configuration dependence of the signal, we may include redundant or un-informative degrees of freedom but increase the dimensionality of the test statistic, further amplifying the cumulative impact of small mismodelling.
Moreover, the simulations used for DESI DR1 are not tuned to match higher-order clustering, which may cause them to underestimate nonlinear gravitational effects. This, in turn, can produce a data–simulation mismatch in the covariance, including in the parity-odd sector.

To mitigate these issues, we incorporate two additional tests. In the first approach, before computing $\chi^2$, we reduce the dimensionality of the 4PCF by selecting from among the 2760 eigenvectors of the analytic covariance matrix those with the lowest noise, that is, with the lowest eigenvalues. In the second approach, following~\cite{Krolewski2024:parity}, we perform two types of cross-correlation tests, hereafter referred to as {\it Type-I} and {\it Type-II}, which isolate the deterministic and stochastic components of the measured statistics, respectively.

\subsection{Defining Auto-correlation and Cross-correlation}

The conventional way to proceed is to map the 4PCF to the scalar quantity $\chi^2$, what we refer to as the {\it auto-correlation} test
\begin{eqnarray}
\label{eqn:chi2-auto}
    \chi^2 = \hat\zeta^{i}\, \mathbb{\hat C}^{-1}_{ij}\, \hat\zeta^{j},
\end{eqnarray}
where $\mathbb{\hat C}$ is an estimate of the true covariance matrix. It can be either derived from simulation or analytic calculation~({\it c.f.} \S\ref{sec:C_analyt}). $\hat\zeta$ is the measured 4PCF from the data, which potentially has the underlying parity-violating signal, $\zeta_{\rm pv}$. The components of the data vector, $\zeta_{\ell_1 \ell_2 \ell_3}(r_1,r_2,r_3)$, are labeled $\zeta^i$. Throughout the paper, we use $\ \hat{\ }\ $ to denote an estimated or measured quantity that includes statistical fluctuations, in contrast to a deterministic term such as a hypothesized signal.

The signal-sensitive {\it Type-I} cross-statistic $\chi^2_{\times}$ is an average over $\calN_\times$ pairs, follows~\cite{Krolewski2024:parity} and was also used in~\cite{Hou2025:DESIeven4PCF}:
\begin{eqnarray}\label{eqn:chi2-cross}
\chi^2_{\times} = \frac{1}{\calN_\times}\sum_{\mu<\nu} \hat\zeta^{\mu,i}\, \mathbb{\hat C}^{-1}_{ij} \, \hat\zeta^{\nu,j},  
\end{eqnarray}
where $\mu$ and $\nu$ indicate patches, that is subregions, of the sky with the normalization factor
\begin{eqnarray}
\calN_\times \equiv N_p (N_p-1)/2,
\end{eqnarray}
with $N_p$ being the number of patches. The normalization is given by the combinatorial factor. {We expect the Type-I cross-statistic to have contributions only from signal: the noise should be uncorrelated between spatially separated patches, whereas the signal is common to all patches~\footnote{For the cross-correlation test, we assume that the signal is shared across different patches. However, if the signal peaks in the squeezed limit on large scales, it may extend beyond the coherence length ($>500\mpch$) of the sky patches considered in this work, in which case the cross-correlation test may not apply.}.}

In addition, we also consider the noise-sensitive {\it Type-II} cross-statistic~\cite{Krolewski2024:parity}, where the potential common signal is removed by taking the difference of two patches
\begin{eqnarray}\label{eqn:chi2-Delta}
\chi^2_{\Delta} = \frac{1}{\calN_\Delta} \sum_{\mu<\nu} (\hat\zeta^{\mu}-\hat\zeta^{\nu})^i\, \mathbb{\hat C}^{-1}_{ij} \, (\hat\zeta^{\mu}-\hat\zeta^{\nu})^j,   
\end{eqnarray}
with $\calN_{\Delta}$ being the normalization factor, which is defined such that the expectation value of the Type-II cross statistics equals the number of degrees of freedom~\cite{Krolewski2024:parity,Hou2025:DESIeven4PCF},
\begin{eqnarray}\label{eqn:normalization}
    \calN_\Delta \equiv (N_p-1)\sum_{\mu}\frac{V_{\rm eff}}{V_{\rm eff}^{\mu}}.
\end{eqnarray}
with the effective volume being
\begin{eqnarray}\label{eqn:V_eff}
    V_{\rm eff} \equiv \frac{V_{\rm fid} N_{\rm dof}}{{\rm Tr}\left[\mathbb{C}^{-1}_{\rm th}\, \mathbb{\hat C}_{\rm mock}\right]},
\end{eqnarray}
where $V_{\rm fid}$ is the fiducial volume of the survey, obtained by fitting the analytic covariance to the \ezmock. $N_{\rm dof}$ is the number of degrees of freedom for the data vector. $\mathbb{C}_{\rm th}$ is the analytic covariance~\cite{Hou2022:AnalytCov}. For each patch, we define the effective volume as $V_{\rm eff}^{\mu} \equiv V_{\rm fid} N_{\rm dof} / {\rm Tr}\left[\mathbb{C}^{-1}_{\rm th} \mathbb{\hat C}^{\mu}_{\rm mock}\right]$, with $\mathbb{\hat C}^{\mu}_{\rm mock}$
being the covariance for the patch $\mu$.
If the analytic covariance accurately approximates the mock covariance ${\mathbb{\hat C}}_{\rm mock}$, then $V_{\rm eff} \approx V_{\rm fid}$. However, differences in number density between patches can lead to deviations.

\subsection{Statistical Properties of the Correlation Tests}
Since our goal is to perform a null-hypothesis test, we focus on comparing the data with the distribution obtained from the mocks. When an analytic prediction for the mocks' distribution is available, we also compare the data against this theoretical expectation. To enable these comparisons, we compute the mean and variance of the relevant statistics. Below, we briefly review the main results as previously derived in~\cite{Krolewski2024:parity, hou2022:parity}. 

We assume that the observed parity-odd 4PCF is composed of the following terms
\begin{eqnarray}\label{eqn:zeta_3_term}
    \hat\zeta = \zeta_{\rm pv} + \hat\zeta_{\rm s} + \hat\epsilon,
\end{eqnarray}
where $\zeta_{\rm pv}$ represents the parity-violating signal, and $\hat{\epsilon}$ corresponds to statistical fluctuations arising from cosmic variance. $\hat\zeta_{\rm s}$ accounts for systematic effects. Thus the covariance due to cosmic variance is given by $\av{{\hat\epsilon}_i{\hat\epsilon}_j}$. While in~\cite{hou2022:parity}, we discussed various scenarios and concluded most systematics do not contaminate parity-odd 4PCF at the signal level,  there could still be residual systematics.  Moreover, the stochastic component of the systematics does not necessarily vanish. The expectation value of the data $\chi^2_{\rm data}$ thus reads  

\begin{eqnarray}
    \av{\chi^2_{\rm data}} &=& \av{\hat\zeta^i\, \mathbb{\hat C}^{-1}_{ij}\, \hat\zeta^j}\\
    &=& \mathbb{\hat C}^{-1}_{ij}\left(\zeta_{\rm pv}^i\zeta_{\rm pv}^j + 2\zeta_{\rm s}^i\zeta_{\rm pv}^j+  \zeta_{\rm s}^i\zeta_{\rm s}^j\right) + {\rm Tr}\left[\mathbb{\hat C}^{-1} \mathbb{C}_{\rm data}\right]\nonumber,
\end{eqnarray}
where $\mathbb{\hat C}^{-1}_{ij} \zeta_{\rm s}^i\zeta_{\rm pv}^j = \mathbb{\hat C}^{-1}_{ij} \zeta_{\rm pv}^i\zeta_{\rm s}^j$, given that the covariance matrix is symmetric $\mathbb{C}_{ij} = \mathbb{C}_{ji}$. The data covariance $\mathbb{C}_{\rm data}$ represents the fluctuations arising from cosmic variance and statistical noise. While a parity-violating signal could in principle contribute to the data covariance, it is not expected to appear at leading order when the parity-odd signal has low signal-to-noise ratio. 

The mocks do not have parity violation and the $\zeta_{\rm pv}$-related terms drop:
\begin{eqnarray}
    \av{\chi^2_{\rm mock}} &=& \av{\hat\zeta^i\, \mathbb{\hat C}^{-1}_{ij}\, \hat\zeta^j}\\
    &=& \mathbb{\hat C}^{-1}_{ij}\left(\zeta_{\rm s}^i\zeta_{\rm s}^j\right) + {\rm Tr}\left[\mathbb{\hat C}^{-1} \mathbb{C}_{\rm mock}\right]\nonumber.
\end{eqnarray}
The difference between the $\chi^2$ from the data and the mocks is given by 
\begin{eqnarray}\label{eqn:chi2_avg_diff_data_mock}
&&\av{\chi^2_{\rm data}} - \av{\chi^2_{\rm mock}}\\ 
&=& \av{\hat\zeta_{\rm data}^i (\mathbb{\hat C}^{-1})_{ij} \hat\zeta_{\rm data}^j}-\, \av{\hat\zeta_{\rm mock}^i (\mathbb{\hat C}^{-1})_{ij} \hat\zeta_{\rm mock}^j}\nonumber\\
&\equiv& \mathbb{\hat C}^{-1}_{ij} \left(\zeta_{\rm pv}^i\zeta_{\rm pv}^j + 2\zeta_{\rm s}^i\zeta_{\rm pv}^j+  \Delta_{\rm ss}^{ij}\right) +\, {\rm Tr}\left[\mathbb{\hat C}^{-1} (\mathbb{C}_{\rm data} - \mathbb{C}_{\rm mock})\right], \nonumber
\end{eqnarray}
with the residual difference in cross-correlation products between data and mock catalogs $\Delta^{ij}_{\rm ab}$ defined as

\begin{eqnarray}\label{eqn:Delta_ab}
    \Delta_{\rm ss}^{ij} \equiv \zeta^{{\rm data},i}_{\rm s} \zeta^{{\rm data},j}_{\rm s} - \zeta^{{\rm mock},i}_{\rm s} \zeta^{{\rm mock},j}_{\rm s}.
\end{eqnarray}
From Eq.~\eqref{eqn:chi2_avg_diff_data_mock}, the difference between $\chi^2$ evaluated on data and on the mocks can arise from a parity-violating signal, from systematics, or from a mismatch between data and mocks at the noise level. These contributions map onto the two types of cross-correlations, which allow us to isolate the signal and the noise-level data–mock mismatch.

Under the assumption that the different sky patches are statistically independent, the expectation value of the signal-sensitive statistic $\av{\chi^2_{\times, {\rm data}}}$ is
\begin{eqnarray}\label{eqn:chi2x_data_avg}
    &&\av{\chi^2_{\times, {\rm data}}}\nonumber\\
    &=& \frac{1}{\calN_\times}\sum_{\mu<\nu} \av{\hat\zeta^{\mu, i}\, \mathbb{\hat C}^{-1}_{ij} \, \hat\zeta^{\nu,j}}\\
    &=& \mathbb{\hat C}^{-1}_{ij}\Bigg\{\zeta_{\rm pv}^i  \zeta_{\rm pv}^j + \frac{1}{\calN_\times}\sum_{\mu<\nu} \left(\zeta_{\rm pv}^{\mu, i}  \zeta_{\rm s}^{\nu,j} +\zeta_{\rm s}^{\mu,i}  \zeta_{\rm pv}^{\nu,j} + \zeta_{\rm s}^{\mu,i}  \zeta_{\rm s}^{\nu,j}\right)\Bigg\}.\nonumber
\end{eqnarray}
Here we presume that the parity-violating signal is identical across different sky regions, while allowing the systematics in the $\mu$-th and $\nu$-th patches to differ in general~\footnote{For parity-violating models with squeezed signal leading to local modulation, the assumption that $\zeta^{\mu}=\zeta^{\nu}$ may not hold.}.
Compared to the derivation in~\cite{Krolewski2024:parity}, the additional systematics-induced term can potentially contaminate the cross-statistic. Notably, the cross-correlation between the signal and the systematics, or between the systematics in different subregions, may lead to a negative correlation. 

The expectation value of the signal-sensitive statistic for mocks reads

\begin{eqnarray}\label{eqn:chi2x_mock_avg}
    \av{\chi^2_{\times, {\rm mock}}} &=& \frac{1}{\calN_\times}\sum_{\mu<\nu} \av{\hat\zeta^{\mu, i}\, \mathbb{\hat C}^{-1}_{ij} \, \hat\zeta^{\nu,j}}\\
    &=& \mathbb{\hat C}^{-1}_{ij}\Bigg\{ \frac{1}{\calN_\times}\sum_{\mu<\nu} \zeta_{\rm s}^{\mu,i}  \zeta_{\rm s}^{\nu,j} + \av{\hat\epsilon^{\mu, i}\hat\epsilon^{\nu, j}}\Bigg\}.\nonumber
\end{eqnarray}
The Type-I cross-correlation can in principle be non-zero in presence of systematics, but as we will show later we did not find clear evidence of contributions from systematics due to fiber assignment at the signal level.
By contrast, the second term arises from correlations in the the cosmic variance $\epsilon$ across different sky patches in mocks. In typical situations, the cosmic variance across patches is uncorrelated and one does not expect this term, {\it i.e.} $\av{\hat\epsilon^{\mu, i}\hat\epsilon^{\nu, j}}=0$ for $\mu\neq \nu$. While it is generally reasonable to assume statistical independence between different sky patches in the data, this assumption does not always hold for the mocks. In such cases, both types of cross-correlation tests may be affected, as we will demonstrate in \S\ref{sec:Result}.

Comparing the difference between  $\chi^2_{\times}$ for data and mocks, we arrive at the following:
\begin{eqnarray}\label{eqn:chi2x_avg_diff_data_mock}
    &&\av{\chi^2_{\times, {\rm data}}} - \av{\chi^2_{\times, {\rm mock}}} \nonumber \\
    &\equiv& \mathbb{\hat C}^{-1}_{ij} \Bigg(\zeta^i_{ {\rm pv, data}} \zeta^j_{{\rm pv, data}} + \tilde\Delta_{\rm ss}^{ij}\nonumber\\
    &&\,+ \frac{1}{\calN_\times}\sum_{\mu<\nu} \left(\zeta^{\mu,i}_{{\rm pv, data}} \zeta^{\nu,j}_{s, {\rm data}} + \zeta^{\mu,i}_{s, {\rm data}} \zeta^{\nu,j}_{{\rm pv, data}}\right)
    \Bigg),
\end{eqnarray}
where the difference between the mean in the data and mocks of the deterministic-component-sensitive cross-correlation indicates the parity-violating signal.
We have defined the patch-wise residual difference in cross-correlation products between data and mock catalogs $\tilde\Delta_{\rm ss}^{ij}$ by
\begin{eqnarray}
\tilde\Delta_{\rm ss}^{ij} \equiv \frac{1}{\calN_\times}\sum_{\mu<\nu} \zeta^{\rm data,\mu}_{\rm s} \zeta^{\rm data, \nu}_{\rm s} - \zeta^{\rm mock,\mu}_{\rm s} \zeta^{\rm mock, \nu}_{\rm s}.      
\end{eqnarray}
When different sky patches share the identical systematics, Eq.~\eqref{eqn:chi2x_avg_diff_data_mock} reduces to the first term of Eq.~\eqref{eqn:chi2_avg_diff_data_mock}. However, with position-dependent systematics, auto- and cross-statistics will yield different results, since auto-correlation statistics compute the 4PCF with systematics applied globally across the entire survey volume, whereas
cross-statistics compute the 4PCF with systematics applied separately to each individual patch before combining results. 
Finally, the expectation value of the noise-sensitive statistic is given as 

\begin{eqnarray}\label{eqn:chi2_Delta_tot_avg}
    \av{\chi^2_{\Delta}} &=& \frac{1}{\calN_\Delta}\sum_{\mu<\nu} \av{ \left(\hat\zeta^{\mu}-\hat\zeta^{\nu}\right)^i\, \mathbb{\hat C}^{-1}_{ij} \, \left(\hat\zeta^{\mu}-\hat\zeta^{\nu}\right)^j}\nonumber\\
    &=&\frac{1}{\calN_\Delta} \mathbb{\hat C}^{-1}_{ij} \sum_{\mu<\nu} \av{\hat\zeta^{\mu,i}\hat\zeta^{\mu,j}}+\av{\hat\zeta^{\nu,i}\hat\zeta^{\nu,j}}\nonumber\\
    &&\hspace{15mm}- \left(\av{\hat\zeta^{\mu,i}\hat\zeta^{\nu,j}}+\av{\hat\zeta^{\nu,i}\hat\zeta^{\mu,j}}\right)\nonumber\\
    &=& \av{\chi^2_{{\Delta_{0}}}} - \frac{2{\cal N}_\times}{\calN_{\Delta}}\av{\chi^2_{\times}},
\end{eqnarray}
where first term $\chi^2_{{\Delta_{0}}}$ captures the   covariance mismatch between the data and the mocks with the expectation value given by 
\begin{eqnarray}\label{eqn:chi2_Delta_avg}
    \av{\chi^2_{{\Delta_{0}}}} 
    &=& \frac{1}{\calN_\Delta}\left\{(N_p-1)\sum_\mu {\rm Tr}\left[\mathbb{\hat C}^{-1} \mathbb{C}^{\mu}\right] \right.\nonumber\\
    &&\,\,\left.+ \sum_{\mu<\nu} \Delta\zeta_s^{\mu\nu,i} \mathbb{\hat C}^{-1}_{ij} \Delta\zeta_s^{\mu\nu,j}\right\},
\end{eqnarray}
where $\mathbb{C}^{\mu}$ is the covariance for the $\mu$-th subpatch. We keep the potential difference in the systematics-induced bias across different patches $\Delta\zeta_{\rm s}^{\mu\nu,i}\equiv \zeta_{\rm s}^{\mu,i}-\zeta_{\rm s}^{\nu,i}$~\footnote{To obtain the second line of Eq.~\eqref{eqn:chi2_Delta_avg}, we used the definition of the trace $\mathrm{Tr}\left[\mathbb{\hat C}^{-1} \mathbb{\hat C}^{\mu}\right]=\sum_i \left[\mathbb{\hat C}^{-1} \mathbb{\hat C}^{\mu}\right]_{ii} = \sum_{ij} (\mathbb{\hat C}^{-1})_{ij} \mathbb{\hat C}^{\mu}_{ij}$.}.
We typically do not expect the second term in Eq.~\eqref{eqn:chi2_Delta_tot_avg} involving $\chi^2_{\times}$ to contribute under the null hypothesis. For the mocks, in the absence of parity-violating signals and assuming that different sky patches are statistically independent, the expected Type-I cross-correlation is zero. For the data, if we adopt the prior that the signal should be positively correlated and that systematics are negligible, the Type-I cross-correlation should be non-negative within the statistical uncertainty.

The difference in the pure data-mock between  $\chi^2_{{\Delta_0}}$ in statistical fluctuation is given as follows:
\begin{eqnarray}
    &&\av{\chi^2_{{\Delta_0}, {\rm data}}} - \av{\chi^2_{{\Delta_0},{\rm mock}}} \nonumber \\
    &=& \frac{N_p-1}{\calN_\Delta} \sum_\mu {\rm Tr}\left[\mathbb{\hat C}^{-1} (\mathbb{C}^{\mu}_{\rm data} - \mathbb{C}^{\mu}_{\rm mock})\right] \\
    &&\,+ \mathbb{\hat C}^{-1}_{ij}\sum_{\mu<\nu} \left[\Delta\zeta_s^{\mu\nu,i}  \Delta\zeta_s^{\mu\nu,j}\Big|_{\rm data} - \Delta\zeta_s^{\mu\nu,i} \Delta\zeta_s^{\mu\nu,j}\Big|_{\rm mock}\right], \nonumber
\end{eqnarray}
which quantifies the mismatch between the statistical fluctuation of the data and mocks, as well as the potential contribution due to the systematics.

As discussed above, the auto correlation is sensitive to data-mock covariance mismatch. To isolate the potential parity-violating signal from the auto correlation,  we introduce a {\it corrected} auto correlation $\chi^2_{\rm data, c}$ by subtracting the data-mock covariance mismatch $\chi^2_{\Delta_0}$ introduced in Eq.~\eqref{eqn:chi2_Delta_tot_avg} from the auto correlation for the $\mu$-th patch $\chi^2_{\rm data,\mu}$. The corrected auto correlation reads
\begin{eqnarray}\label{eqn:chi2_delta_c}
\chi^2_{\rm data, c} &\equiv& \sum_{\mu}^{N_p}\chi^2_{\rm data,\mu} - N_p \left({\chi^2_{{\Delta_{0}}, \rm data}} - {\chi^2_{{\Delta_0}, \rm mock}} \right)\nonumber\\
&=&\sum_{\mu}^{N_p}\chi^2_{\rm data,\mu} - N_p\Big({\chi^2_{\Delta, {\rm data}}} + \frac{2\calN_\times}{\calN_{\Delta}} {\chi^2_{\times, \rm data}}\\
&&\quad- \av{\chi^2_{\Delta, {\rm  mock}}} - \frac{2\calN_\times}{\calN_{\Delta}} \av{\chi^2_{\times, \rm mock}}\Big),\nonumber
\end{eqnarray}
where the auto-correlation in the first term can be a summed contribution from $N_p$ sky patches. Accordingly, when subtracting the data–mock covariance mismatch, we also account for this $N_p$ prefactor.

In the following \S\ref{sec:Result}, we will perform analyses using both auto correlation $\chi^2$ and two types of cross correlation $\chi^2_\times$ and $\chi^2_\Delta$. Since the auto-correlation can be sensitive to data–mock covariance mismatch, the cross-correlations provide a complementary diagnostic: $\chi^2_\times$ isolates deterministic contributions such as a parity signal, whereas $\chi^2_\Delta$ isolates stochastic components such as noise, allowing us to disentangle their contributions to the overall significance. 

\section{Results}
\label{sec:Result}

In this section, we present the detection significance based on the full auto-correlation statistic $\chi^2$, as defined in Eq.~\eqref{eqn:chi2-auto} as well as the two types of cross correlations. We first present results based on the analytic covariance for all three statistics, then repeat the analysis with the hybrid covariance. With the analytic covariance, we additionally test sensitivity to sky dependence. With the hybrid covariance, we assess scale dependence.

Unless otherwise stated, as a default choice, we consider 23 combinations of $\ell_1, \ell_2,$ and $\ell_3$, and 120 combinations of $r_1, r_2,$ and $r_3$, yielding a data vector of dimensionality 2760. 

\subsection{Analytic Covariance}
\label{sec:C_analyt}
The analytic covariance used in this paper is computed following~\cite{Hou2022:AnalytCov}. We fit the analytic covariance for the number density and the volume for each sky patch to \ezmock. A summary of the fitted fiducial volume $V_{\rm fid}$ and effective number density $\bar{n}_g$ can be found in Table~\ref{tab:statistic}.
\vspace{5mm}

\begin{figure}
    \centering
    \includegraphics[width=0.85\linewidth]{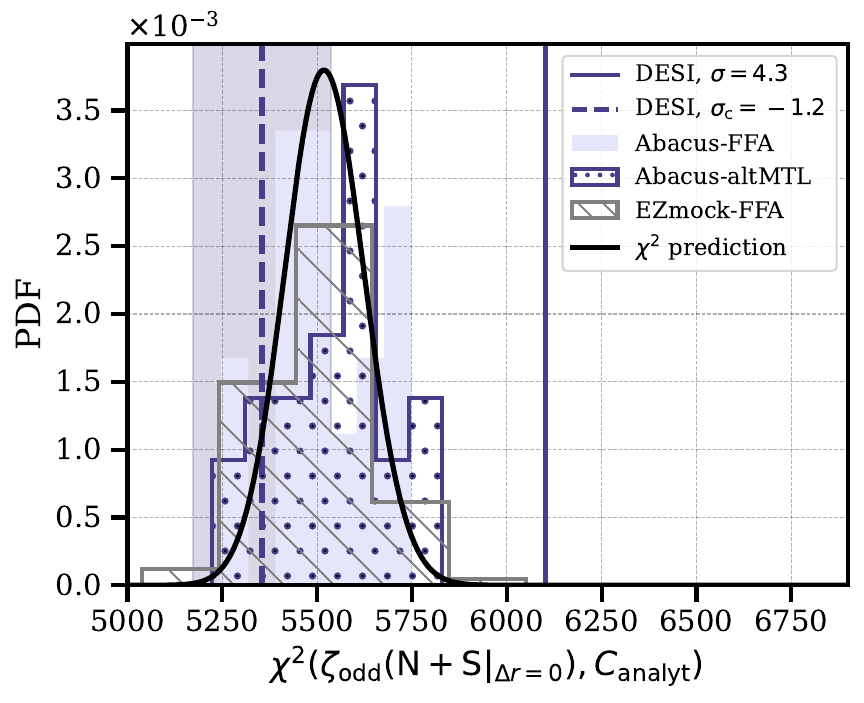}
    \caption{Distribution of the auto-correlation $\chi^2$ for the parity-odd connected 4PCF, using the full DESI DR1 LRG sample and combining the NGC and SGC, with each using the corresponding covariance. The panel shows a comparison between simulations and data, as well as the detection significance of the parity-odd 4PCF using the analytic covariance matrix.  
    We used the range $20\, \mpch <r<160\,\mpch$ without additional radial cuts ($\Delta r=0$). The vertical lines indicate the statistics for the DESI data. The dashed line represents the correction derived from the cross-statistics in Eq.~\eqref{eqn:chi2_delta_c}, with the uncertainty shown as the shaded region, estimated from the \ezmock. The filled colored histograms correspond to \abacus-FFA; the circled histogram corresponds to \abacus-altMTL; the hashed histogram corresponds to \ezmock-FFA. The black curves show the theoretical $\chi^2$ distribution. Additionally, the distribution of the altMTL mocks and the data have been calibrated by multiplying by the ratios deriving from the fiber implementations in the FFA and altMTL mocks, whereas the \abacus\ mocks are rescaled for the volume effect.}
    \label{fig:chi2_analyt_NS_odd_dr0}
\end{figure}

\noindent{\it \textbf{Auto-Correlation with $\mathbb{C}_{\rm analyt}$} \,\,}
We start the discussion with the auto correlation tests. Fig.~\ref{fig:chi2_analyt_NS_odd_dr0} presents $\chi^2$ for the full DESI DR1 LRG sample, combining both the NGC and SGC regions. Here we use the analytic covariance matrix tuned to \ezmock\ with the range $20\, \mpch <r<160\,\mpch$  using ten equally spaced radial bins. This range and spacing is used in all the following analyses~\footnote{We use \textsc{scipy.linalg.inv} for matrix inversion, which relies on \textsc{LAPACK Fortran} library. For symmetric positive-definite matrices, it uses Cholesky-based algorithm and is generally backward-stable for a $\calO(10^3)\times\calO(10^3)$ matrix. Numerically, we found the condition number defined by the ratio of the maximum to the mininum eigenvalues to be $\lambda_{\rm max} / \lambda_{\rm min}\approx \calO(10^4)$ and the relative matrix reconstruction error $\|\mathbb{C}^{-1} \mathbb{C} - \mathbb{I}\| / \| \mathbb{I} \|\approx \calO(10^{-14})$. For readers interested in the structure of the normalized correlation matrix, we note that it is visually identical to Figs. 7 and 8 of~\cite{hou2022:parity}}. 
The histograms represent the distributions from the \ezmock\ (hatched grey), \abacus-FFA (filled purple), and \abacus-altMTL (circled purple).
The solid vertical line shows the measurement from the DESI data. The dashed line represents the correction derived from the cross-statistics in Eq.~\eqref{eqn:chi2_delta_c}, with the uncertainty shown by the shaded region, estimated from the $1\sigma$ deviation of the \ezmock.

For the mocks, there are two major effects to consider: the fiber assignment effect and the volume effect. As we discussed in \S\ref{sec:simulations}, the FFA fiber assignment scheme used in the \ezmock\ mocks tends to underestimate the covariance matrix. To address this, we compute the ratio of the traces of the covariance matrices from the \abacus\ mocks with two different fiber assignment schemes, and use this ratio to correct the covariance estimate, $R={\text{Tr}\left[\mathbb{\hat C}_{\rm FFA}\right]}/{\text{Tr}\left[\mathbb{\hat C}_{\rm altMTL}\right]}$. Accordingly, the $\chi^2$ obtained with the altMTL fiber assignment needs to be rescaled by $R$ to account for the difference in fluctuations between the two fiber-assignment schemes. The ratio is around 0.93 for NGC and 0.81 for SGC (see also Fig.~\ref{fig:ratio_cov_NS_odd_abacus} in Appendix \ref{sec:incomplete_stat_fluctuations}).  

The volume effect applies only to the \abacus\ mocks. While the replication scheme allows covering the full DESI DR1 geometry, it does not introduce new, independent $k$-modes, so it leads to an overestimation of the fluctuations. We correct for this by rescaling with the ratio of the effective volume of the \abacus-FFA covariance to that of the analytic covariance tuned to match the \ezmock-FFA, $V^{\rm abcs}_{\rm eff} / V_{\rm eff}$, where $V^{\rm abcs}_{\rm eff}$ and $V_{\rm eff}$ are given in Table~\ref{tab:statistic}.


For the data measurement, as discussed in \S\ref{sec:methodology}, the auto-correlation receives deterministic contributions from underlying signals or systematics, as well as stochastic contributions arising from mismatches in the statistical fluctuations between the data and the mocks. In particular, the latter could be misinterpreted as a genuine detection. Therefore, here we apply the correction derived from the cross statistic to correct for the auto correlation given in Eq.~\eqref{eqn:chi2_delta_c}. The width of the shaded region around the dashed vertical line in Fig.~\ref{fig:chi2_analyt_NS_odd_dr0} is derived from the \ezmock. We compare the data to the \abacus-altMTL mock to quantify the data-mock covariance mismatch and set $\av{\chi^2_{\times, \rm mock}}=\av{\chi^2_{\times, \rm abcs-altmtl}}$. Here we also consider the volume replication effect and absorb it into the normalization coefficient $\calN_{\Delta, {\rm abcs}}=\sum_\mu V_{\rm eff} / V_{\rm eff}^{\mu, {\rm abcs}}$ (see Table~\ref{tab:statistic}).
If the mocks accurately described the intrinsic fluctuations in the data, the second term in Eq.~\eqref{eqn:chi2_delta_c} vanishes. In the presence of a data–mock covariance mismatch, we find that the inferred auto-correlation amplitude is reduced.


\begin{figure*}
    \centering
    \begin{subfigure}[b]{0.31\textwidth}
    \includegraphics[width=0.9\linewidth]{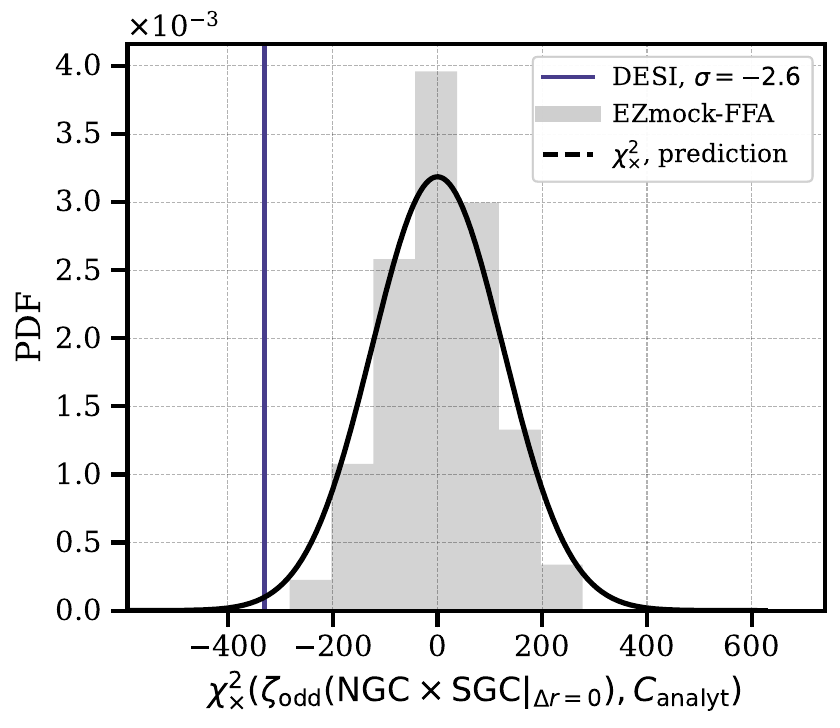}
    \end{subfigure}
    \begin{subfigure}[b]{0.3\textwidth}
    \includegraphics[width=0.9\linewidth]{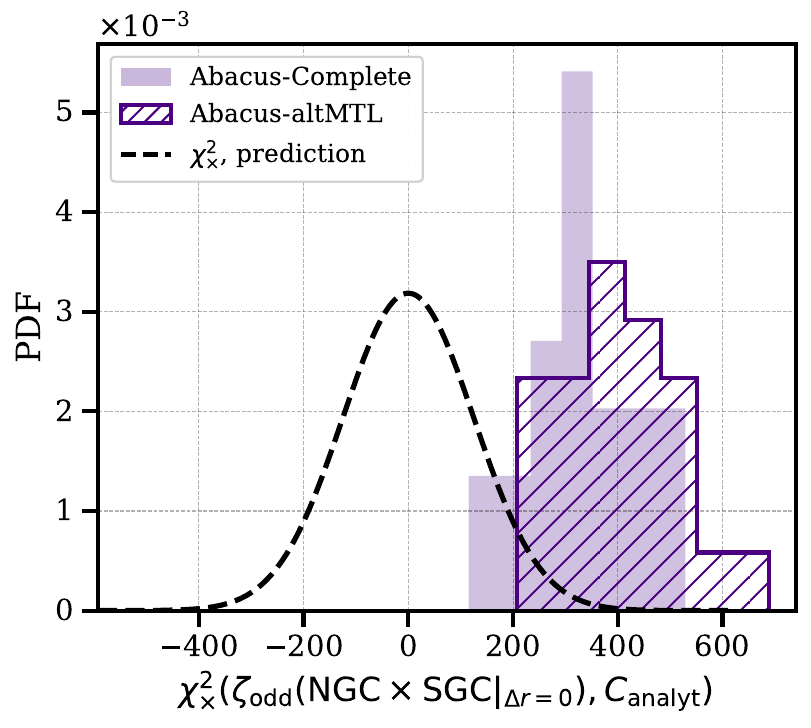}
    \end{subfigure}
    \begin{subfigure}[b]{0.3\textwidth}
    \includegraphics[width=0.9\linewidth]{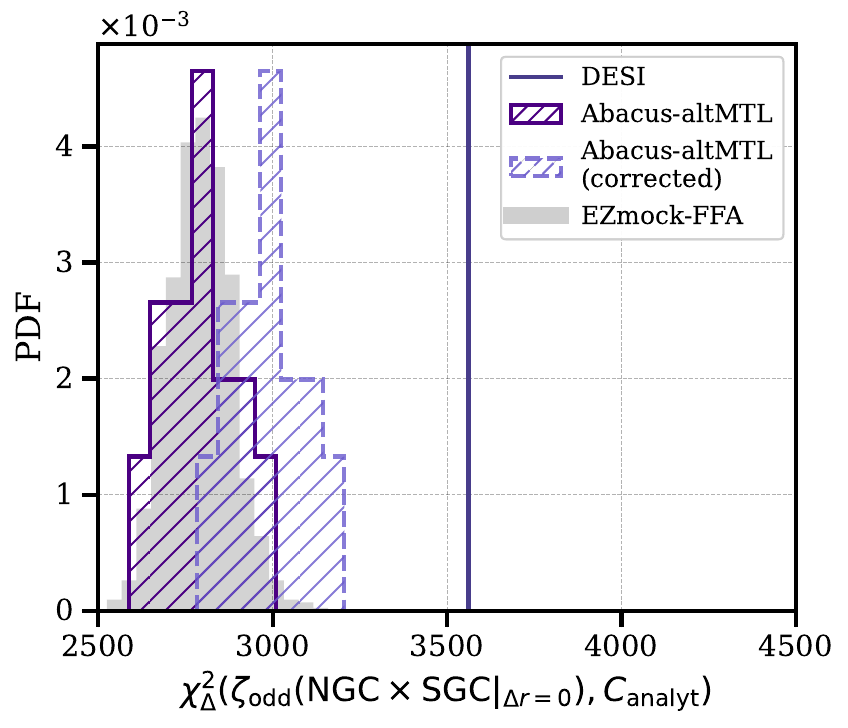}
    \end{subfigure}
    \caption{Distribution of both types of cross-correlation statistics ($\chi^2_\times$ (left and middle) and $\chi^2_{\Delta}$  (right)) for the parity-odd 4PCF between the NGC and SGC of the DESI DR1 LRG sample. This figure compares simulations with data using the analytic covariance matrix. For all the panels, the filled grey histograms are for the \ezmock-FFA, the purple histograms are for the \abacus\ mocks, and the black solid or dashed curves denote the theoretical $\chi^2_\times$ distribution, modeled as a Gaussian centered at zero with the width predicted by Eq.~\eqref{eqn:var_chi2x}. {\it Left:} The Vertical line indicates the DESI data statistics, which is consistent with zero with a negative value. {\it Middle:} filled purple histograms correspond to the \abacus-Complete mocks, and the hatched histogram represents the \abacus-altMTL mocks. We find that the \abacus\ mocks do not center around zero due to the limited volume effect (for more discussion see below). {\it Right:} The Vertical line again indicates the DESI data statistics. The filled grey histograms correspond to the \ezmock-FFA, while the purple histograms show the \abacus\ mocks. The dark hatched histogram is for \abacus-altMTL without applying correction. The light dashed histogram is for the corrected \abacus-altMTL using Eq.~\eqref{eqn:chi2_Delta_tot_avg} given the shifted \abacus\ center in the Type-I cross correlation shown in the middle panel.}
    \label{fig:chi2_x_analyt_NS_odd_dr0}
\end{figure*}

\vspace{5mm}

\noindent{\it \textbf{Cross Correlation $\mathbb{C}_{\rm analyt}$} \,\,}
Fig.~\ref{fig:chi2_x_analyt_NS_odd_dr0} shows the two types of cross correlations for the DESI DR1 LRG sample between NGC and SGC. Vertical line indicates the $\chi^2_\times$ value obtained from the DESI data and the histograms shows the distributions from the mocks. In the left panel, the filled grey histograms correspond to the \ezmock-FFA mocks, the filled purple histograms to the \textsc{Abacus}-Complete mocks, and the hatched histogram to the \textsc{Abacus}-altMTL mocks. 

The theoretically predicted distribution of the signal-sensitive cross statistic is shown as the black curve, with a zero center with the variance predicted as (for the derivation see~\cite{Hou2025:DESIeven4PCF, Krolewski2024:parity}): 

\begin{eqnarray}\label{eqn:var_chi2x}
    \text{Var}(\chi^2_{\times, {\rm null}}) &=& \frac{1}{\calN_\times^2} \sum_{\mu< \nu} \text{Tr} \left[\mathbb{\hat C}^{-1}\, \mathbb{C}^{\mu}\, \mathbb{\hat C}^{-1}\, \mathbb{C}^{\nu}\right] \nonumber\\
    &\approx& \frac{N_{\rm dof}}{\calN_\times^2}  \sum_{\mu< \nu}  \frac{V^2_{\rm fid}}{V^\mu_{\rm eff} V^\nu_{\rm eff}},
\end{eqnarray}
with $V_{\rm eff}^{\mu}$ being the effective volume for each patch, defined in Eq.~\eqref{eqn:V_eff}. Here, we choose $V_{\rm fid}$ to correspond to the patch with the larger volume between the $\mu$ and $\nu$ patches. 

In the left panel of Fig.~\ref{fig:chi2_x_analyt_NS_odd_dr0}, we show the Type-I cross correlation. The \ezmock-FFA distribution is centered around zero, with its width reasonably well described by the black theoretical curve. The data are also consistent with zero, albeit with a negative value. 

The \abacus\ mocks exhibit a non-zero mean as shown in the middle panel of Fig.~\ref{fig:chi2_x_analyt_NS_odd_dr0}. This offset arises because the NGC and SGC patches are built from the same underlying density realization and therefore not fully statistically independent. This is the scenario where we discussed in Eq.~\eqref{eqn:chi2x_mock_avg} that the correlation in the cosmic variance do not vanish, leading to a biased contribution with non-zero expectation value of the Type-II correlation for the \abacus\ mocks.
Although this correlation is visible in Fig.~\ref{fig:chi2_x_analyt_NS_odd_dr0}, it is small in practice: it is negligible in the parity-even sector, where the signal dominates, and only yields a variance discrepancy in the noise-dominated parity-odd sector~\cite{Hou2025:DESIeven4PCF}. Finally, because the Complete and altMTL samples share the same central value, the systematic contribution in Eq.~\eqref{eqn:chi2x_mock_avg} is consistent with zero and cannot account for the offset.

The right panel of Fig.~\ref{fig:chi2_x_analyt_NS_odd_dr0} shows the Type-II cross-correlation. The filled gray histogram is the \ezmock-FFA baseline; the darker and lighter purple histograms are the \abacus-altMTL results before and after correction, respectively. As noted above, the \abacus\ NGC and SGC are not statistically independent, and they lead to biases in the Type-II $\chi^2_\Delta$. To isolate the data-mock covariance  mismatch term $\chi^2_{\Delta_0}$ in Eq.~\eqref{eqn:chi2_Delta_tot_avg}, we subtract the total Type-II correlation by the Type-I correlation for the \abacus-altMTL, with the normalization $\calN_{\Delta, \rm abcs}$. This normalization absorbs the volume effect of the \abacus\ mocks listed in Table~\ref{tab:statistic}. Since the effective volume and the resulting $\calN_{\Delta}$ are estimated from the \ezmock, we apply an additional $2\%$ normalization to the \abacus-altMTL~\footnote{This $2\%$ correction directly propagates into the corrected $\chi^2_{\rm c}$. However, it has only a marginal impact on the final conclusion and would simply reduce the $\chi^2_{\rm c}$ value slightly further to -2.4$\sigma$.}.  

We additionally note that the data–mock covariance mismatch in the parity-odd sector is approximately $16\%$, comparable to the approximately $10\%$ mismatch previously found in the parity-even sector~\cite{Hou2025:DESIeven4PCF}.

\begin{figure*}
    \centering
    \begin{subfigure}[b]{0.33\textwidth}
    \includegraphics[width=0.99\linewidth]{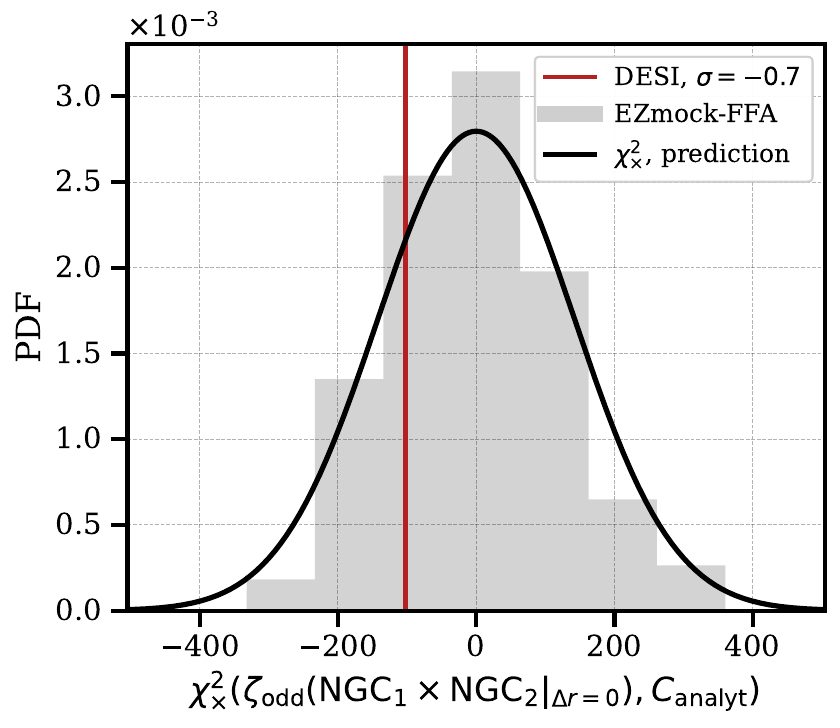}
    \end{subfigure}
    \hfill
    \begin{subfigure}[b]{0.32\textwidth}
    \includegraphics[width=0.99\linewidth]{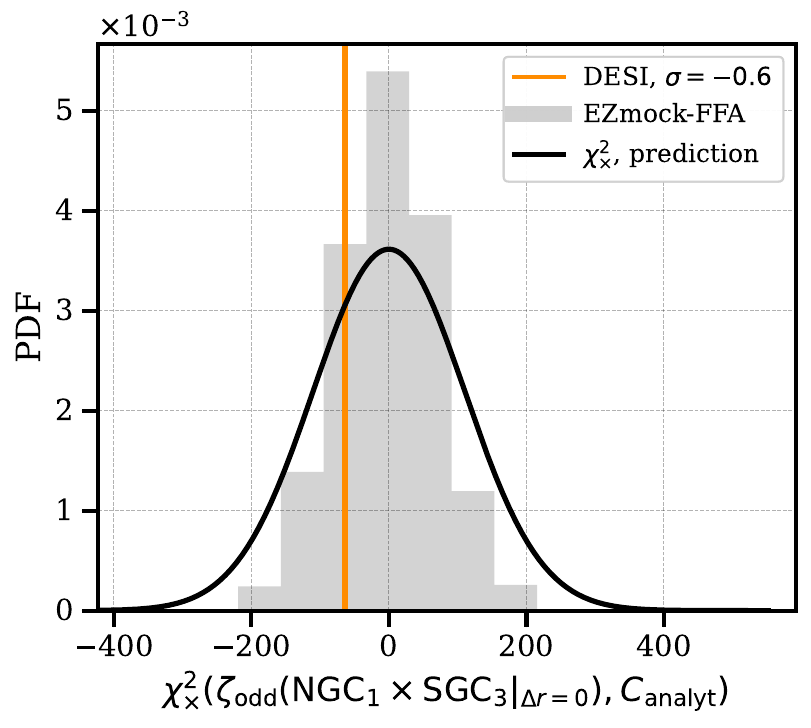}
    \end{subfigure}
    \begin{subfigure}[b]{0.32\textwidth}
    \includegraphics[width=0.99\linewidth]{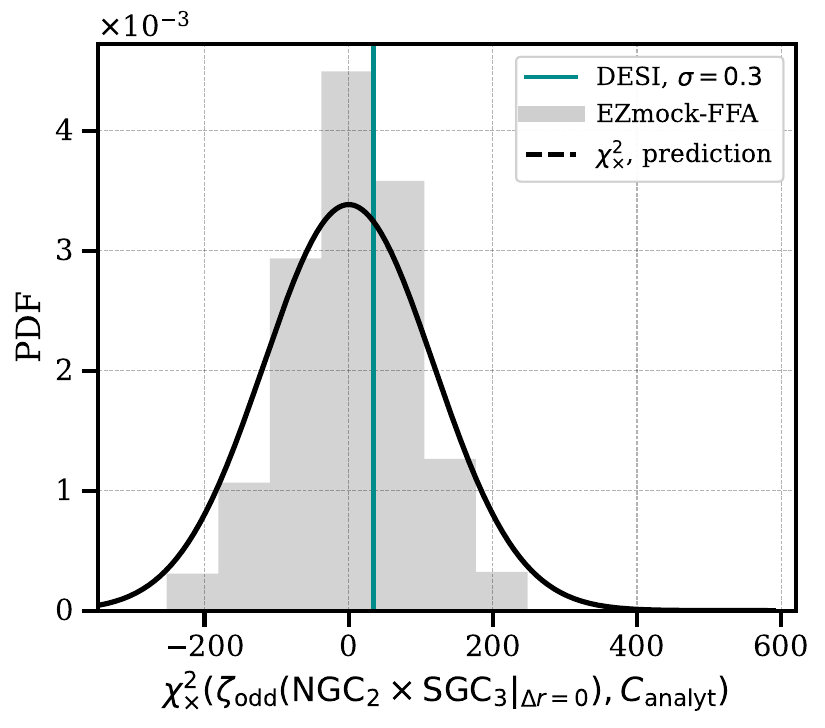}
    \end{subfigure}
    \caption{Distribution of the Type-I cross-correlation for the parity-odd 4PCF, using the full DESI DR1 LRG sample for three patches: NGC-1, NGC-2, and SGC-3, with the full analytic covariance matrix. Vertical lines indicate the DESI data statistics, and the filled grey histograms show the \ezmock-FFA distribution. The black curves denote the theoretical $\chi^2_\times$ distribution, modeled as a Gaussian centered at zero with the width predicted by Eq.~\eqref{eqn:var_chi2x}. }
    \label{fig:chi2_x_analyt_N1N2S3_odd_dr0}
\end{figure*}

\vspace{5mm}
\noindent{\it \textbf{Sky Dependence}\,\,} 
As a further consistency check, we further show the Type-I cross statistics for the three patches: NGC-1, NGC-2, and SGC-3 in Fig.~\ref{fig:chi2_x_analyt_N1N2S3_odd_dr0}. In each panel, the vertical lines represent the statistic derived from the DESI data, while the filled grey histograms show the distribution obtained from the \ezmock-FFA simulations. The black curves correspond to the theoretical prediction for the $\chi^2_\times$ distribution, modeled as a Gaussian centered at zero with variance given by Eq.~\eqref{eqn:var_chi2x}. For different combination of sky patches, we also find that the data to be consistent with zeros. While we do not show them here, we also find the correlation between the NGC and SGC for the \abacus-altMTL mocks, in particular in the NGC1-SGC3 combination.  

\begin{figure*}
    \centering
    \includegraphics[width=0.99\linewidth]{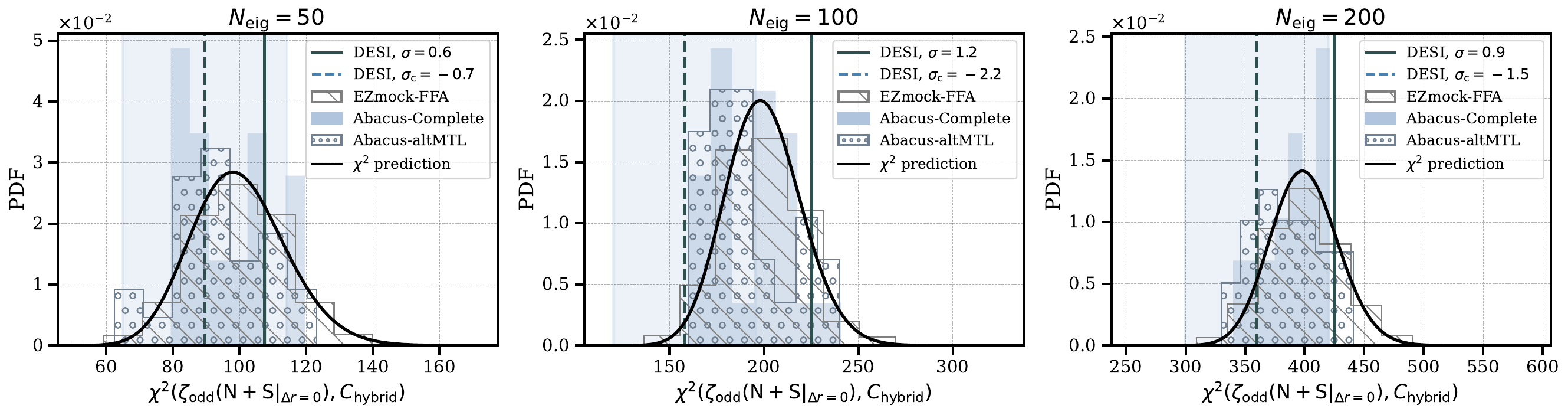}
    \caption{Distribution of the auto-correlation $\chi^2$ statistic for the parity-odd 4PCF measured from the full DESI DR1 LRG sample, combining the NGC and SGC. Each sub-panel corresponds to a different number of eigenvalues used in the data compression: $N_{\rm eig} = {50, 100, 200}$. The analysis covers the range $20\, \mpch < r < 160\, \mpch$ without additional radial cuts ($\Delta r=0$), with both the data vector and the corresponding \ezmock-FFA-based covariance matrix compressed by selecting the eigenvectors associated with the lowest noise. The dashed line represents the correction derived from the cross-statistics in Eq.~\eqref{eqn:chi2_delta_c}, with the uncertainty shown as the shaded region, estimated from the \ezmock; filled purple histograms show the distribution from the \abacus-Complete mocks; circled histograms correspond to the \abacus-altMTL mocks; hatched histograms correspond to the \ezmock-FFA mocks; and black curves present the theoretical predictions for a $\chi^2$ distribution. All distributions are rescaled by the ratio of the fiber implementations in the FFA and altMTL, reducing the $\chi^2$ values by 15\%–20\%. Overall, the DESI data are found to be consistent with zero signal.}
    \label{fig:chi2_hybrid_NS_odd_dr0}
\end{figure*}

\begin{figure*}
    \centering
    \includegraphics[width=0.99\linewidth]{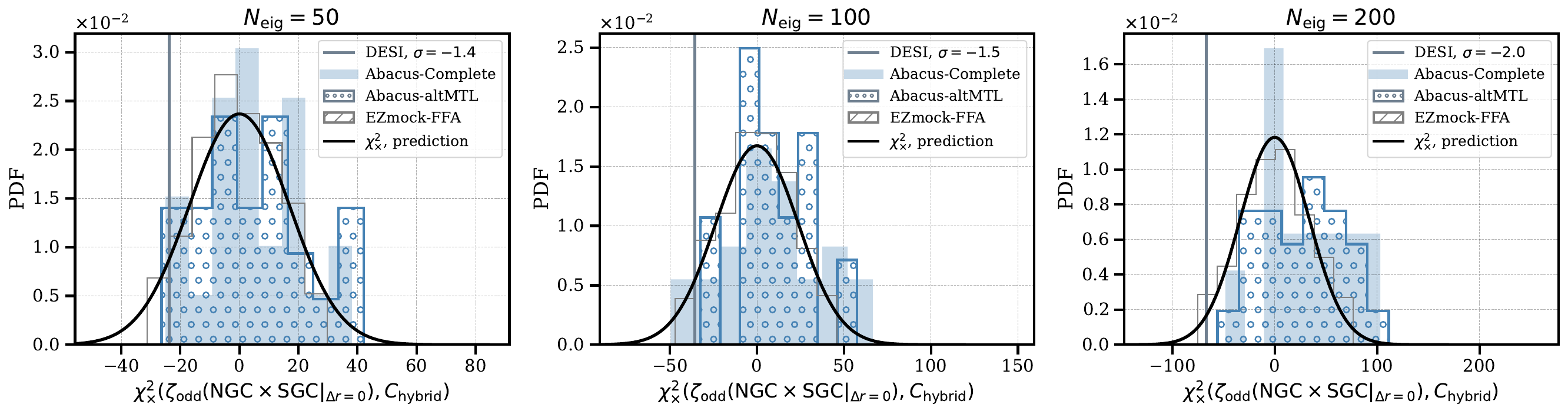}
    \caption{Distribution of the cross-correlation $\chi^2_\times$ statistic for the parity-odd 4PCF, measured from the full DESI DR1 LRG sample using the hybrid covariance, combining the NGC and SGC. Each sub-panel corresponds to a different number of eigenvalues used in the data compression: $N_{\rm eig} = {50, 100, 200}$. Vertical lines indicate the value measured from the DESI data; filled grey histograms show the distribution from the \abacus-Complete mocks; circled histograms correspond to the \abacus-altMTL mocks. The black curves show the theoretical predictions for the cross statistic under the null hypothesis, assuming a Gaussian distribution centered at zero with a variance given by Eq.~\eqref{eqn:var_chi2x}. Overall, we find the measurement to be consistent with zeros from the DR1 sample.}
    \label{fig:chi2_x_hybrid_NS_odd_dr0}
\end{figure*} 

\subsection{Hybrid Covariance}

One potential source of contamination in detection significance estimates is the mismatch in covariance between mocks and data, an effect that can be amplified by the large dimensionality of the data vector. In addition to performing cross-correlation tests, an effective way to mitigate this issue is to reduce the size of the data vector. 
To this end, we use the analytic covariance matrix to define a compression scheme for the data vector. The analytic covariance matrix can be decomposed as
\begin{eqnarray}
    \mathbb{C}_{\rm th} = U \Lambda U^{\rm T},
\end{eqnarray}
with $\Lambda$ being the diagonal matrix of eigenvalues with the diagonal entries $\lambda_i$, and $U$ the orthogonal matrix of eigenvectors. The 4PCF coefficients can be projected onto the eigenbasis of the analytic covariance as 
\begin{eqnarray}
    \tilde\zeta = U^{\rm T}\zeta,
\end{eqnarray}
we then select the $N_{\rm eig}$ components that have the lowest statistical noise $\lambda_i^{-1/2}$. 

This compression method enables us to construct a hybrid covariance matrix by decomposing the 4PCFs measured from simulations into the eigenbasis of the analytic covariance matrix. We select the top $N_{\rm eig}$ eigenmodes with the lowest noise and use the corresponding rotated data vectors $\tilde\zeta$ to build a covariance matrix in the compressed basis $\hat{\mathbb{C}}_{{\rm hybrid}, ij} = \av{\tilde\zeta_i, \tilde\zeta_j}$. 
Using 1000 \ezmock-FFA mocks, this procedure allows us to generate hybrid covariance matrices with up to 200 degrees of freedom.

\vspace{5mm}

\noindent{\it \textbf{Auto-Correlation with $\mathbb{C}_{\rm hybrid}$} \,\,}
Fig.~\ref{fig:chi2_hybrid_NS_odd_dr0} shows the distribution of the auto-correlation $\chi^2$ statistic for the parity-odd 4PCF, measured using the combined NGC and SGC from the full DESI DR1 LRG sample. Each panel here corresponds to a different number of eigenvalues ($N_{\rm eig} = {50, 100, 200}$) used for data compression. The analysis is performed over the range $20\, \mpch < r < 160\, \mpch$. In each panel, the vertical line marks the value obtained from the DESI data. The filled purple histograms depict the $\chi^2$ distribution from the \abacus-Complete mocks, while the hatched histograms show results from the \abacus-altMTL mocks. The black curves indicate the theoretical predictions for a $\chi^2$ distribution. 

Since we applied the data compression scheme, the covariance matrix is primarily based on the \ezmock-FFA realizations. As discussed in \S\ref{sec:incomplete_stat_fluctuations}, the FFA fiber assignment implementation leads to an underestimation of the covariance compared to the altMTL implementation. To account for this, we rescale all distributions by the ratio of the covariance estimates from the FFA to the altMTL samples, which results in a $15\%$–$20\%$ reduction in the $\chi^2$ values. 

Additionally, we apply the Hartlap factor $M \equiv (N_{\rm m}-N_{\rm b}-2)/(N_{\rm m}-1)$ to de-bias the inverse of the sample covariance, following~\cite{Hartlap2007:hartlap}.  Overall, we find that the results from the DESI DR1 sample are all consistent with zero and agree with the expected data-mock covariance mismatch.

\vspace{5mm}

\noindent{\it \textbf{Cross Correlation with $\mathbb{C}_{\rm hybrid}$} \,\,}
Fig.\ref{fig:chi2_x_hybrid_NS_odd_dr0} presents the distribution of the cross-correlation $\chi^2_\times$ statistic for the parity-odd 4PCF, measured from the combined NGC and SGC regions of the full DESI DR1 LRG sample using the hybrid covariance. Each panel of the figure corresponds to a different number of eigenvalues ($N_{\rm eig} = {50, 100, 200}$) used in the data compression. In each panel, the vertical line marks the $\chi^2_\times$ value measured from the DESI data. The filled grey histograms represent the distribution obtained from the \abacus-Complete mocks, while the circled histograms correspond to the \abacus-altMTL mocks. For reference, the black curves show the expected distribution of the cross statistic under the null hypothesis, assuming a Gaussian distribution centered at zero with variance given by Eq.~\eqref{eqn:var_chi2x}. 

Compared to the case using the analytic covariance matrix, where the full dimensionality of the data vector is retained, the excess observed in the \abacus\ mocks is significantly reduced. This $10\%$ effect only begins to appear when $N_{\rm eig}=200$, at which point we also start to see a shift in the center of the \abacus-altMTL mock distributions. These findings reinforce the interpretation that the excess around the center of the \abacus\ distribution in Fig.~\ref{fig:chi2_x_analyt_NS_odd_dr0}, obtained with the analytic covariance, mainly arises from the cumulative effect of statistical noise per degree of freedom.

\begin{figure*}
    \centering
    \includegraphics[width=0.99\linewidth]{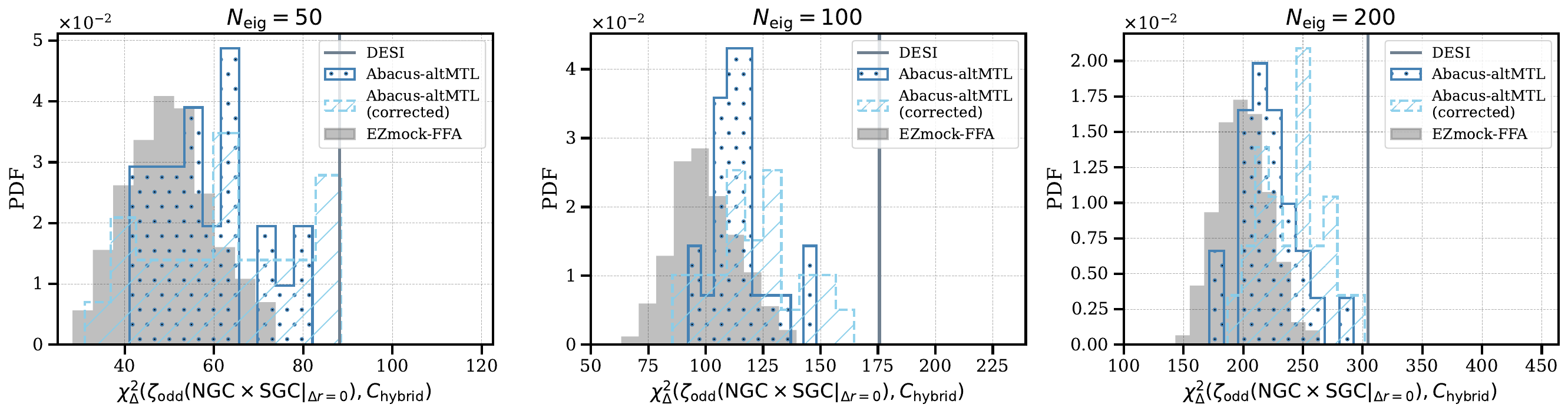}
    \caption{Distribution of the $\chi^2_\Delta$ statistic for the parity-odd 4PCF, using the full DESI DR1 LRG sample, combining the NGC and SGC. Each plot corresponds to the number of eigenvalues $N_{\rm eig}=\{50,100, 200\}$. Both the data vector and the corresponding \ezmock-FFA-based covariance matrix are compressed accordingly. The vertical lines represent the statistics for the DESI data; the filled gray histograms correspond to the \textsc{Abacus}-Complete mocks. The histogram marked by circles corresponds to the \textsc{Abacus}-altMTL mocks.}
    \label{fig:chi2_delta_hybrid_NS_odd_dr0}
\end{figure*}

To complete the analysis, we perform the Type-II cross-correlation $\chi^2_{\Delta}$ for the DESI DR1 LRG sample between NGC and SGC as shown in Fig.~\ref{fig:chi2_delta_hybrid_NS_odd_dr0}. As before, each plot here corresponds to the number of eigenvalues $N_{\rm eig}=\{50,100, 200\}$. 
Both the data vector and the corresponding \ezmock-FFA-based covariance matrix are compressed accordingly. The vertical lines represent the statistics for the DESI data, the filled gray histograms correspond to the \textsc{Abacus}-Complete mocks, the histogram marked with circles corresponds to the \abacus-altMTL mocks. Here, we also apply the fiber correction factor and the correction factor found in the correlation in the \abacus\ mocks (see Eq.~\ref{eqn:chi2_delta_c}). 

After applying the correction to the Type-II cross statistics, the agreement between the data and the mocks improves, although some residual discrepancy remains, especially as $N_{\rm eig}$ increases. To account for this, we still to include this effect when computing the auto-correlation, as shown in Fig.~\ref{fig:chi2_hybrid_NS_odd_dr0}.

\begin{figure}
    \centering
    \includegraphics[width=0.95\linewidth]{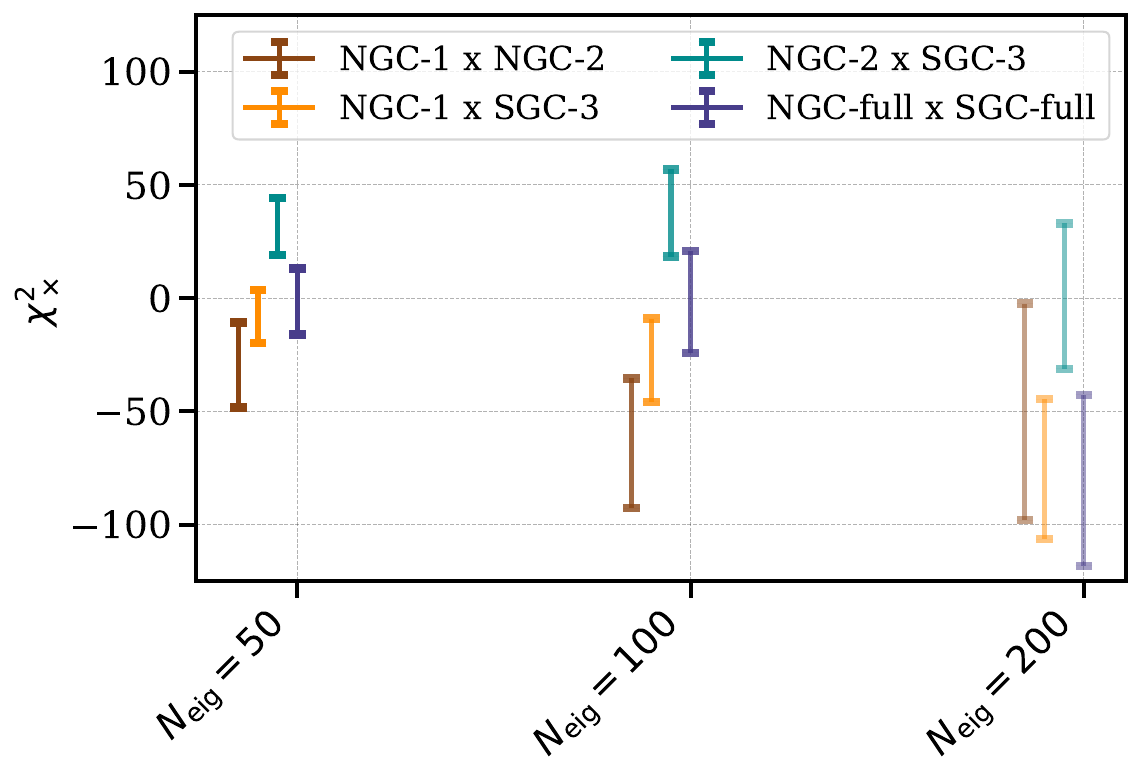}
    \caption{Type-I cross correlation $\chi^2_\times$ with hybrid covariance. Here we apply additional radial cuts $\Delta r=14\,\mpch$ to enforce a minimum separation between the radial bins. Different colored error bars denote the combination of sky regions, whereas the size of the errorbars are derived from the \ezmock. They are grouped into different choices of eigenvalues $N_{\rm eig}=\{50,100,200\}$.}.
    \label{fig:chi2_x_hybrid_N1N2S3_odd_dr14}
\end{figure}

\vspace{5mm}
\noindent{\it \textbf{Scale Dependence} \,\,}
Fig.~\ref{fig:chi2_x_hybrid_N1N2S3_odd_dr14} summarizes the detection significance of the Type-I cross-correlation in the DESI DR1 LRG sample. Colored error bars denote different combinations of sky regions, and each panel corresponds to a choice of the number of eigenmodes, $N_{\rm eig} = \{50, 100, 200\}$. We impose an additional radial cut of $\Delta r = 14,\mpch$ to enforce a minimum separation between radial bins, ensuring that galaxies contributing to the measurement do not populate adjacent bins.

\section{Summary}\label{sec:summary}
In this paper, we presented measurements of the parity-odd 4PCF on the DESI DR1 LRG sample and quantified its detection significance. We carried out analyses using both the full analytic covariance matrix with the uncompressed data vector, as well as a compressed data vector in combination with a hybrid covariance matrix derived from both simulations and analytic calculations. We found that, when using the full analytic covariance without corrections, the auto-correlation could exhibit significance levels as high as $4\sigma$. However, this apparent excess is reduced after correcting for the expected data-mock covariance mismatch derived from the cross-correlations between different sky patches. Overall, our results indicate that the signal is consistent with zero.

In our analysis, we identified two main effects: the volume effect and the fiber assignment effect. Moreover, the limited volume of the \abacus\ mocks leads to additional artifacts in the cross-correlation calculations among different sky patches, which ultimately require an extra correction to the significance derived from the auto-correlation test. As a result, in contrast to~\cite{Slepian2025:DESIodd4PCF}, we do not find any tension when comparing different analyses methods.  

In addition, we tested various analysis setups, including scale dependence by imposing additional radial bin cuts, partitioning the sky into different patches and testing various combinations, and performing both auto- and cross-correlation tests. In particular, the two cross statistics allow us to isolate the impact of general systematics by comparing the data to the \abacus-complete mocks. We find that systematics have subdominant impact on the signal. 

In the following, we highlight several additional points that are important for the interpretation of our results:

\begin{enumerate}
    \item[(i).] Several key factors play a role in the DESI DR1 sample in contrast to the SDSS BOSS sample~\cite{Dawson2013}. The BOSS sample achieved a completeness exceeding 90\%, the DR1 sample has an average completeness of only about 50\%. Combined with the low signal-to-noise ratio anticipated in the parity-odd 4PCF, this lower completeness increases the sample’s susceptibility to fiber assignment effects. Furthermore, the limited volume of the \abacus\ simulation presents additional challenges. Together, these factors make the calibration and correction of the mocks and covariance matrix particularly difficult but also essential. 
    \item[(ii).] In our analysis we observe a negative Type-I cross-correlation. The value remains consistent with zero at the $3\sigma$ level, indicating that it is compatible with statistical fluctuations. In addition to statistical uncertainty, deterministic contributions from observational systematics may also play a role. While our tests suggest that fiber assignment is unlikely to be the dominant factor, other residual effects, such as spatial variations in the number density across sky patches, cannot be excluded. It will therefore be important to revisit this test in future data releases as number-density uniformity and systematic control improve. 
    \item[(iii).] The cross-correlation test further assumes that the signal is shared across different patches. If a large-scale modulation exists beyond the coherence length of the sky patches considered in this work ($>500,\mpch$), neither type of cross-correlation test necessarily applies. In particular, the data–mock mismatch observed in the Type-II statistic can arise from this large-scale modulation, and our analysis cannot rule out this scenario. However, verifying the impact of this large-scale modulation requires rigorous testing, such as verifying that the Type-II statistic is not contaminated by {\it e.g.} position-dependent systematics.
    \item[(iv).] Compared to~\cite{Slepian2025:DESIodd4PCF}, there are two main differences: the choice of radial binning and the inclusion of the Type-II cross-correlation. Despite the different binning choices, we find broadly consistent results for the uncorrected auto-correlation and the Type-I cross-correlation. However,~\cite{Slepian2025:DESIodd4PCF} did not report the Type-II cross-correlation. By contrast, in our analysis, including the Type-II cross correlation alleviates the previously reported ``tension''.
\end{enumerate}

These insights may inform analyses of future DESI data releases, as well as other ongoing or upcoming galaxy surveys, such as Euclid~\cite{Euclid2011:WhitePaper} and Roman~\cite{Wang2022:RomanHLSS}.

\section*{Acknowledgments}
We thank Drew Jamieson and Eiichiro Komatsu for insightful discussions. We thank DESI internal referees for their comments on the manuscript as well as all the feedback we received during the collaboration wide review. JH thanks Will Handley and his group for their continued support. JH thanks~{\hypersetup{urlcolor=black}\href{https://github.com/Moctobers/Acknowledgement/blob/main/fox_in_office.jpg}{Jue Fox}} for his office support. JH has received funding from the European Union’s Horizon 2020 research and innovation program under the Marie Sk\l{}odowska-Curie grant agreement No. 101025187 and was funded by Deutsche Forschungsgemeinschaft (DFG) -- Project number 554476934. We acknowledge
the organizers and participants of {\hypersetup{urlcolor=Blue}\href{https://parity.cosmodiscussion.com}{\it Parity Violation from Home 2025}} for helpful feedback.
We acknowledge UFIT Research Computing for providing computational resources and support that have contributed to the research results reported in this publication. 
The work of RNC was supported in part by the Director, Office of Science, Office of High Energy Physics, of the U.S. Department of Energy under contract No. DE-AC02-05CH11231.

This material is based upon work supported by the U.S. Department of Energy (DOE), Office of Science, Office of High-Energy Physics, under Contract No. DE–AC02–05CH11231, and by the National Energy Research Scientific Computing Center, a DOE Office of Science User Facility under the same contract. Additional support for DESI was provided by the U.S. National Science Foundation (NSF), Division of Astronomical Sciences under Contract No. AST-0950945 to the NSF’s National Optical-Infrared Astronomy Research Laboratory; the Science and Technology Facilities Council of the United Kingdom; the Gordon and Betty Moore Foundation; the Heising-Simons Foundation; the French Alternative Energies and Atomic Energy Commission (CEA); the National Council of Humanities, Science and Technology of Mexico (CONAHCYT); the Ministry of Science, Innovation and Universities of Spain (MICIU/AEI/10.13039/501100011033), and by the DESI Member Institutions: \url{https://www.desi.lbl.gov/collaborating-institutions}. Any opinions, findings, and conclusions or recommendations expressed in this material are those of the author(s) and do not necessarily reflect the views of the U. S. National Science Foundation, the U. S. Department of Energy, or any of the listed funding agencies.

The authors are honored to be permitted to conduct scientific research on I'oligam Du'ag (Kitt Peak), a mountain with particular significance to the Tohono O’odham Nation.

\section{Data Availability}
Data from the plots in this paper is available on Zenodo as part of DESI’s Data Management Plan~\footnote{\url{https://doi.org/10.5281/zenodo.17753485}}.

\appendix
\section{Spherical Harmonics, Helicity Basis, and Scalar Triple Product}
\label{eqn:normalization_triprod}
The parity-odd scalar observables requires a pseudoscalar, which is the essential quantity in isolating the parity information. This quantity can be expressed in the harmonic space, but very often, we also see it arises when performing the polarization decomposition. Here we make the connection explicitly by expressing the scalar triple product in the harmonic basis.

The four-point function of density field in $k$-space, the trispectrum, can be decomposed into a parity-odd part $T_+$ and a parity-odd part $T_-$, where both $T(\bfk)$ and $T(-\bfk)$ are well defined 
\begin{eqnarray}\label{eqn:4PCF_odd}
T(\bfk_1, \bfk_2, \bfk_3) = T_{+}(\bfk_1, \bfk_2, \bfk_3) + i\,T_{-}(\bfk_1, \bfk_2, \bfk_3),
\end{eqnarray}
where $T_+$ and $T_-$ are real functions. Due to the translational symmetry and the momentum conservation, the sum of the four momenta is zero $\sum_{i=1}^4 \bfk_i = 0$. 

A direct measurement of the four-point correlation function does not inherently distinguish between parity-even and parity-odd contributions. To isolate parity-specific information, it is necessary to express the four-point function in a basis that explicitly separates the two components.
In Fourier space, the parity-odd part can be parametrized as
\begin{eqnarray}
\hspace{-5mm} T_-(\bfk_1, \bfk_2, \bfk_3) = \bfk_1 \cdot (\bfk_2 \times \bfk_3) \, \tau_-(k_1, k_2, k_3, k_4, K, \tilde K),
\end{eqnarray}
with the scalar triple product $\bfk_1 \cdot (\bfk_2 \times \bfk_3)$ being a pseudoscalar, constructed from a polar vector and a pseudovector. $K=|\bfk_1+\bfk_2|$ and $\tilde K=|\bfk_1+\bfk_3|$ are two diagonals of the tetrahedra.$\tau_-$ is a parity-even function that in the most general case can have six degrees of freedom. 

Now we consider a rank-$n$ quantity ${F}_{i_1..i_n}$, this quantity can be decomposed as~\cite{Dai2012:TAM}
\begin{eqnarray}
{F}_{i_1..i_n}(\bfk) = \sum_{\ell,m} \sum_{\lambda = 0, \pm1} F_{\ell m}^{\lambda}(k) \; {}_\lambda Y_{\ell m}(\hat{{k}}) \, \hat\epsilon^{\lambda}_{i_1..i_n}(\hat{{k}}),  
\end{eqnarray}
where $\lambda$ denotes the helicity state, ${}_\lambda Y_{\ell m}(\hat{{k}})$ is the spin-weighted spherical harmonics, and $\hat\epsilon^{\lambda}_{i_1..i_n}$ is the rank-$n$ polarization state,
\begin{eqnarray}
\hat\epsilon^{\lambda}_{i_1..i_n} = \sum_{i_1,..,i_n\in\{0, \pm 1\}}\calC_{i_1..i_n}^{\lambda}\,\hat{\rme}_{i_1}  \otimes \ldots \otimes \hat{\rme}_{i_n},
\end{eqnarray}
where $\sum_n i_n = \lambda$, and $\calC_{j_1..j_n}^{\lambda}$ is the Clebsch Gordan coefficient, $\hat{\rme}_{i_n}$ is the helicity basis.
Here we make a global coordinate choice for the helicity basis
\begin{eqnarray}
\hat \rme_{\pm 1} = \frac{\hat x \pm i\hat y}{\sqrt2},\qquad
\hat \rme_{0} = \hat z,
\end{eqnarray}
with the longitudinal mode align with the quantization axis used to define the spherical harmonics. Alternatively, we can explicitly write the Cartesian components as
\begin{eqnarray}
\hat \rme_{\pm 1} = \Big(\tfrac{1}{\sqrt2},\; \pm\tfrac{i}{\sqrt2},\; 0\Big),\qquad
\hat \rme_{0} = (0,0,1).    
\end{eqnarray}
For rank-1 tensor, the polarization state is simply $\hat\epsilon^{\lambda} = \hat{\rme}_{\lambda}$ with the Clebsch Gordan coefficient being one.
In particular, when ${F}_i = k_i\equiv\bfk$ as a rank-1 tensor, we have $\ell=1$. Since we made a global choice of the coordinate for the helicity basis and the spherical harmonics, and chose them to be aligned with each other, the helicity state $\lambda$ becomes degenerate with the $m$ label in the spherical coordinate, and we are free to set $\lambda=m$. Therefore, the wave vector can be expressed as
\begin{eqnarray}
\bfk 
&=& \sum_{m = 0, \pm1} F_{1m}^m(k) \;  Y_{1 m}(\hat{{k}}) \, \hat\rme_m(\hat{{k}})\nonumber\\
&=& \sqrt{\frac{4\pi}{3}}\sum_{m = 0, \pm1} k \;  Y_{1 m}(\hat{{k}}) \, \hat\rme_m(\hat{{k}})
\end{eqnarray}
in the second line above we normalize the coefficient $F_{1m}^m$ such that $|\hk|^2 = 1$.  
The scalar triple product can thus be written as
\begin{eqnarray}\label{eqn:scalar_triple}
&&\bfk_1 \cdot (\bfk_2 \times \bfk_3)\nonumber\\
&=&  \sum_{m_1 m_2 m_3}  \prod_{a=1}^3\rmk_{a} Y_{1 m_a}(\hk_a) \, \epsilon_{ijk}
\hat\rme_{m_1}^i \hat\rme_{m_2}^j  \hat\rme_{m_3}^k 
\end{eqnarray}
where $\epsilon_{ijk}
\hat\rme_{m_1}^i \hat\rme_{m_2}^j  \hat\rme_{m_3}^k = \left<0,0| 1, m_1; 1, m_2; 1, m_3\right>$ is an invariant scalar and is proportional to the 3-$j$ symbol
\begin{eqnarray}
 \epsilon_{ijk}
\hat\rme_{m_1}^i \hat\rme_{m_2}^j  \hat\rme_{m_3}^k = C \six{1}{1}{1}{m_1}{m_2}{m_3},   
\end{eqnarray}
letting $m_1=1$, $m_2=0$, $m_3=-1$, we have
\begin{eqnarray}
\epsilon_{ijk}\,\hat \rme_{+1}^i \hat \rme_{0}^j \hat \rme_{-1}^k = i.
\end{eqnarray}
Meanwhile the Wigner 3-$j$ for that choice is
\begin{eqnarray}
\six{1}{1}{1}{1}{0}{-1} = -1/\sqrt{6},    
\end{eqnarray}
therefore, the coefficient $C = i / (-1/\sqrt{6})=-i\sqrt{6}$. Inserting this coefficient into Eq.~\eqref{eqn:scalar_triple}, the triple product can be simplified as
\begin{eqnarray}
&&\bfk_1 \cdot (\bfk_2 \times \bfk_3)\nonumber\\
&=& -{i} \sqrt{6}\left(\frac{4\pi}{3}\right)^{3 / 2} \prod_{a=1}^3 k_a Y_{1 m_a}(\hk_a) \sum_{m's}\six{1}{1}{1}{m_1}{m_2}{m_3} \nonumber
\end{eqnarray}

\section{Mock Calibration}
\label{sec:calibration}
As discussed in \S\ref{sec:Result}, there are two major effects in the mocks: the volume effect and the fiber assignment effect. In this section, we demonstrate the consequences in the mocks when correction or calibration is not properly applied. We highlight what biases or discrepancies may arise, thereby emphasizing the importance of accurate calibration and correction.

\begin{figure}
    \centering
    \includegraphics[width=0.8\linewidth]{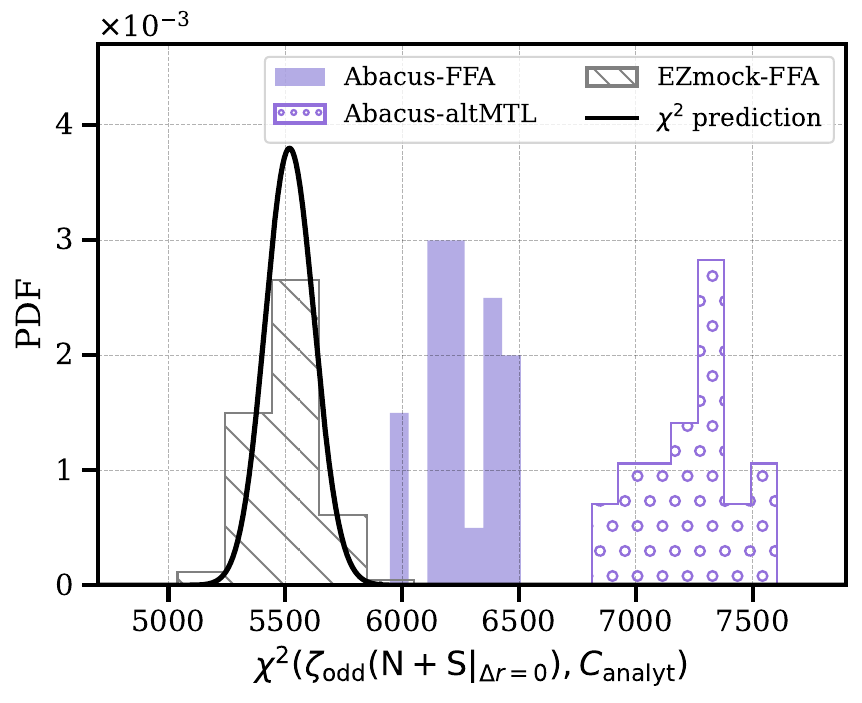}
    \caption{Distribution of the auto correlation $\chi^2$ for different mocks without calibration. In contrast to Fig.~\ref{fig:chi2_analyt_NS_odd_dr0}, the centers of the distributions can shift significantly if calibration is not properly applied. This plot demonstrates the importance of proper calibrations of the mocks.}
    \label{fig:chi2_NS_no_calib}
\end{figure}

Fig.~\ref{fig:chi2_NS_no_calib} shows the distribution of different mocks, analogous to Fig.~\ref{fig:chi2_analyt_NS_odd_dr0}, but without applying calibrations to the mocks. The \ezmock\ distribution (grey histogram) remains centered around the number of degrees of freedom expected for a $\chi^2$ distribution (black curve). This is not surprising, as the analytic covariance matrix was calibrated using the \ezmock\ as reference. 

In contrast, we find that the \abacus\ mocks are centered at higher values than the expected $\chi^2$ distribution. This is because the \abacus\ mocks are generated by replicating the same mock volume multiple times, resulting in a lack of independent $k$-modes and an inaccurate estimate of the true covariance. Consequently, the variance is overestimated. In the absence of a true signal, the center of the distribution is entirely dominated by this inflated variance and shifts the center to higher values.

To correct for these effects, we first apply the fiber assignment correction by multiplying the distribution by the ratio $ {\rm Tr}[\mathbb{C}_{\rm{FFA}}]/{\rm Tr}[\mathbb{C}_{{\rm altMTL}}]$  from the covariance matrix of the FFA-based mocks to those of the altMTL mocks. Next, we apply the volume correction by computing the ratio between the covariance matrix derived from the \abacus\ and the \ezmock-derived analytic covariance matrix. After these calibration steps, we find all the mock distributions center around the expected $\chi^2$ distribution.

\begin{figure*}
    \centering
    \begin{subfigure}{0.4\textwidth}
        \centering
        \includegraphics[width=\linewidth]{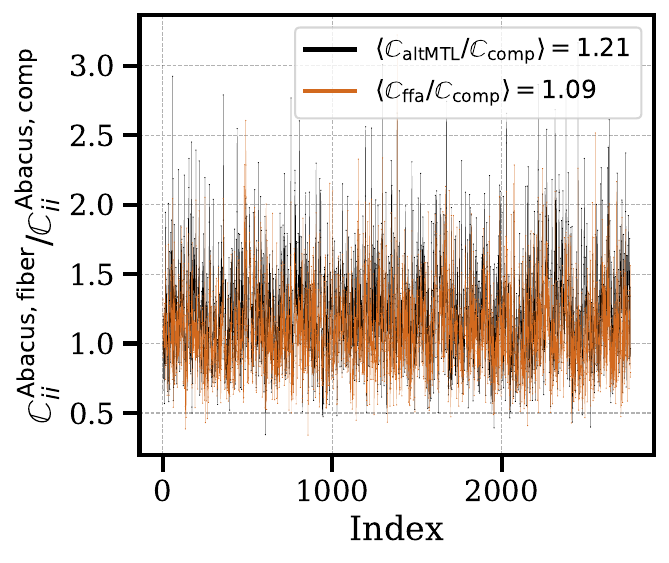}
    \end{subfigure}
    \begin{subfigure}{0.39\textwidth}
        \centering
        \includegraphics[width=\linewidth]{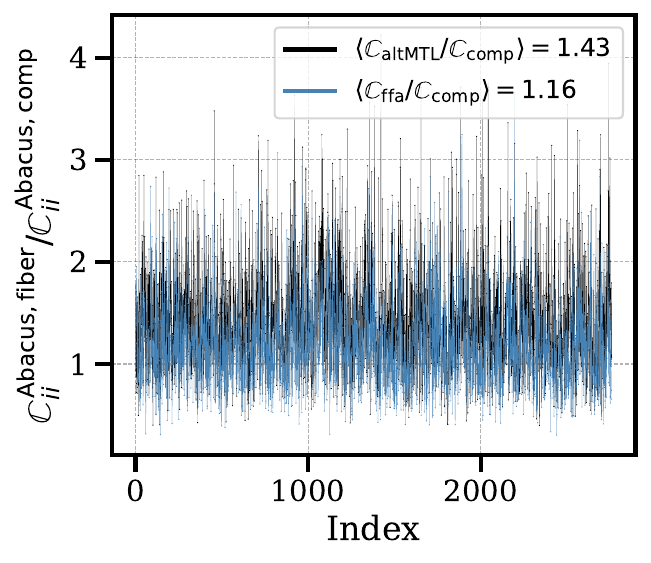}
    \end{subfigure}    
    \caption{The figure compares statistical fluctuations in Abacus mocks under different fiber-assignment schemes by plotting the ratio of diagonal covariance elements for fiber-assigned mocks to those from complete mocks. In the left panel (NGC), the ratios show that fiber assignment increases the diagonal covariance by about 9\% for FFA (orange) and 30\% for altMTL (black). The right panel presents the same ratio for the SGC, where the increases are larger: about 16\% for FFA and 43\% for altMTL. }
    \label{fig:ratio_cov_NS_odd_abacus}
\end{figure*}

\section{Impact of Incompleteness due to Statistical Fluctuations}
\label{sec:incomplete_stat_fluctuations}
As seen in Fig.~\ref{fig:footprint}, there is a large variation in the galaxy number density in the DR1 sample. This is essentially due to the limited number of fibers and during the early stage of the survey, not every galaxy gets the chance to be assigned with a fiber, therefore, the sample exhibits low completeness and is therefore sensitive to the details of fiber assignment implementation.
When constructing the DESI mocks, two fiber assignment implementation schemes were applied: the fast fiber assignment (FFA) and the alternate Merged Target Ledger (altMTL). We described the implementation of fiber assignment in~\cite{Hou2025:DESIeven4PCF}. Further details can be found in~\cite{Myers2023:DESItargetSelection, Bianchi2024:DESIfiber, Lasker2025:DESIFiberAssign}. As a brief summary: altMTL is close to the realistic fiber assignment run on the data, while FFA is an emulator-based algorithm that learns the mapping between an input and a \textsc{fiberassign}-processed galaxy catalog.

For the parity-even 4PCF, we previously found that FFA underestimates the covariance relative to a more realistic algorithm; Fig.~\ref{fig:ratio_cov_NS_odd_abacus} extends this comparison to the parity-odd sector. The left panel shows, for the NGC, the ratio of diagonal covariance elements in Abacus mocks with fiber assignment to those in the complete sample. We find the average ratio between the Abacus altMTL to the complete sample (black) to be ${\rm Tr}[\mathbb{C}_{\rm altMTL}] / {\rm Tr}[\mathbb{C}_{\rm comp}]\approx 1.21$, whereas the average ratio between the Abacus FFA to the complete sample (orange) to be ${\rm Tr}[\mathbb{C}_{\rm FFA}] / {\rm Tr}[\mathbb{C}_{\rm comp}]\approx 1.09$. The right panel presents the same comparison for the SGC and shows an even stronger deviation from the complete mocks, with ${\rm Tr}[\mathbb{C}_{\rm altMTL}] / {\rm Tr}[\mathbb{C}_{\rm comp}]\approx 1.43$ and ${\rm Tr}[\mathbb{C}_{\rm FFA}] / {\rm Tr}[\mathbb{C}_{\rm comp}]\approx 1.16$, potentially due to the lower completeness in the SGC.

An under-estimated covariance in the FFA mocks was also found for the two-point functions~\cite{DESI2024:SampleDefinition,ForeroSanchez2025:DESIcovariance}. Since the analytic covariance was fit with respect to the FFA-based \ezmock, the under-estimated covariance due to the fiber assignment need to be accounted for by applying an additional correction factor to our significance, and leaving impact on the detection significance as well as in the mock calibration (see discussions in \S\ref{sec:Result} and Appendix~\ref{sec:calibration}).

\section{Testing Gaussianity of the Likelihood}
\label{subsec:nonGaussian_likelihood}

In this section, we test the validity of the Gaussian likelihood assumption for the 4PCF. This assumption is generally motivated by the central limit theorem, since each element of the binned data vector represents an average over many Fourier- or position-space modes, their joint distribution is expected to approach Gaussianity. However, it is important to explicitly verify this assumption, particularly for higher-order statistics. 

Following~\cite{Philcox2021boss4pcf}, we repeat the same test as in our companion parity-even paper~\cite{Hou2025:DESIeven4PCF}, but now for the parity-odd 4PCF. We transform the data vector into an uncorrelated basis using the Cholesky decomposition of the covariance matrix, $\hat{\mathbb{C}} = LL^\top$. This “whitening’’ procedure removes linear correlations, such that the transformed components should follow independent standard normal distributions if the Gaussian assumption holds. Any systematic deviation from this behavior indicates the presence of non-Gaussianity either in the data or in the estimated covariance.

\begin{figure*}
    \centering
    \begin{subfigure}{0.45\textwidth}
        \centering
        \includegraphics[width=\linewidth]{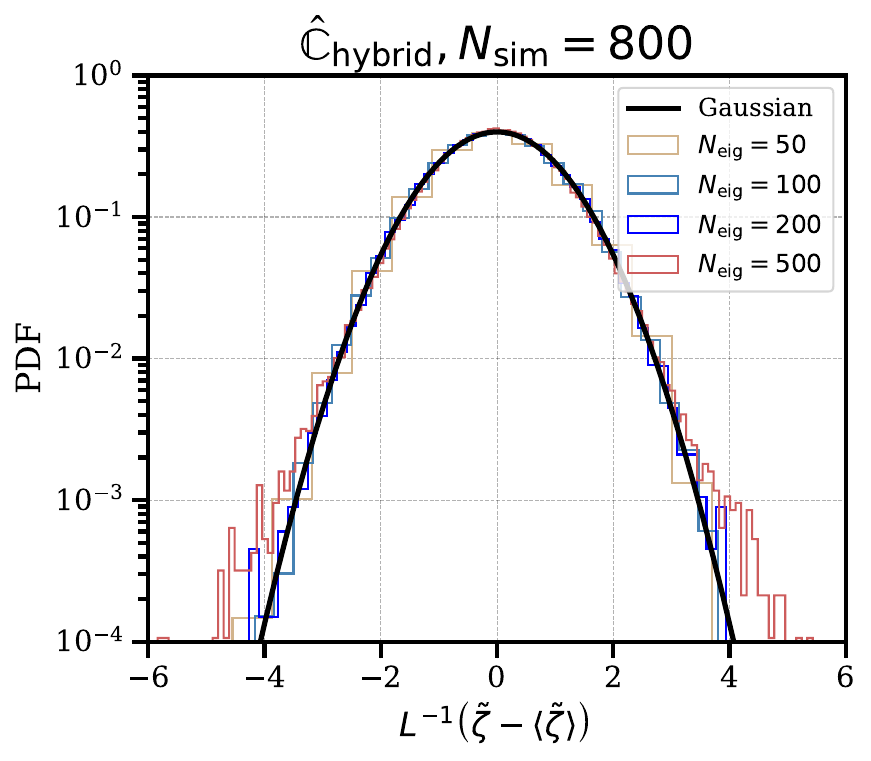}
    \end{subfigure}
    \begin{subfigure}{0.45\textwidth}
        \centering
        \includegraphics[width=\linewidth]{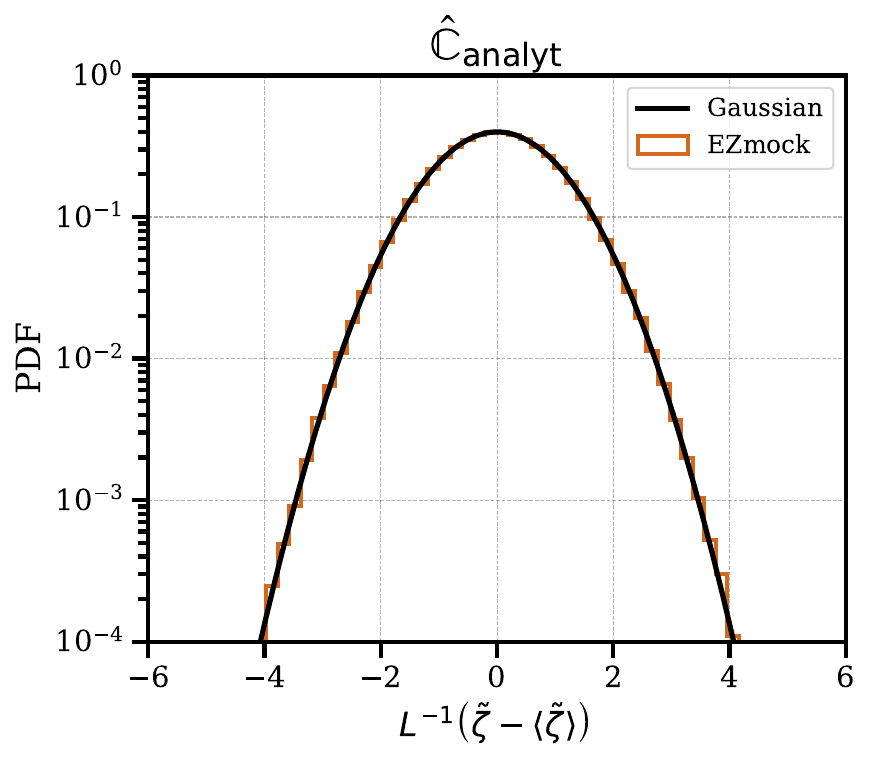}
    \end{subfigure}    
    \caption{The histograms show the distribution of the normalized components of the data vector in mode space. The black dashed line denotes the standard normal distribution $\mathcal{N}(0,1)$. In the left, the mode space normalization is performed using the inverse Cholesky decomposition $L^{-1}$ of the hybrid covariance matrix $\hat{\mathbb{C}}_{\rm hybrid} = LL^\top$, where $\hat{\mathbb{C}}_{\rm hybrid}$ is obtained from the \ezmock. Here, we split the mock into two sets: one subset is used to estimate the covariance matrix, and the other to generate the test data vectors. We use 800 mocks for covariance estimation and 200 as test data. We vary different choices of $N_{\rm eig}$ (indicated by color). When the condition $N_{\rm cov} \gg N_{\rm eig}$ is not well satisfied, the normalized data vector shows deviations from Gaussianity, indicating that the Gaussian likelihood assumption breaks down in this regime. In the right panel, we apply a Cholesky decomposition to the analytic covariance matrix and use the resulting Cholesky factor to normalize the unprojected data vector from the \ezmock. Despite various simplified assumptions in the analytic covariance, we find good agreement with the Gaussian likelihood.}
    \label{fig:NG_PDF_Ncov}
\end{figure*}

Fig.~\ref{fig:NG_PDF_Ncov} shows the distribution of the normalized components across all eigenmodes and simulations for different numbers of degrees of freedom.  The black dashed line denotes the standard normal distribution $\mathcal{N}(0,\sigma_{\rm std})$. In the left plot, The mode-space normalization uses the inverse Cholesky factor $L^{-1}$ of the hybrid covariance matrix $\hat{\mathbb{C}}_{\rm hybrid} = LL^\top$, based on the \ezmock. In the left panel, 800 mocks are used for covariance estimation and 200 for testing. We find that when not sufficiently large enough of simulation for covariance estimation such that $N_{\rm cov} \gg N_{\rm eig}$ is not well satisfied, as in the case of $N_{\rm eig}=500$, the normalized data vector shows a slight deviation from Gaussianity, indicating a breakdown of the Gaussian likelihood assumption in this regime. 
In the right panel, we repeat the test with the analytic covariance matrix $\hat{\mathbb{C}}_{\rm analyt} = LL^\top$, and use the Cholesky factor $L$ to normalise the data vectors in the mode space. Despite the simplified treatment of survey geometry and sample incompleteness, the analytic covariance still shows good agreement with a Gaussian likelihood. 

\clearpage
\bibliographystyle{mod-apsrev4-2}
\bibliography{ref.bib} 

@ARTICLE{Philcox2023:CMB_temperature,
       author = {{Philcox}, Oliver H.~E.},
        title = "{Do the CMB Temperature Fluctuations Conserve Parity?}",
      journal = {arXiv e-prints},
         year = 2023,
        month = mar,
          eid = {arXiv:2303.12106},
        pages = {arXiv:2303.12106},
archivePrefix = {arXiv},
       eprint = {2303.12106},
 primaryClass = {astro-ph.CO},
       adsurl = {https://ui.adsabs.harvard.edu/abs/2023arXiv230312106P}
}

@ARTICLE{Philcox2024:CMB_polarization,
       author = {{Philcox}, Oliver H.~E. and {Shiraishi}, Maresuke},
        title = "{Testing parity symmetry with the polarized cosmic microwave background}",
      journal = {\prd},
         year = 2024,
        month = apr,
       volume = {109},
       number = {8},
          eid = {083514},
        pages = {083514},
          doi = {10.1103/PhysRevD.109.083514},
archivePrefix = {arXiv},
       eprint = {2308.03831},
 primaryClass = {astro-ph.CO},
       adsurl = {https://ui.adsabs.harvard.edu/abs/2024PhRvD.109h3514P}
}

@ARTICLE{Jenks2023:parametrizeParity,
       author = {{Jenks}, Leah and {Choi}, Lyla and {Lagos}, Macarena and {Yunes}, Nicol{\'a}s},
        title = "{Parametrized parity violation in gravitational wave propagation}",
      journal = {\prd},
         year = 2023,
        month = aug,
       volume = {108},
       number = {4},
          eid = {044023},
        pages = {044023},
          doi = {10.1103/PhysRevD.108.044023},
archivePrefix = {arXiv},
       eprint = {2305.10478},
 primaryClass = {gr-qc},
       adsurl = {https://ui.adsabs.harvard.edu/abs/2023PhRvD.108d4023J}
}

@ARTICLE{Krolewski2024:parity,
       author = {{Krolewski}, Alex and {May}, Simon and {Smith}, Kendrick and {Hopkins}, Hans},
        title = "{No evidence for parity violation in BOSS}",
      journal = {arXiv e-prints},
         year = 2024,
        month = jul,
          eid = {arXiv:2407.03397},
        pages = {arXiv:2407.03397},
          doi = {10.48550/arXiv.2407.03397},
archivePrefix = {arXiv},
       eprint = {2407.03397},
 primaryClass = {astro-ph.CO},
       adsurl = {https://ui.adsabs.harvard.edu/abs/2024arXiv240703397K}
}

@ARTICLE{philcox_parity,
       author = {{Philcox}, Oliver H.~E.},
        title = "{Probing parity violation with the four-point correlation function of BOSS galaxies}",
      journal = {\prd},
         year = 2022,
        month = sep,
       volume = {106},
       number = {6},
          eid = {063501},
        pages = {063501},
          doi = {10.1103/PhysRevD.106.063501},
archivePrefix = {arXiv},
       eprint = {2206.04227},
 primaryClass = {astro-ph.CO},
       adsurl = {https://ui.adsabs.harvard.edu/abs/2022PhRvD.106f3501P}
}

@ARTICLE{Cho2025:Axion,
       author = {{Cho}, Hing-Tong and {Ng}, Kin-Wang},
        title = "{Four-point correlation functions in axion inflation}",
      journal = {arXiv e-prints},
         year = 2025,
        month = jun,
          eid = {arXiv:2506.02331},
        pages = {arXiv:2506.02331},
          doi = {10.48550/arXiv.2506.02331},
archivePrefix = {arXiv},
       eprint = {2506.02331},
 primaryClass = {hep-ph},
       adsurl = {https://ui.adsabs.harvard.edu/abs/2025arXiv250602331C}
}

@ARTICLE{Hou2025:DESIeven4PCF,
       author = {{Hou}, J. and {Cahn}, R.~N. and {Aguilar}, J. and {Ahlen}, S. and {Bianchi}, D. and {Brooks}, D. and {Claybaugh}, T. and {Doel}, P. and {Ferraro}, S. and {Forero-Romero}, J.~E. and {Gazta{\~n}aga}, E. and {Le Guillou}, L. and {Gutierrez}, G. and {Honscheid}, K. and {Huterer}, D. and {Ishak}, M. and {Joyce}, R. and {Juneau}, S. and {Kehoe}, R. and {Kirkby}, D. and {Kisner}, T. and {Kremin}, A. and {Lamman}, C. and {Landriau}, M. and {de la Macorra}, A. and {Manera}, M. and {de Mattia}, A. and {Miquel}, R. and {Mueller}, E. and {Nadathur}, S. and {Niz}, G. and {Percival}, W.~J. and {Prada}, F. and {P{\'e}rez-R{\`a}fols}, I. and {Ross}, A.~J. and {Rossi}, G. and {Sanchez}, E. and {Schlegel}, D. and {Schubnell}, M. and {Seo}, H. and {Silber}, J. and {Slepian}, Z. and {Sprayberry}, D. and {Tarl{\'e}}, G. and {Weaver}, B.~A. and {Zou}, H.},
        title = "{Study of the Connected Four-Point Correlation Function of Galaxies from DESI Data Release 1 Luminous Red Galaxy Sample}",
      journal = {arXiv e-prints},
         year = 2025,
        month = aug,
          eid = {arXiv:2508.09070},
        pages = {arXiv:2508.09070},
          doi = {10.48550/arXiv.2508.09070},
archivePrefix = {arXiv},
       eprint = {2508.09070},
 primaryClass = {astro-ph.CO},
       adsurl = {https://ui.adsabs.harvard.edu/abs/2025arXiv250809070H}
}

@ARTICLE{Reinhard2024:axion,
       author = {{Reinhard}, Matthew and {Slepian}, Zachary and {Hou}, Jiamin and {Greco}, Alessandro},
        title = "{Full Parity-Violating Trispectrum in Axion Inflation: Reduction to Low-D Integrals}",
      journal = {arXiv e-prints},
         year = 2024,
        month = dec,
          eid = {arXiv:2412.16037},
        pages = {arXiv:2412.16037},
          doi = {10.48550/arXiv.2412.16037},
archivePrefix = {arXiv},
       eprint = {2412.16037},
 primaryClass = {astro-ph.CO},
       adsurl = {https://ui.adsabs.harvard.edu/abs/2024arXiv241216037R}
}

@ARTICLE{Shiim2025:vector_fossil,
       author = {{Shim}, Junsup and {Pen}, Ue-Li and {Yu}, Hao-Ran and {Okumura}, Teppei},
        title = "{Probing Vector Chirality in the Early Universe}",
      journal = {\prl},
         year = 2025,
        month = oct,
       volume = {135},
       number = {14},
          eid = {141002},
        pages = {141002},
          doi = {10.1103/ym2n-lzts},
archivePrefix = {arXiv},
       eprint = {2406.06080},
 primaryClass = {astro-ph.CO},
       adsurl = {https://ui.adsabs.harvard.edu/abs/2025PhRvL.135n1002S}
}

@ARTICLE{Zhu2025:tensor_fossil,
       author = {{Zhu}, Hong-Ming and {Pen}, Ue-Li},
        title = "{Systematic Analysis of Parity-Violating Modes}",
      journal = {\prl},
         year = 2025,
        month = sep,
       volume = {135},
       number = {11},
          eid = {111003},
        pages = {111003},
          doi = {10.1103/8s52-x3r2},
archivePrefix = {arXiv},
       eprint = {2409.11400},
 primaryClass = {astro-ph.CO},
       adsurl = {https://ui.adsabs.harvard.edu/abs/2025PhRvL.135k1003Z}
}

@ARTICLE{Dai2012:TAM,
       author = {{Dai}, Liang and {Kamionkowski}, Marc and {Jeong}, Donghui},
        title = "{Total angular momentum waves for scalar, vector, and tensor fields}",
      journal = {\prd},
         year = 2012,
        month = dec,
       volume = {86},
       number = {12},
          eid = {125013},
        pages = {125013},
          doi = {10.1103/PhysRevD.86.125013},
archivePrefix = {arXiv},
       eprint = {1209.0761},
 primaryClass = {astro-ph.CO},
       adsurl = {https://ui.adsabs.harvard.edu/abs/2012PhRvD..86l5013D}
}

@ARTICLE{Jamieson2025:AxionLattice,
       author = {{Jamieson}, Drew and {Caravano}, Angelo and {Komatsu}, Eiichiro},
        title = "{Primordial Power Spectrum and Bispectrum from Lattice Simulations of Axion-U(1) Inflation}",
      journal = {arXiv e-prints},
         year = 2025,
        month = jul,
          eid = {arXiv:2507.22285},
        pages = {arXiv:2507.22285},
          doi = {10.48550/arXiv.2507.22285},
archivePrefix = {arXiv},
       eprint = {2507.22285},
 primaryClass = {astro-ph.CO},
       adsurl = {https://ui.adsabs.harvard.edu/abs/2025arXiv250722285J}
}

@ARTICLE{BAO2025:anatomy,
       author = {{Bao}, Yunjia and {Wang}, Lian-Tao and {Xianyu}, Zhong-Zhi and {Zhong}, Yi-Ming},
        title = "{Anatomy of Parity-violating Trispectra in Galaxy Surveys}",
      journal = {arXiv e-prints},
         year = 2025,
        month = apr,
          eid = {arXiv:2504.02931},
        pages = {arXiv:2504.02931},
          doi = {10.48550/arXiv.2504.02931},
archivePrefix = {arXiv},
       eprint = {2504.02931},
 primaryClass = {astro-ph.CO},
       adsurl = {https://ui.adsabs.harvard.edu/abs/2025arXiv250402931B}
}

@ARTICLE{Jaraba2025:AstrometryParity,
       author = {{Jaraba}, Santiago and {Kuroyanagi}, Sachiko and {Liang}, Qiuyue and {Lin}, Meng-Xiang and {Trodden}, Mark},
        title = "{First astrometric constraints on parity-violation in the gravitational wave background}",
      journal = {\jcap},
         year = 2025,
        month = aug,
       volume = {2025},
       number = {8},
          eid = {057},
        pages = {057},
          doi = {10.1088/1475-7516/2025/08/057},
archivePrefix = {arXiv},
       eprint = {2505.18085},
 primaryClass = {astro-ph.CO},
       adsurl = {https://ui.adsabs.harvard.edu/abs/2025JCAP...08..057J}
}

@article{Cahn202010,
       author = {{Cahn}, Robert N. and {Slepian}, Zachary},
        title = "{Isotropic N-point basis functions and their properties}",
      journal = {J. Phys. A},
         year = 2023,
        month = aug,
       volume = {56},
       number = {32},
        pages = {325204},
          doi = {10.1088/1751-8121/acdfc4},
archivePrefix = {arXiv},
       eprint = {2010.14418},
 primaryClass = {astro-ph.CO},
       adsurl = {https://ui.adsabs.harvard.edu/abs/2023JPhA...56F5204C}
}

@ARTICLE{Philcox2021boss4pcf,
       author = {{Philcox}, Oliver H.~E. and {Hou}, Jiamin and {Slepian}, Zachary},
        title = "{A First Detection of the Connected 4-Point Correlation Function of Galaxies Using the BOSS CMASS Sample}",
      journal = {arXiv e-prints},
         year = 2021,
        month = aug,
archivePrefix = {arXiv},
       eprint = {2108.01670},
 primaryClass = {astro-ph.CO},
       adsurl = {https://ui.adsabs.harvard.edu/abs/2021arXiv210801670P}
}

@ARTICLE{Dawson2013,
   author = {{Dawson}, K.~S. and {Schlegel}, D.~J. and {Ahn}, C.~P. and {Anderson}, S.~F. and 
	{Aubourg}, {\'E}. and {Bailey}, S. and {Barkhouser}, R.~H. and 
	{Bautista}, J.~E. and {Beifiori}, A. and {Berlind}, A.~A. and 
	{Bhardwaj}, V. and {Bizyaev}, D. and {Blake}, C.~H. and {Blanton}, M.~R. and 
	{Blomqvist}, M. and {Bolton}, A.~S. and {Borde}, A. and {Bovy}, J. and 
	{Brandt}, W.~N. and {Brewington}, H. and {Brinkmann}, J. and 
	{Brown}, P.~J. and {Brownstein}, J.~R. and {Bundy}, K. and {Busca}, N.~G. and 
	{Carithers}, W. and {Carnero}, A.~R. and {Carr}, M.~A. and {Chen}, Y. and 
	{Comparat}, J. and {Connolly}, N. and {Cope}, F. and {Croft}, R.~A.~C. and 
	{Cuesta}, A.~J. and {da Costa}, L.~N. and {Davenport}, J.~R.~A. and 
	{Delubac}, T. and {de Putter}, R. and {Dhital}, S. and {Ealet}, A. and 
	{Ebelke}, G.~L. and {Eisenstein}, D.~J. and {Escoffier}, S. and 
	{Fan}, X. and {Filiz Ak}, N. and {Finley}, H. and {Font-Ribera}, A. and 
	{G{\'e}nova-Santos}, R. and {Gunn}, J.~E. and {Guo}, H. and 
	{Haggard}, D. and {Hall}, P.~B. and {Hamilton}, J.-C. and {Harris}, B. and 
	{Harris}, D.~W. and {Ho}, S. and {Hogg}, D.~W. and {Holder}, D. and 
	{Honscheid}, K. and {Huehnerhoff}, J. and {Jordan}, B. and {Jordan}, W.~P. and 
	{Kauffmann}, G. and {Kazin}, E.~A. and {Kirkby}, D. and {Klaene}, M.~A. and 
	{Kneib}, J.-P. and {Le Goff}, J.-M. and {Lee}, K.-G. and {Long}, D.~C. and 
	{Loomis}, C.~P. and {Lundgren}, B. and {Lupton}, R.~H. and {Maia}, M.~A.~G. and 
	{Makler}, M. and {Malanushenko}, E. and {Malanushenko}, V. and 
	{Mandelbaum}, R. and {Manera}, M. and {Maraston}, C. and {Margala}, D. and 
	{Masters}, K.~L. and {McBride}, C.~K. and {McDonald}, P. and 
	{McGreer}, I.~D. and {McMahon}, R.~G. and {Mena}, O. and {Miralda-Escud{\'e}}, J. and 
	{Montero-Dorta}, A.~D. and {Montesano}, F. and {Muna}, D. and 
	{Myers}, A.~D. and {Naugle}, T. and {Nichol}, R.~C. and {Noterdaeme}, P. and 
	{Nuza}, S.~E. and {Olmstead}, M.~D. and {Oravetz}, A. and {Oravetz}, D.~J. and 
	{Owen}, R. and {Padmanabhan}, N. and {Palanque-Delabrouille}, N. and 
	{Pan}, K. and {Parejko}, J.~K. and {P{\^a}ris}, I. and {Percival}, W.~J. and 
	{P{\'e}rez-Fournon}, I. and {P{\'e}rez-R{\`a}fols}, I. and {Petitjean}, P. and 
	{Pfaffenberger}, R. and {Pforr}, J. and {Pieri}, M.~M. and {Prada}, F. and 
	{Price-Whelan}, A.~M. and {Raddick}, M.~J. and {Rebolo}, R. and 
	{Rich}, J. and {Richards}, G.~T. and {Rockosi}, C.~M. and {Roe}, N.~A. and 
	{Ross}, A.~J. and {Ross}, N.~P. and {Rossi}, G. and {Rubi{\~n}o-Martin}, J.~A. and 
	{Samushia}, L. and {S{\'a}nchez}, A.~G. and {Sayres}, C. and 
	{Schmidt}, S.~J. and {Schneider}, D.~P. and {Sc{\'o}ccola}, C.~G. and 
	{Seo}, H.-J. and {Shelden}, A. and {Sheldon}, E. and {Shen}, Y. and 
	{Shu}, Y. and {Slosar}, A. and {Smee}, S.~A. and {Snedden}, S.~A. and 
	{Stauffer}, F. and {Steele}, O. and {Strauss}, M.~A. and {Streblyanska}, A. and 
	{Suzuki}, N. and {Swanson}, M.~E.~C. and {Tal}, T. and {Tanaka}, M. and 
	{Thomas}, D. and {Tinker}, J.~L. and {Tojeiro}, R. and {Tremonti}, C.~A. and 
	{Vargas Maga{\~n}a}, M. and {Verde}, L. and {Viel}, M. and {Wake}, D.~A. and 
	{Watson}, M. and {Weaver}, B.~A. and {Weinberg}, D.~H. and {Weiner}, B.~J. and 
	{West}, A.~A. and {White}, M. and {Wood-Vasey}, W.~M. and {Yeche}, C. and 
	{Zehavi}, I. and {Zhao}, G.-B. and {Zheng}, Z.},
    title = "{The Baryon Oscillation Spectroscopic Survey of SDSS-III}",
  journal = {\aj},
archivePrefix = "arXiv",
   eprint = {1208.0022},
     year = 2013,
    month = jan,
   volume = 145,
      eid = {10},
    pages = {10},
      doi = {10.1088/0004-6256/145/1/10},
   adsurl = {https://ui.adsabs.harvard.edu/abs/2013AJ....145...10D}
}

@ARTICLE{DESI2013:Snowmass,
       author = {{Levi}, Michael and {Bebek}, Chris and {Beers}, Timothy and {Blum}, Robert and {Cahn}, Robert and {Eisenstein}, Daniel and {Flaugher}, Brenna and {Honscheid}, Klaus and {Kron}, Richard and {Lahav}, Ofer and {McDonald}, Patrick and {Roe}, Natalie and {Schlegel}, David and {representing the DESI collaboration}},
        title = "{The DESI Experiment, a whitepaper for Snowmass 2013}",
      journal = {arXiv e-prints},
         year = 2013,
        month = aug,
          eid = {arXiv:1308.0847},
        pages = {arXiv:1308.0847},
          doi = {10.48550/arXiv.1308.0847},
archivePrefix = {arXiv},
       eprint = {1308.0847},
 primaryClass = {astro-ph.CO},
       adsurl = {https://ui.adsabs.harvard.edu/abs/2013arXiv1308.0847L}
}

@ARTICLE{DESI2016:Science,
       author = {{DESI Collaboration} and {Aghamousa}, Amir and {Aguilar}, Jessica and {Ahlen}, Steve and {Alam}, Shadab and {Allen}, Lori E. and {Allende Prieto}, Carlos and {Annis}, James and {Bailey}, Stephen and {Balland}, Christophe and {Ballester}, Otger and {Baltay}, Charles and {Beaufore}, Lucas and {Bebek}, Chris and {Beers}, Timothy C. and {Bell}, Eric F. and {Bernal}, Jos{\'e} Luis and {Besuner}, Robert and {Beutler}, Florian and {Blake}, Chris and {Bleuler}, Hannes and {Blomqvist}, Michael and {Blum}, Robert and {Bolton}, Adam S. and {Briceno}, Cesar and {Brooks}, David and {Brownstein}, Joel R. and {Buckley-Geer}, Elizabeth and {Burden}, Angela and {Burtin}, Etienne and {Busca}, Nicolas G. and {Cahn}, Robert N. and {Cai}, Yan-Chuan and {Cardiel-Sas}, Laia and {Carlberg}, Raymond G. and {Carton}, Pierre-Henri and {Casas}, Ricard and {Castander}, Francisco J. and {Cervantes-Cota}, Jorge L. and {Claybaugh}, Todd M. and {Close}, Madeline and {Coker}, Carl T. and {Cole}, Shaun and {Comparat}, Johan and {Cooper}, Andrew P. and {Cousinou}, M. -C. and {Crocce}, Martin and {Cuby}, Jean-Gabriel and {Cunningham}, Daniel P. and {Davis}, Tamara M. and {Dawson}, Kyle S. and {de la Macorra}, Axel and {De Vicente}, Juan and {Delubac}, Timoth{\'e}e and {Derwent}, Mark and {Dey}, Arjun and {Dhungana}, Govinda and {Ding}, Zhejie and {Doel}, Peter and {Duan}, Yutong T. and {Ealet}, Anne and {Edelstein}, Jerry and {Eftekharzadeh}, Sarah and {Eisenstein}, Daniel J. and {Elliott}, Ann and {Escoffier}, St{\'e}phanie and {Evatt}, Matthew and {Fagrelius}, Parker and {Fan}, Xiaohui and {Fanning}, Kevin and {Farahi}, Arya and {Farihi}, Jay and {Favole}, Ginevra and {Feng}, Yu and {Fernandez}, Enrique and {Findlay}, Joseph R. and {Finkbeiner}, Douglas P. and {Fitzpatrick}, Michael J. and {Flaugher}, Brenna and {Flender}, Samuel and {Font-Ribera}, Andreu and {Forero-Romero}, Jaime E. and {Fosalba}, Pablo and {Frenk}, Carlos S. and {Fumagalli}, Michele and {Gaensicke}, Boris T. and {Gallo}, Giuseppe and {Garcia-Bellido}, Juan and {Gaztanaga}, Enrique and {Pietro Gentile Fusillo}, Nicola and {Gerard}, Terry and {Gershkovich}, Irena and {Giannantonio}, Tommaso and {Gillet}, Denis and {Gonzalez-de-Rivera}, Guillermo and {Gonzalez-Perez}, Violeta and {Gott}, Shelby and {Graur}, Or and {Gutierrez}, Gaston and {Guy}, Julien and {Habib}, Salman and {Heetderks}, Henry and {Heetderks}, Ian and {Heitmann}, Katrin and {Hellwing}, Wojciech A. and {Herrera}, David A. and {Ho}, Shirley and {Holland}, Stephen and {Honscheid}, Klaus and {Huff}, Eric and {Hutchinson}, Timothy A. and {Huterer}, Dragan and {Hwang}, Ho Seong and {Illa Laguna}, Joseph Maria and {Ishikawa}, Yuzo and {Jacobs}, Dianna and {Jeffrey}, Niall and {Jelinsky}, Patrick and {Jennings}, Elise and {Jiang}, Linhua and {Jimenez}, Jorge and {Johnson}, Jennifer and {Joyce}, Richard and {Jullo}, Eric and {Juneau}, St{\'e}phanie and {Kama}, Sami and {Karcher}, Armin and {Karkar}, Sonia and {Kehoe}, Robert and {Kennamer}, Noble and {Kent}, Stephen and {Kilbinger}, Martin and {Kim}, Alex G. and {Kirkby}, David and {Kisner}, Theodore and {Kitanidis}, Ellie and {Kneib}, Jean-Paul and {Koposov}, Sergey and {Kovacs}, Eve and {Koyama}, Kazuya and {Kremin}, Anthony and {Kron}, Richard and {Kronig}, Luzius and {Kueter-Young}, Andrea and {Lacey}, Cedric G. and {Lafever}, Robin and {Lahav}, Ofer and {Lambert}, Andrew and {Lampton}, Michael and {Landriau}, Martin and {Lang}, Dustin and {Lauer}, Tod R. and {Le Goff}, Jean-Marc and {Le Guillou}, Laurent and {Le Van Suu}, Auguste and {Lee}, Jae Hyeon and {Lee}, Su-Jeong and {Leitner}, Daniela and {Lesser}, Michael and {Levi}, Michael E. and {L'Huillier}, Benjamin and {Li}, Baojiu and {Liang}, Ming and {Lin}, Huan and {Linder}, Eric and {Loebman}, Sarah R. and {Luki{\'c}}, Zarija and {Ma}, Jun and {MacCrann}, Niall and {Magneville}, Christophe and {Makarem}, Laleh and {Manera}, Marc and {Manser}, Christopher J. and {Marshall}, Robert and {Martini}, Paul and {Massey}, Richard and {Matheson}, Thomas and {McCauley}, Jeremy and {McDonald}, Patrick and {McGreer}, Ian D. and {Meisner}, Aaron and {Metcalfe}, Nigel and {Miller}, Timothy N. and {Miquel}, Ramon and {Moustakas}, John and {Myers}, Adam and {Naik}, Milind and {Newman}, Jeffrey A. and {Nichol}, Robert C. and {Nicola}, Andrina and {Nicolati da Costa}, Luiz and {Nie}, Jundan and {Niz}, Gustavo and {Norberg}, Peder and {Nord}, Brian and {Norman}, Dara and {Nugent}, Peter and {O'Brien}, Thomas and {Oh}, Minji and {Olsen}, Knut A.~G.},
        title = "{The DESI Experiment Part I: Science,Targeting, and Survey Design}",
      journal = {arXiv e-prints},
         year = 2016,
        month = oct,
          eid = {arXiv:1611.00036},
        pages = {arXiv:1611.00036},
          doi = {10.48550/arXiv.1611.00036},
archivePrefix = {arXiv},
       eprint = {1611.00036},
 primaryClass = {astro-ph.IM},
       adsurl = {https://ui.adsabs.harvard.edu/abs/2016arXiv161100036D}
}

@ARTICLE{DESI2016:InstrumentDesign,
       author = {{DESI Collaboration} and {Aghamousa}, Amir and {Aguilar}, Jessica and {Ahlen}, Steve and {Alam}, Shadab and {Allen}, Lori E. and {Allende Prieto}, Carlos and {Annis}, James and {Bailey}, Stephen and {Balland}, Christophe and {Ballester}, Otger and {Baltay}, Charles and {Beaufore}, Lucas and {Bebek}, Chris and {Beers}, Timothy C. and {Bell}, Eric F. and {Bernal}, Jos{\'e} Luis and {Besuner}, Robert and {Beutler}, Florian and {Blake}, Chris and {Bleuler}, Hannes and {Blomqvist}, Michael and {Blum}, Robert and {Bolton}, Adam S. and {Briceno}, Cesar and {Brooks}, David and {Brownstein}, Joel R. and {Buckley-Geer}, Elizabeth and {Burden}, Angela and {Burtin}, Etienne and {Busca}, Nicolas G. and {Cahn}, Robert N. and {Cai}, Yan-Chuan and {Cardiel-Sas}, Laia and {Carlberg}, Raymond G. and {Carton}, Pierre-Henri and {Casas}, Ricard and {Castander}, Francisco J. and {Cervantes-Cota}, Jorge L. and {Claybaugh}, Todd M. and {Close}, Madeline and {Coker}, Carl T. and {Cole}, Shaun and {Comparat}, Johan and {Cooper}, Andrew P. and {Cousinou}, M. -C. and {Crocce}, Martin and {Cuby}, Jean-Gabriel and {Cunningham}, Daniel P. and {Davis}, Tamara M. and {Dawson}, Kyle S. and {de la Macorra}, Axel and {De Vicente}, Juan and {Delubac}, Timoth{\'e}e and {Derwent}, Mark and {Dey}, Arjun and {Dhungana}, Govinda and {Ding}, Zhejie and {Doel}, Peter and {Duan}, Yutong T. and {Ealet}, Anne and {Edelstein}, Jerry and {Eftekharzadeh}, Sarah and {Eisenstein}, Daniel J. and {Elliott}, Ann and {Escoffier}, St{\'e}phanie and {Evatt}, Matthew and {Fagrelius}, Parker and {Fan}, Xiaohui and {Fanning}, Kevin and {Farahi}, Arya and {Farihi}, Jay and {Favole}, Ginevra and {Feng}, Yu and {Fernandez}, Enrique and {Findlay}, Joseph R. and {Finkbeiner}, Douglas P. and {Fitzpatrick}, Michael J. and {Flaugher}, Brenna and {Flender}, Samuel and {Font-Ribera}, Andreu and {Forero-Romero}, Jaime E. and {Fosalba}, Pablo and {Frenk}, Carlos S. and {Fumagalli}, Michele and {Gaensicke}, Boris T. and {Gallo}, Giuseppe and {Garcia-Bellido}, Juan and {Gaztanaga}, Enrique and {Pietro Gentile Fusillo}, Nicola and {Gerard}, Terry and {Gershkovich}, Irena and {Giannantonio}, Tommaso and {Gillet}, Denis and {Gonzalez-de-Rivera}, Guillermo and {Gonzalez-Perez}, Violeta and {Gott}, Shelby and {Graur}, Or and {Gutierrez}, Gaston and {Guy}, Julien and {Habib}, Salman and {Heetderks}, Henry and {Heetderks}, Ian and {Heitmann}, Katrin and {Hellwing}, Wojciech A. and {Herrera}, David A. and {Ho}, Shirley and {Holland}, Stephen and {Honscheid}, Klaus and {Huff}, Eric and {Hutchinson}, Timothy A. and {Huterer}, Dragan and {Hwang}, Ho Seong and {Illa Laguna}, Joseph Maria and {Ishikawa}, Yuzo and {Jacobs}, Dianna and {Jeffrey}, Niall and {Jelinsky}, Patrick and {Jennings}, Elise and {Jiang}, Linhua and {Jimenez}, Jorge and {Johnson}, Jennifer and {Joyce}, Richard and {Jullo}, Eric and {Juneau}, St{\'e}phanie and {Kama}, Sami and {Karcher}, Armin and {Karkar}, Sonia and {Kehoe}, Robert and {Kennamer}, Noble and {Kent}, Stephen and {Kilbinger}, Martin and {Kim}, Alex G. and {Kirkby}, David and {Kisner}, Theodore and {Kitanidis}, Ellie and {Kneib}, Jean-Paul and {Koposov}, Sergey and {Kovacs}, Eve and {Koyama}, Kazuya and {Kremin}, Anthony and {Kron}, Richard and {Kronig}, Luzius and {Kueter-Young}, Andrea and {Lacey}, Cedric G. and {Lafever}, Robin and {Lahav}, Ofer and {Lambert}, Andrew and {Lampton}, Michael and {Landriau}, Martin and {Lang}, Dustin and {Lauer}, Tod R. and {Le Goff}, Jean-Marc and {Le Guillou}, Laurent and {Le Van Suu}, Auguste and {Lee}, Jae Hyeon and {Lee}, Su-Jeong and {Leitner}, Daniela and {Lesser}, Michael and {Levi}, Michael E. and {L'Huillier}, Benjamin and {Li}, Baojiu and {Liang}, Ming and {Lin}, Huan and {Linder}, Eric and {Loebman}, Sarah R. and {Luki{\'c}}, Zarija and {Ma}, Jun and {MacCrann}, Niall and {Magneville}, Christophe and {Makarem}, Laleh and {Manera}, Marc and {Manser}, Christopher J. and {Marshall}, Robert and {Martini}, Paul and {Massey}, Richard and {Matheson}, Thomas and {McCauley}, Jeremy and {McDonald}, Patrick and {McGreer}, Ian D. and {Meisner}, Aaron and {Metcalfe}, Nigel and {Miller}, Timothy N. and {Miquel}, Ramon and {Moustakas}, John and {Myers}, Adam and {Naik}, Milind and {Newman}, Jeffrey A. and {Nichol}, Robert C. and {Nicola}, Andrina and {Nicolati da Costa}, Luiz and {Nie}, Jundan and {Niz}, Gustavo and {Norberg}, Peder and {Nord}, Brian and {Norman}, Dara and {Nugent}, Peter and {O'Brien}, Thomas and {Oh}, Minji and {Olsen}, Knut A.~G.},
        title = "{The DESI Experiment Part II: Instrument Design}",
      journal = {arXiv e-prints},
         year = 2016,
        month = oct,
          eid = {arXiv:1611.00037},
        pages = {arXiv:1611.00037},
          doi = {10.48550/arXiv.1611.00037},
archivePrefix = {arXiv},
       eprint = {1611.00037},
 primaryClass = {astro-ph.IM},
       adsurl = {https://ui.adsabs.harvard.edu/abs/2016arXiv161100037D}
}

@ARTICLE{DESI2022:InstrumentOverview,
       author = {{DESI Collaboration} and {Abareshi}, B. and {Aguilar}, J. and {Ahlen}, S. and {Alam}, Shadab and {Alexander}, David M. and {Alfarsy}, R. and {Allen}, L. and {Allende Prieto}, C. and {Alves}, O. and {Ameel}, J. and {Armengaud}, E. and {Asorey}, J. and {Aviles}, Alejandro and {Bailey}, S. and {Balaguera-Antol{\'\i}nez}, A. and {Ballester}, O. and {Baltay}, C. and {Bault}, A. and {Beltran}, S.~F. and {Benavides}, B. and {BenZvi}, S. and {Berti}, A. and {Besuner}, R. and {Beutler}, Florian and {Bianchi}, D. and {Blake}, C. and {Blanc}, P. and {Blum}, R. and {Bolton}, A. and {Bose}, S. and {Bramall}, D. and {Brieden}, S. and {Brodzeller}, A. and {Brooks}, D. and {Brownewell}, C. and {Buckley-Geer}, E. and {Cahn}, R.~N. and {Cai}, Z. and {Canning}, R. and {Capasso}, R. and {Carnero Rosell}, A. and {Carton}, P. and {Casas}, R. and {Castander}, F.~J. and {Cervantes-Cota}, J.~L. and {Chabanier}, S. and {Chaussidon}, E. and {Chuang}, C. and {Circosta}, C. and {Cole}, S. and {Cooper}, A.~P. and {da Costa}, L. and {Cousinou}, M. -C. and {Cuceu}, A. and {Davis}, T.~M. and {Dawson}, K. and {de la Cruz-Noriega}, R. and {de la Macorra}, A. and {de Mattia}, A. and {Della Costa}, J. and {Demmer}, P. and {Derwent}, M. and {Dey}, A. and {Dey}, B. and {Dhungana}, G. and {Ding}, Z. and {Dobson}, C. and {Doel}, P. and {Donald-McCann}, J. and {Donaldson}, J. and {Douglass}, K. and {Duan}, Y. and {Dunlop}, P. and {Edelstein}, J. and {Eftekharzadeh}, S. and {Eisenstein}, D.~J. and {Enriquez-Vargas}, M. and {Escoffier}, S. and {Evatt}, M. and {Fagrelius}, P. and {Fan}, X. and {Fanning}, K. and {Fawcett}, V.~A. and {Ferraro}, S. and {Ereza}, J. and {Flaugher}, B. and {Font-Ribera}, A. and {Forero-Romero}, J.~E. and {Frenk}, C.~S. and {Fromenteau}, S. and {G{\"a}nsicke}, B.~T. and {Garcia-Quintero}, C. and {Garrison}, L. and {Gazta{\~n}aga}, E. and {Gerardi}, F. and {Gil-Mar{\'\i}n}, H. and {Gontcho A Gontcho}, S. and {Gonzalez-Morales}, Alma X. and {Gonzalez-de-Rivera}, G. and {Gonzalez-Perez}, V. and {Gordon}, C. and {Graur}, O. and {Green}, D. and {Grove}, C. and {Gruen}, D. and {Gutierrez}, G. and {Guy}, J. and {Hahn}, C. and {Harris}, S. and {Herrera}, D. and {Herrera-Alcantar}, Hiram K. and {Honscheid}, K. and {Howlett}, C. and {Huterer}, D. and {Ir{\v{s}}i{\v{c}}}, V. and {Ishak}, M. and {Jelinsky}, P. and {Jiang}, L. and {Jimenez}, J. and {Jing}, Y.~P. and {Joyce}, R. and {Jullo}, E. and {Juneau}, S. and {Kara{\c{c}}ayl{\i}}, N.~G. and {Karamanis}, M. and {Karcher}, A. and {Karim}, T. and {Kehoe}, R. and {Kent}, S. and {Kirkby}, D. and {Kisner}, T. and {Kitaura}, F. and {Koposov}, S.~E. and {Kov{\'a}cs}, A. and {Kremin}, A. and {Krolewski}, Alex and {L'Huillier}, B. and {Lahav}, O. and {Lambert}, A. and {Lamman}, C. and {Lan}, Ting-Wen and {Landriau}, M. and {Lane}, S. and {Lang}, D. and {Lange}, J.~U. and {Lasker}, J. and {Le Guillou}, L. and {Leauthaud}, A. and {Le Van Suu}, A. and {Levi}, Michael E. and {Li}, T.~S. and {Magneville}, C. and {Manera}, M. and {Manser}, Christopher J. and {Marshall}, B. and {Martini}, Paul and {McCollam}, W. and {McDonald}, P. and {Meisner}, Aaron M. and {Mena-Fern{\'a}ndez}, J. and {Meneses-Rizo}, J. and {Mezcua}, M. and {Miller}, T. and {Miquel}, R. and {Montero-Camacho}, P. and {Moon}, J. and {Moustakas}, J. and {Mueller}, E. and {Mu{\~n}oz-Guti{\'e}rrez}, Andrea and {Myers}, Adam D. and {Nadathur}, S. and {Najita}, J. and {Napolitano}, L. and {Neilsen}, E. and {Newman}, Jeffrey A. and {Nie}, J.~D. and {Ning}, Y. and {Niz}, G. and {Norberg}, P. and {Noriega}, Hern{\'a}n E. and {O'Brien}, T. and {Obuljen}, A. and {Palanque-Delabrouille}, N. and {Palmese}, A. and {Zhiwei}, P. and {Pappalardo}, D. and {PENG}, X. and {Percival}, W.~J. and {Perruchot}, S. and {Pogge}, R. and {Poppett}, C. and {Porredon}, A. and {Prada}, F. and {Prochaska}, J. and {Pucha}, R. and {P{\'e}rez-Fern{\'a}ndez}, A. and {P{\'e}rez-R{\`a}fols}, I. and {Rabinowitz}, D. and {Raichoor}, A.},
        title = "{Overview of the Instrumentation for the Dark Energy Spectroscopic Instrument}",
      journal = {\aj},
         year = 2022,
        month = nov,
       volume = {164},
       number = {5},
          eid = {207},
        pages = {207},
          doi = {10.3847/1538-3881/ac882b},
archivePrefix = {arXiv},
       eprint = {2205.10939},
 primaryClass = {astro-ph.IM},
       adsurl = {https://ui.adsabs.harvard.edu/abs/2022AJ....164..207D}
}

@ARTICLE{Myers2023:DESItargetSelection,
       author = {{Myers}, Adam D. and {Moustakas}, John and {Bailey}, Stephen and {Weaver}, Benjamin A. and {Cooper}, Andrew P. and {Forero-Romero}, Jaime E. and {Abolfathi}, Bela and {Alexander}, David M. and {Brooks}, David and {Chaussidon}, Edmond and {Chuang}, Chia-Hsun and {Dawson}, Kyle and {Dey}, Arjun and {Dey}, Biprateep and {Dhungana}, Govinda and {Doel}, Peter and {Fanning}, Kevin and {Gazta{\~n}aga}, Enrique and {Gontcho A Gontcho}, Satya and {Gonzalez-Morales}, Alma X. and {Hahn}, ChangHoon and {Herrera-Alcantar}, Hiram K. and {Honscheid}, Klaus and {Ishak}, Mustapha and {Karim}, Tanveer and {Kirkby}, David and {Kisner}, Theodore and {Koposov}, Sergey E. and {Kremin}, Anthony and {Lan}, Ting-Wen and {Landriau}, Martin and {Lang}, Dustin and {Levi}, Michael E. and {Magneville}, Christophe and {Napolitano}, Lucas and {Martini}, Paul and {Meisner}, Aaron and {Newman}, Jeffrey A. and {Palanque-Delabrouille}, Nathalie and {Percival}, Will and {Poppett}, Claire and {Prada}, Francisco and {Raichoor}, Anand and {Ross}, Ashley J. and {Schlafly}, Edward F. and {Schlegel}, David and {Schubnell}, Michael and {Tan}, Ting and {Tarle}, Gregory and {Wilson}, Michael J. and {Y{\`e}che}, Christophe and {Zhou}, Rongpu and {Zhou}, Zhimin and {Zou}, Hu},
        title = "{The Target-selection Pipeline for the Dark Energy Spectroscopic Instrument}",
      journal = {\aj},
         year = 2023,
        month = feb,
       volume = {165},
       number = {2},
          eid = {50},
        pages = {50},
          doi = {10.3847/1538-3881/aca5f9},
archivePrefix = {arXiv},
       eprint = {2208.08518},
 primaryClass = {astro-ph.IM},
       adsurl = {https://ui.adsabs.harvard.edu/abs/2023AJ....165...50M}
}

@ARTICLE{Bianchi2024:DESIfiber,
       author = {{Bianchi}, D. and {Hanif}, M.~M. S and {Carnero Rosell}, A. and {Lasker}, J. and {Ross}, A.~J. and {Pinon}, M. and {de Mattia}, A. and {White}, M. and {Ahlen}, S. and {Bailey}, S. and {Brooks}, D. and {Burtin}, E. and {Chaussidon}, E. and {Claybaugh}, T. and {Cole}, S. and {de la Macorra}, A. and {Ferraro}, S. and {Font-Ribera}, A. and {Forero-Romero}, J.~E. and {Gazta{\~n}aga}, E. and {Gontcho}, S. Gontcho A and {Gutierrez}, G. and {Guy}, J. and {Hahn}, C. and {Honscheid}, K. and {Howlett}, C. and {Juneau}, S. and {Kirkby}, D. and {Kisner}, T. and {Kremin}, A. and {Landriau}, M. and {Le Guillou}, L. and {Levi}, M.~E. and {McDonald}, P. and {Meisner}, A. and {Miquel}, R. and {Moustakas}, J. and {Palanque-Delabrouille}, N. and {Percival}, W.~J. and {Prada}, F. and {P{\'e}rez-R{\`a}fols}, I. and {Raichoor}, A. and {Rossi}, G. and {Sanchez}, E. and {Schlegel}, D. and {Schubnell}, M. and {Sharples}, R. and {Silber}, J. and {Sprayberry}, D. and {Tarl{\'e}}, G. and {Vargas-Maga{\~n}a}, M. and {Weaver}, B.~A. and {Zarrouk}, P. and {Zhou}, R. and {Zou}, H.},
        title = "{Characterization of DESI fiber assignment incompleteness effect on 2-point clustering and mitigation methods for DR1 analysis}",
      journal = {arXiv e-prints},
         year = 2024,
        month = nov,
          eid = {arXiv:2411.12025},
        pages = {arXiv:2411.12025},
          doi = {10.48550/arXiv.2411.12025},
archivePrefix = {arXiv},
       eprint = {2411.12025},
 primaryClass = {astro-ph.CO},
       adsurl = {https://ui.adsabs.harvard.edu/abs/2024arXiv241112025B}
}

@ARTICLE{DESI_Corrector_2024,
       author = {{Miller}, Timothy N. and {Doel}, Peter and {Gutierrez}, Gaston and {Besuner}, Robert and {Brooks}, David and {Gallo}, Giuseppe and {Heetderks}, Henry and {Jelinsky}, Patrick and {Kent}, Stephen M. and {Lampton}, Michael and {Levi}, Michael E. and {Liang}, Ming and {Meisner}, Aaron and {Sholl}, Michael J. and {Silber}, Joseph Harry and {Sprayberry}, David and {Aguilar}, Jessica Nicole and {de la Macorra}, Axel and {Eisenstein}, Daniel and {Fanning}, Kevin and {Font-Ribera}, Andreu and {Gazta{\~n}aga}, Enrique and {Gontcho A Gontcho}, Satya and {Honscheid}, Klaus and {Jimenez}, Jorge and {Joyce}, Dick and {Kehoe}, Robert and {Kisner}, Theodore and {Kremin}, Anthony and {Landriau}, Martin and {Le Guillou}, Laurent and {Magneville}, Christophe and {Martini}, Paul and {Miquel}, Ramon and {Moustakas}, John and {Nie}, Jundan and {Percival}, Will and {Poppett}, Claire and {Prada}, Francisco and {Rossi}, Graziano and {Schlegel}, David and {Schubnell}, Michael and {Seo}, Hee-Jong and {Sharples}, Ray and {Tarl{\'e}}, Gregory and {Vargas-Maga{\~n}a}, Mariana and {Zhou}, Zhimin and {the DESI Collaboration}},
        title = "{The Optical Corrector for the Dark Energy Spectroscopic Instrument}",
      journal = {\aj},
         year = 2024,
        month = aug,
       volume = {168},
       number = {2},
          eid = {95},
        pages = {95},
          doi = {10.3847/1538-3881/ad45fe},
archivePrefix = {arXiv},
       eprint = {2306.06310},
 primaryClass = {astro-ph.IM},
       adsurl = {https://ui.adsabs.harvard.edu/abs/2024AJ....168...95M}
}

@ARTICLE{DESI_Fiber_System_2024,
       author = {{Poppett}, Claire and {Tyas}, Luke and {Aguilar}, J. and {Bebek}, Christopher and {Bramall}, D. and {Claybaugh}, T. and {Edelstein}, J. and {Fagrelius}, P. and {Heetderks}, H. and {Jelinsky}, P. and {Jelinsky}, S. and {Lafever}, Robin and {Lambert}, A. and {Lampton}, M. and {Levi}, Michael E. and {Martini}, P. and {Rockosi}, C. and {Schmoll}, J. and {Sharples}, Ray M. and {Sirk}, Martin and {Wishnow}, Edward and {Yu}, Jiaxi and {Ahlen}, S. and {Bault}, A. and {BenZvi}, S. and {Brooks}, D. and {Cole}, S. and {de la Macorra}, A. and {Dey}, Arjun and {Doel}, P. and {Fanning}, K. and {Font-Ribera}, A. and {Forero-Romero}, J.~E. and {Gazta{\~n}aga}, E. and {Gontcho A Gontcho}, S. and {Gonzalez-Morales}, A.~X. and {Hahn}, C. and {Honscheid}, K. and {Jimenez}, J. and {Juneau}, S. and {Kirkby}, D. and {Kremin}, A. and {Landriau}, M. and {Le Guillou}, L. and {Manera}, M. and {Meisner}, A. and {Miquel}, R. and {Moustakas}, J. and {Mueller}, E. and {Mu{\~n}oz-Guti{\'e}rrez}, A. and {Myers}, A.~D. and {Nie}, J. and {Niz}, G. and {Palanque-Delabrouille}, N. and {Percival}, W.~J. and {Prada}, F. and {Rabinowitz}, D. and {Rezaie}, M. and {Rossi}, G. and {Sanchez}, E. and {Schlafly}, Edward F. and {Schlegel}, D. and {Schubnell}, M. and {Seo}, H. and {Sprayberry}, D. and {Tarl{\'e}}, G. and {Vargas-Maga{\~n}a}, M. and {Weaver}, B.~A. and {Zhou}, R.},
        title = "{Overview of the Fiber System for the Dark Energy Spectroscopic Instrument}",
      journal = {\aj},
         year = 2024,
        month = dec,
       volume = {168},
       number = {6},
          eid = {245},
        pages = {245},
          doi = {10.3847/1538-3881/ad76a4},
       adsurl = {https://ui.adsabs.harvard.edu/abs/2024AJ....168..245P}
}

@ARTICLE{DESI_spectroscopic_pipeline_2023,
       author = {{Guy}, J. and {Bailey}, S. and {Kremin}, A. and {Alam}, Shadab and {Alexander}, D.~M. and {Allende Prieto}, C. and {BenZvi}, S. and {Bolton}, A.~S. and {Brooks}, D. and {Chaussidon}, E. and {Cooper}, A.~P. and {Dawson}, K. and {de la Macorra}, A. and {Dey}, A. and {Dey}, Biprateep and {Dhungana}, G. and {Eisenstein}, D.~J. and {Font-Ribera}, A. and {Forero-Romero}, J.~E. and {Gazta{\~n}aga}, E. and {Gontcho A Gontcho}, S. and {Green}, D. and {Honscheid}, K. and {Ishak}, M. and {Kehoe}, R. and {Kirkby}, D. and {Kisner}, T. and {Koposov}, Sergey E. and {Lan}, Ting-Wen and {Landriau}, M. and {Le Guillou}, L. and {Levi}, Michael E. and {Magneville}, C. and {Manser}, Christopher J. and {Martini}, P. and {Meisner}, Aaron M. and {Miquel}, R. and {Moustakas}, J. and {Myers}, Adam D. and {Newman}, Jeffrey A. and {Nie}, Jundan and {Palanque-Delabrouille}, N. and {Percival}, W.~J. and {Poppett}, C. and {Prada}, F. and {Raichoor}, A. and {Ravoux}, C. and {Ross}, A.~J. and {Schlafly}, E.~F. and {Schlegel}, D. and {Schubnell}, M. and {Sharples}, Ray M. and {Tarl{\'e}}, Gregory and {Weaver}, B.~A. and {Y{\'e}che}, Christophe and {Zhou}, Rongpu and {Zhou}, Zhimin and {Zou}, H.},
        title = "{The Spectroscopic Data Processing Pipeline for the Dark Energy Spectroscopic Instrument}",
      journal = {\aj},
         year = 2023,
        month = apr,
       volume = {165},
       number = {4},
          eid = {144},
        pages = {144},
          doi = {10.3847/1538-3881/acb212},
archivePrefix = {arXiv},
       eprint = {2209.14482},
 primaryClass = {astro-ph.IM},
       adsurl = {https://ui.adsabs.harvard.edu/abs/2023AJ....165..144G}
}

@ARTICLE{ForeroSanchez2025:DESIcovariance,
       author = {{Forero-S{\'a}nchez}, D. and {Rashkovetskyi}, M. and {Alves}, O. and {de Mattia}, A. and {Padmanabhan}, N. and {Seo}, H. and {Nadathur}, S. and {Ross}, A.~J. and {Gil-Mar{\'\i}n}, H. and {Zarrouk}, P. and {Yu}, J. and {Ding}, Z. and {Andrade}, U. and {Chen}, X. and {Garcia-Quintero}, C. and {Mena-Fern{\'a}ndez}, J. and {Ahlen}, S. and {Bianchi}, D. and {Brooks}, D. and {Burtin}, E. and {Chaussidon}, E. and {Claybaugh}, T. and {Cole}, S. and {de la Macorra}, A. and {Enriquez-Vargas}, M. and {Gazta{\~n}aga}, E. and {Gutierrez}, G. and {Honscheid}, K. and {Howlett}, C. and {Kisner}, T. and {Landriau}, M. and {Le Guillou}, L. and {Levi}, M.~E. and {Miquel}, R. and {Moustakas}, J. and {Palanque-Delabrouille}, N. and {Percival}, W.~J. and {P{\'e}rez-R{\`a}fols}, I. and {Rossi}, G. and {Sanchez}, E. and {Schlegel}, D. and {Schubnell}, M. and {Sprayberry}, D. and {Tarl{\'e}}, G. and {Vargas-Maga{\~n}a}, M. and {Weaver}, B.~A. and {Zou}, H.},
        title = "{Analytical and EZmock covariance validation for the DESI 2024 results}",
      journal = {\jcap},
         year = 2025,
        month = apr,
       volume = {2025},
       number = {4},
          eid = {055},
        pages = {055},
          doi = {10.1088/1475-7516/2025/04/055},
archivePrefix = {arXiv},
       eprint = {2411.12027},
 primaryClass = {astro-ph.CO},
       adsurl = {https://ui.adsabs.harvard.edu/abs/2025JCAP...04..055F}
}

@ARTICLE{DESI2024:SampleDefinition,
       author = {{DESI Collaboration} and {Adame}, A.~G. and {Aguilar}, J. and {Ahlen}, S. and {Alam}, S. and {Alexander}, D.~M. and {Alvarez}, M. and {Alves}, O. and {Anand}, A. and {Andrade}, U. and {Armengaud}, E. and {Avila}, S. and {Aviles}, A. and {Awan}, H. and {Bailey}, S. and {Baltay}, C. and {Bault}, A. and {Behera}, J. and {BenZvi}, S. and {Beutler}, F. and {Bianchi}, D. and {Blake}, C. and {Blum}, R. and {Brieden}, S. and {Brodzeller}, A. and {Brooks}, D. and {Brown}, Z. and {Buckley-Geer}, E. and {Burtin}, E. and {Calderon}, R. and {Canning}, R. and {Carnero Rosell}, A. and {Cereskaite}, R. and {Cervantes-Cota}, J.~L. and {Chabanier}, S. and {Chaussidon}, E. and {Chaves-Montero}, J. and {Chen}, S. and {Chen}, X. and {Claybaugh}, T. and {Cole}, S. and {Cuceu}, A. and {Davis}, T.~M. and {Dawson}, K. and {de la Macorra}, A. and {de Mattia}, A. and {Deiosso}, N. and {Demina}, R. and {Dey}, A. and {Dey}, B. and {Ding}, Z. and {Doel}, P. and {Edelstein}, J. and {Eftekharzadeh}, S. and {Eisenstein}, D.~J. and {Elliott}, A. and {Fagrelius}, P. and {Fanning}, K. and {Ferraro}, S. and {Ereza}, J. and {Findlay}, N. and {Flaugher}, B. and {Font-Ribera}, A. and {Forero-S{\'a}nchez}, D. and {Forero-Romero}, J.~E. and {Frenk}, C.~S. and {Garcia-Quintero}, C. and {Gazta{\~n}aga}, E. and {Gil-Mar{\'\i}n}, H. and {Gontcho}, S. Gontcho A and {Gonzalez-Morales}, A.~X. and {Gonzalez-Perez}, V. and {Gordon}, C. and {Green}, D. and {Gruen}, D. and {Gsponer}, R. and {Gutierrez}, G. and {Guy}, J. and {Hadzhiyska}, B. and {Hahn}, C. and {Hanif}, M.~M. S and {Herrera-Alcantar}, H.~K. and {Honscheid}, K. and {Hou}, J. and {Howlett}, C. and {Huterer}, D. and {Ir{\v{s}}i{\v{c}}}, V. and {Ishak}, M. and {Juneau}, S. and {Kara{\c{c}}ayl{\i}}, N.~G. and {Kehoe}, R. and {Kent}, S. and {Kirkby}, D. and {Kitaura}, F. -S. and {Kong}, H. and {Kremin}, A. and {Krolewski}, A. and {Lai}, Y. and {Lan}, T. -W. and {Landriau}, M. and {Lang}, D. and {Lasker}, J. and {Le Goff}, J.~M. and {Le Guillou}, L. and {Leauthaud}, A. and {Levi}, M.~E. and {Li}, T.~S. and {Lodha}, K. and {Magneville}, C. and {Manera}, M. and {Margala}, D. and {Martini}, P. and {Maus}, M. and {McDonald}, P. and {Medina-Varela}, L. and {Meisner}, A. and {Mena-Fern{\'a}ndez}, J. and {Miquel}, R. and {Moon}, J. and {Moore}, S. and {Moustakas}, J. and {Mudur}, N. and {Mueller}, E. and {Mu{\~n}oz-Guti{\'e}rrez}, A. and {Myers}, A.~D. and {Nadathur}, S. and {Napolitano}, L. and {Neveux}, R. and {Newman}, J.~A. and {Nguyen}, N.~M. and {Nie}, J. and {Niz}, G. and {Noriega}, H.~E. and {Padmanabhan}, N. and {Paillas}, E. and {Palanque-Delabrouille}, N. and {Pan}, J. and {Penmetsa}, S. and {Percival}, W.~J. and {Pieri}, M.~M. and {Pinon}, M. and {Poppett}, C. and {Porredon}, A. and {Prada}, F. and {P{\'e}rez-Fern{\'a}ndez}, A. and {P{\'e}rez-R{\`a}fols}, I. and {Rabinowitz}, D. and {Raichoor}, A. and {Ram{\'\i}rez-P{\'e}rez}, C. and {Ramirez-Solano}, S. and {Rashkovetskyi}, M. and {Ravoux}, C. and {Rezaie}, M. and {Rich}, J. and {Rocher}, A. and {Rockosi}, C. and {Roe}, N.~A. and {Rosado-Marin}, A. and {Ross}, A.~J. and {Rossi}, G. and {Ruggeri}, R. and {Ruhlmann-Kleider}, V. and {Samushia}, L. and {Sanchez}, E. and {Saulder}, C. and {Schlafly}, E.~F. and {Schlegel}, D. and {Scholte}, D. and {Schubnell}, M. and {Seo}, H. and {Sharples}, R. and {Silber}, J. and {Slosar}, A. and {Smith}, A. and {Sprayberry}, D. and {Tan}, T. and {Tarl{\'e}}, G. and {Trusov}, S. and {Vaisakh}, R. and {Valcin}, D. and {Valdes}, F. and {Vargas-Maga{\~n}a}, M. and {Verde}, L. and {Walther}, M. and {Wang}, B. and {Wang}, M.~S. and {Weaver}, B.~A. and {Weaverdyck}, N. and {Wechsler}, R.~H. and {Weinberg}, D.~H. and {White}, M. and {Wilson}, M.~J. and {Yu}, J. and {Yu}, Y. and {Yuan}, S. and {Y{\`e}che}, C. and {Zaborowski}, E.~A. and {Zarrouk}, P. and {Zhang}, H. and {Zhao}, C.},
        title = "{DESI 2024 II: Sample Definitions, Characteristics, and Two-point Clustering Statistics}",
      journal = {arXiv e-prints},
         year = 2024,
        month = nov,
          eid = {arXiv:2411.12020},
        pages = {arXiv:2411.12020},
          doi = {10.48550/arXiv.2411.12020},
archivePrefix = {arXiv},
       eprint = {2411.12020},
 primaryClass = {astro-ph.CO},
       adsurl = {https://ui.adsabs.harvard.edu/abs/2024arXiv241112020D}
}

@ARTICLE{Lasker2025:DESIFiberAssign,
       author = {{Lasker}, J. and {Carnero Rosell}, A. and {Myers}, A.~D. and {Ross}, A.~J. and {Bianchi}, D. and {Hanif}, M.~M.~S. and {Kehoe}, R. and {de Mattia}, A. and {Napolitano}, L. and {Percival}, W.~J. and {Staten}, R. and {Aguilar}, J. and {Ahlen}, S. and {Bigwood}, L. and {Brooks}, D. and {Claybaugh}, T. and {Cole}, S. and {de la Macorra}, A. and {Ding}, Z. and {Doel}, P. and {Fanning}, K. and {Forero-Romero}, J.~E. and {Gazta{\~n}aga}, E. and {Gontcho A Gontcho}, S. and {Gutierrez}, G. and {Honscheid}, K. and {Howlett}, C. and {Juneau}, S. and {Kremin}, A. and {Landriau}, M. and {Le Guillou}, L. and {Levi}, M.~E. and {Manera}, M. and {Meisner}, A. and {Miquel}, R. and {Moustakas}, J. and {Mueller}, E. and {Nie}, J. and {Niz}, G. and {Oh}, M. and {Palanque-Delabrouille}, N. and {Poppett}, C. and {Prada}, F. and {Rezaie}, M. and {Rossi}, G. and {Sanchez}, E. and {Schlegel}, D. and {Schubnell}, M. and {Seo}, H. and {Sprayberry}, D. and {Tarl{\'e}}, G. and {Vargas-Maga{\~n}a}, M. and {Weaver}, B.~A. and {Wilson}, Michael J. and {Zheng}, Y. and {DESI Collaboration}},
        title = "{Production of alternate realizations of DESI fiber assignment for unbiased clustering measurement in data and simulations}",
      journal = {\jcap},
         year = 2025,
        month = jan,
       volume = {2025},
       number = {1},
          eid = {127},
        pages = {127},
          doi = {10.1088/1475-7516/2025/01/127},
archivePrefix = {arXiv},
       eprint = {2404.03006},
 primaryClass = {astro-ph.CO},
       adsurl = {https://ui.adsabs.harvard.edu/abs/2025JCAP...01..127L}
}

@ARTICLE{Wang2022:RomanHLSS,
       author = {{Wang}, Yun and {Zhai}, Zhongxu and {Alavi}, Anahita and {Massara}, Elena and {Pisani}, Alice and {Benson}, Andrew and {Hirata}, Christopher M. and {Samushia}, Lado and {Weinberg}, David H. and {Colbert}, James and {Dor{\'e}}, Olivier and {Eifler}, Tim and {Heinrich}, Chen and {Ho}, Shirley and {Krause}, Elisabeth and {Padmanabhan}, Nikhil and {Spergel}, David and {Teplitz}, Harry I.},
        title = "{The High Latitude Spectroscopic Survey on the Nancy Grace Roman Space Telescope}",
      journal = {\apj},
         year = 2022,
        month = mar,
       volume = {928},
       number = {1},
          eid = {1},
        pages = {1},
          doi = {10.3847/1538-4357/ac4973},
archivePrefix = {arXiv},
       eprint = {2110.01829},
 primaryClass = {astro-ph.CO},
       adsurl = {https://ui.adsabs.harvard.edu/abs/2022ApJ...928....1W}
}

@ARTICLE{Ng2023:AmplBirefrigenceGWTC3,
       author = {{Ng}, Thomas C.~K. and {Isi}, Maximiliano and {Wong}, Kaze W.~K. and {Farr}, Will M.},
        title = "{Constraining gravitational wave amplitude birefringence with GWTC-3}",
      journal = {\prd},
         year = 2023,
        month = oct,
       volume = {108},
       number = {8},
        pages = {084068},
          doi = {10.1103/PhysRevD.108.084068},
archivePrefix = {arXiv},
       eprint = {2305.05844},
 primaryClass = {gr-qc},
       adsurl = {https://ui.adsabs.harvard.edu/abs/2023PhRvD.108h4068N}
}

@ARTICLE{Inomata2024:ORF,
       author = {{Inomata}, Keisuke and {Kamionkowski}, Marc and {Toral}, Celia M. and {Taylor}, Stephen R.},
        title = "{Overlap reduction functions for pulsar timing arrays and astrometry}",
      journal = {\prd},
         year = 2024,
        month = sep,
       volume = {110},
       number = {6},
          eid = {063547},
        pages = {063547},
          doi = {10.1103/PhysRevD.110.063547},
archivePrefix = {arXiv},
       eprint = {2406.00096},
 primaryClass = {astro-ph.CO},
       adsurl = {https://ui.adsabs.harvard.edu/abs/2024PhRvD.110f3547I}
}

@ARTICLE{Slepian2025:DESIodd4PCF,
       author = {{Slepian}, Zachary and {Krolewski}, Alex and {Greco}, Alessandro and {May}, Simon and {Ortola Leonard}, William and {Kamalinejad}, Farshad and {Chellino}, Jessica and {Reinhard}, Matthew and {Fernandez}, Elena and {Prada}, Francisco and {Ahlen}, Steven and {Bianchi}, Davide and {Brooks}, David and {Claybaugh}, Todd and {de la Macorra}, Axel and {de Mattia}, Arnaud and {Dey}, Biprateep and {Doel}, Peter and {Gaztanaga}, Enrique and {Gutierrez}, Gaston and {Honscheid}, Klaus and {Huterer}, Dragan and {Joyce}, Dick and {Kehoe}, Robert and {Kirkby}, David and {Kisner}, Theodore and {Landriau}, Martin and {Le Guillou}, Laurent and {Manera}, Marc and {Meisner}, Aaron and {Miquel}, Ramon and {Nadathur}, Seshadri and {Percival}, Will and {Ross}, Ashley and {Sanchez}, Eusebio and {Schlegel}, David and {Schubnell}, Michael and {Seo}, Hee-Jong and {Silber}, Joseph and {Sprayberry}, David and {Tarle}, Gregory},
        title = "{Measurement of Parity-Violating Modes of the Dark Energy Spectroscopic Instrument (DESI) Year 1 Luminous Red Galaxies' 4-Point Correlation Function}",
      journal = {arXiv e-prints},
         year = 2025,
        month = aug,
          eid = {arXiv:2508.09133},
        pages = {arXiv:2508.09133},
          doi = {10.48550/arXiv.2508.09133},
archivePrefix = {arXiv},
       eprint = {2508.09133},
 primaryClass = {astro-ph.CO},
       adsurl = {https://ui.adsabs.harvard.edu/abs/2025arXiv250809133S}
}

@ARTICLE{Minami2020:Birefringence,
       author = {{Minami}, Yuto and {Komatsu}, Eiichiro},
        title = "{New Extraction of the Cosmic Birefringence from the Planck 2018 Polarization Data}",
      journal = {\prl},
         year = 2020,
        month = nov,
       volume = {125},
       number = {22},
          eid = {221301},
        pages = {221301},
          doi = {10.1103/PhysRevLett.125.221301},
archivePrefix = {arXiv},
       eprint = {2011.11254},
 primaryClass = {astro-ph.CO},
       adsurl = {https://ui.adsabs.harvard.edu/abs/2020PhRvL.125v1301M}
}

@ARTICLE{Chuang2015:EZmock,
       author = {{Chuang}, Chia-Hsun and {Kitaura}, Francisco-Shu and {Prada}, Francisco and {Zhao}, Cheng and {Yepes}, Gustavo},
        title = "{EZmocks: extending the Zel'dovich approximation to generate mock galaxy catalogues with accurate clustering statistics}",
      journal = {\mnras},
         year = 2015,
        month = jan,
       volume = {446},
       number = {3},
        pages = {2621-2628},
          doi = {10.1093/mnras/stu2301},
archivePrefix = {arXiv},
       eprint = {1409.1124},
 primaryClass = {astro-ph.CO},
       adsurl = {https://ui.adsabs.harvard.edu/abs/2015MNRAS.446.2621C}
}

@ARTICLE{Zhao2021:EZmock,
       author = {{Zhao}, Cheng and {Chuang}, Chia-Hsun and {Bautista}, Julian and {de Mattia}, Arnaud and {Raichoor}, Anand and {Ross}, Ashley J. and {Hou}, Jiamin and {Neveux}, Richard and {Tao}, Charling and {Burtin}, Etienne and {Dawson}, Kyle S. and {de la Torre}, Sylvain and {Gil-Mar{\'\i}n}, H{\'e}ctor and {Kneib}, Jean-Paul and {Percival}, Will J. and {Rossi}, Graziano and {Tamone}, Am{\'e}lie and {Tinker}, Jeremy L. and {Zhao}, Gong-Bo and {Alam}, Shadab and {Mueller}, Eva-Maria},
        title = "{The completed SDSS-IV extended Baryon Oscillation Spectroscopic Survey: 1000 multi-tracer mock catalogues with redshift evolution and systematics for galaxies and quasars of the final data release}",
      journal = {\mnras},
         year = 2021,
        month = may,
       volume = {503},
       number = {1},
        pages = {1149-1173},
          doi = {10.1093/mnras/stab510},
archivePrefix = {arXiv},
       eprint = {2007.08997},
 primaryClass = {astro-ph.CO},
       adsurl = {https://ui.adsabs.harvard.edu/abs/2021MNRAS.503.1149Z}
}

@ARTICLE{Maksimova2021:Abacus,
       author = {{Maksimova}, Nina A. and {Garrison}, Lehman H. and {Eisenstein}, Daniel J. and {Hadzhiyska}, Boryana and {Bose}, Sownak and {Satterthwaite}, Thomas P.},
        title = "{ABACUSSUMMIT: a massive set of high-accuracy, high-resolution N-body simulations}",
      journal = {\mnras},
         year = 2021,
        month = dec,
       volume = {508},
       number = {3},
        pages = {4017-4037},
          doi = {10.1093/mnras/stab2484},
archivePrefix = {arXiv},
       eprint = {2110.11398},
 primaryClass = {astro-ph.CO},
       adsurl = {https://ui.adsabs.harvard.edu/abs/2021MNRAS.508.4017M}
}

@ARTICLE{Garrison2019:Abacus,
       author = {{Garrison}, Lehman H. and {Eisenstein}, Daniel J. and {Pinto}, Philip A.},
        title = "{A high-fidelity realization of the Euclid code comparison N-body simulation with ABACUS}",
      journal = {\mnras},
         year = 2019,
        month = may,
       volume = {485},
       number = {3},
        pages = {3370-3377},
          doi = {10.1093/mnras/stz634},
archivePrefix = {arXiv},
       eprint = {1810.02916},
 primaryClass = {astro-ph.CO},
       adsurl = {https://ui.adsabs.harvard.edu/abs/2019MNRAS.485.3370G}
}

@ARTICLE{Yuan2024:DESIAbacusHODLRG,
       author = {{Yuan}, Sihan and {Zhang}, Hanyu and {Ross}, Ashley J. and {Donald-McCann}, Jamie and {Hadzhiyska}, Boryana and {Wechsler}, Risa H. and {Zheng}, Zheng and {Alam}, Shadab and {Gonzalez-Perez}, Violeta and {Aguilar}, Jessica Nicole and {Ahlen}, Steven and {Bianchi}, Davide and {Brooks}, David and {de la Macorra}, Axel and {Fanning}, Kevin and {Forero-Romero}, Jaime E. and {Honscheid}, Klaus and {Ishak}, Mustapha and {Kehoe}, Robert and {Lasker}, James and {Landriau}, Martin and {Manera}, Marc and {Martini}, Paul and {Meisner}, Aaron and {Miquel}, Ramon and {Moustakas}, John and {Nadathur}, Seshadri and {Newman}, Jeffrey A. and {Nie}, Jundan and {Percival}, Will and {Poppett}, Claire and {Rocher}, Antoine and {Rossi}, Graziano and {Sanchez}, Eusebio and {Samushia}, Lado and {Schubnell}, Michael and {Seo}, Hee-Jong and {Tarl{\'e}}, Gregory and {Weaver}, Benjamin Alan and {Yu}, Jiaxi and {Zhou}, Zhimin and {Zou}, Hu},
        title = "{The DESI one-per cent survey: exploring the halo occupation distribution of luminous red galaxies and quasi-stellar objects with ABACUSSUMMIT}",
      journal = {\mnras},
         year = 2024,
        month = may,
       volume = {530},
       number = {1},
        pages = {947-965},
          doi = {10.1093/mnras/stae359},
archivePrefix = {arXiv},
       eprint = {2306.06314},
 primaryClass = {astro-ph.CO},
       adsurl = {https://ui.adsabs.harvard.edu/abs/2024MNRAS.530..947Y}
}

@ARTICLE{DESI_Survey_Operation_2023,
       author = {{Schlafly}, Edward F. and {Kirkby}, David and {Schlegel}, David J. and {Myers}, Adam D. and {Raichoor}, Anand and {Dawson}, Kyle and {Aguilar}, Jessica and {Allende Prieto}, Carlos and {Bailey}, Stephen and {BenZvi}, Segev and {Bermejo-Climent}, Jose and {Brooks}, David and {de la Macorra}, Axel and {Dey}, Arjun and {Doel}, Peter and {Fanning}, Kevin and {Font-Ribera}, Andreu and {Forero-Romero}, Jaime E. and {Garc{\'\i}a-Bellido}, Juan and {Gontcho A Gontcho}, Satya and {Guy}, Julien and {Hahn}, ChangHoon and {Honscheid}, Klaus and {Ishak}, Mustapha and {Juneau}, St{\'e}phanie and {Kehoe}, Robert and {Kisner}, Theodore and {Kremin}, Anthony and {Landriau}, Martin and {Lang}, Dustin A. and {Lasker}, James and {Levi}, Michael E. and {Magneville}, Christophe and {Manser}, Christopher J. and {Martini}, Paul and {Meisner}, Aaron M. and {Miquel}, Ramon and {Moustakas}, John and {Newman}, Jeffrey A. and {Nie}, Jundan and {Palanque-Delabrouille}, Nathalie. and {Percival}, Will J. and {Poppett}, Claire and {Rockosi}, Constance and {Ross}, Ashley J. and {Rossi}, Graziano and {Tarl{\'e}}, Gregory and {Weaver}, Benjamin A. and {Y{\`e}che}, Christophe and {Zhou}, Rongpu and {DESI Collaboration}},
        title = "{Survey Operations for the Dark Energy Spectroscopic Instrument}",
      journal = {\aj},
         year = 2023,
        month = dec,
       volume = {166},
       number = {6},
          eid = {259},
        pages = {259},
          doi = {10.3847/1538-3881/ad0832},
archivePrefix = {arXiv},
       eprint = {2306.06309},
 primaryClass = {astro-ph.CO},
       adsurl = {https://ui.adsabs.harvard.edu/abs/2023AJ....166..259S}
}

@ARTICLE{DESI_DR1_2025,
       author = {{DESI Collaboration} and {Abdul-Karim}, M. and {Adame}, A.~G. and {Aguado}, D. and {Aguilar}, J. and {Ahlen}, S. and {Alam}, S. and {Aldering}, G. and {Alexander}, D.~M. and {Alfarsy}, R. and {Allen}, L. and {Allende Prieto}, C. and {Alves}, O. and {Anand}, A. and {Andrade}, U. and {Armengaud}, E. and {Avila}, S. and {Aviles}, A. and {Awan}, H. and {Bailey}, S. and {Baleato Lizancos}, A. and {Ballester}, O. and {Bault}, A. and {Bautista}, J. and {BenZvi}, S. and {Beraldo e Silva}, L. and {Bermejo-Climent}, J.~R. and {Beutler}, F. and {Bianchi}, D. and {Blake}, C. and {Blum}, R. and {Bolton}, A.~S. and {Bonici}, M. and {Brieden}, S. and {Brodzeller}, A. and {Brooks}, D. and {Buckley-Geer}, E. and {Burtin}, E. and {Canning}, R. and {Carnero Rosell}, A. and {Carr}, A. and {Carrilho}, P. and {Casas}, L. and {Castander}, F.~J. and {Cereskaite}, R. and {Cervantes-Cota}, J.~L. and {Chaussidon}, E. and {Chaves-Montero}, J. and {Chen}, S. and {Chen}, X. and {Claybaugh}, T. and {Cole}, S. and {Cooper}, A.~P. and {Cousinou}, M. -C. and {Cuceu}, A. and {Davis}, T.~M. and {Dawson}, K.~S. and {de Belsunce}, R. and {de la Cruz}, R. and {de la Macorra}, A. and {de Mattia}, A. and {Deiosso}, N. and {Della Costa}, J. and {Demina}, R. and {Demirbozan}, U. and {DeRose}, J. and {Dey}, A. and {Dey}, B. and {Ding}, J. and {Ding}, Z. and {Doel}, P. and {Douglass}, K. and {Dowicz}, M. and {Ebina}, H. and {Edelstein}, J. and {Eisenstein}, D.~J. and {Elbers}, W. and {Emas}, N. and {Escoffier}, S. and {Fagrelius}, P. and {Fan}, X. and {Fanning}, K. and {Fawcett}, V.~A. and {Forero-S\'anchez}, E. and {Ferraro}, S. and {Findlay}, N. and {Font-Ribera}, A. and {Forero-Romero}, J.~E. and {Forero-S\'anchez}, D. and {Frenk}, C.~S. and {G\textbackslash''ansicke}, B.~T. and {Galbany}, L. and {Garc\textbackslash'ia-Bellido}, J. and {Garcia-Quintero}, C. and {Garrison}, L.~H. and {Gazta\~naga}, E. and {Gil-Mar\textbackslash'in}, H. and {Gnedin}, O.~Y. and {Gontcho}, S. Gontcho A and {Gonzalez-Morales}, A.~X. and {Gonzalez-Perez}, V. and {Gordon}, C. and {Graur}, O. and {Green}, D. and {Gruen}, D. and {Gsponer}, R. and {Guandalin}, C. and {Gutierrez}, G. and {Guy}, J. and {Hahn}, C. and {Han}, J.~J. and {Han}, J. and {He}, S. and {Herrera-Alcantar}, H.~K. and {Honscheid}, K. and {Hou}, J. and {Howlett}, C. and {Huterer}, D. and {Ir\v{s}i\v{c}}, V. and {Ishak}, M. and {Jacques}, A. and {Jimenez}, J. and {Jing}, Y.~P. and {Joachimi}, B. and {Joudaki}, S. and {Joyce}, R. and {Jullo}, E. and {Juneau}, S. and {Kara{\c{c}}ayl{\i}}, N.~G. and {Karim}, T. and {Kehoe}, R. and {Kent}, S. and {Khederlarian}, A. and {Kirkby}, D. and {Kisner}, T. and {Kitaura}, F. -S. and {Kizhuprakkat}, N. and {Kong}, H. and {Koposov}, S.~E. and {Kremin}, A. and {Krolewski}, A. and {Lahav}, O. and {Lai}, Y. and {Lamman}, C. and {Lan}, T. -W. and {Landriau}, M. and {Lang}, D. and {Lange}, J.~U. and {Lasker}, J. and {Le Goff}, J.~M. and {Le Guillou}, L. and {Leauthaud}, A. and {Levi}, M.~E. and {Li}, S. and {Li}, T.~S. and {Lodha}, K. and {Lokken}, M. and {Luo}, Y. and {Magneville}, C. and {Manera}, M. and {Manser}, C.~J. and {Margala}, D. and {Martini}, P. and {Maus}, M. and {McCullough}, J. and {McDonald}, P. and {Medina}, G.~E. and {Medina-Varela}, L. and {Meisner}, A. and {Mena-Fern\textbackslash'andez}, J. and {Menegas}, A. and {Mezcua}, M. and {Miquel}, R. and {Montero-Camacho}, P. and {Moon}, J. and {Moustakas}, J. and {Mu\~noz-Guti\'errez}, A. and {Mu\~noz-Santos}, D. and {Myers}, A.~D. and {Myles}, J. and {Nadathur}, S. and {Najita}, J. and {Napolitano}, L. and {Newman}, J.~A. and {Nikakhtar}, F. and {Nikutta}, R. and {Niz}, G. and {Noriega}, H.~E. and {Padmanabhan}, N. and {Paillas}, E. and {Palanque-Delabrouille}, N. and {Palmese}, A. and {Pan}, J. and {Pan}, Z. and {Parkinson}, D. and {Peacock}, J. and {Percival}, W.~J. and {P\'erez-Fern\'andez}, A. and {P{\'e}rez-R\`afols}, I. and {Peterson}, P.},
        title = "{Data Release 1 of the Dark Energy Spectroscopic Instrument}",
      journal = {arXiv e-prints},
         year = 2025,
        month = mar,
          eid = {arXiv:2503.14745},
        pages = {arXiv:2503.14745},
          doi = {10.48550/arXiv.2503.14745},
archivePrefix = {arXiv},
       eprint = {2503.14745},
 primaryClass = {astro-ph.CO},
       adsurl = {https://ui.adsabs.harvard.edu/abs/2025arXiv250314745D}
}

@ARTICLE{DESI_DR1_cosmology_2024,
       author = {{DESI Collaboration} and {Adame}, A.~G. and {Aguilar}, J. and {Ahlen}, S. and {Alam}, S. and {Alexander}, D.~M. and {Allende Prieto}, C. and {Alvarez}, M. and {Alves}, O. and {Anand}, A. and {Andrade}, U. and {Armengaud}, E. and {Avila}, S. and {Aviles}, A. and {Awan}, H. and {Bahr-Kalus}, B. and {Bailey}, S. and {Baltay}, C. and {Bault}, A. and {Behera}, J. and {BenZvi}, S. and {Beutler}, F. and {Bianchi}, D. and {Blake}, C. and {Blum}, R. and {Bonici}, M. and {Brieden}, S. and {Brodzeller}, A. and {Brooks}, D. and {Buckley-Geer}, E. and {Burtin}, E. and {Calderon}, R. and {Canning}, R. and {Carnero Rosell}, A. and {Cereskaite}, R. and {Cervantes-Cota}, J.~L. and {Chabanier}, S. and {Chaussidon}, E. and {Chaves-Montero}, J. and {Chebat}, D. and {Chen}, S. and {Chen}, X. and {Claybaugh}, T. and {Cole}, S. and {Cuceu}, A. and {Davis}, T.~M. and {Dawson}, K. and {de la Macorra}, A. and {de Mattia}, A. and {Deiosso}, N. and {Dey}, A. and {Dey}, B. and {Ding}, Z. and {Doel}, P. and {Edelstein}, J. and {Eftekharzadeh}, S. and {Eisenstein}, D.~J. and {Elbers}, W. and {Elliott}, A. and {Fagrelius}, P. and {Fanning}, K. and {Ferraro}, S. and {Ereza}, J. and {Findlay}, N. and {Flaugher}, B. and {Font-Ribera}, A. and {Forero-S{\'a}nchez}, D. and {Forero-Romero}, J.~E. and {Frenk}, C.~S. and {Garcia-Quintero}, C. and {Garrison}, L.~H. and {Gazta{\~n}aga}, E. and {Gil-Mar{\'\i}n}, H. and {Gontcho}, S. Gontcho A and {Gonzalez-Morales}, A.~X. and {Gonzalez-Perez}, V. and {Gordon}, C. and {Green}, D. and {Gruen}, D. and {Gsponer}, R. and {Gutierrez}, G. and {Guy}, J. and {Hadzhiyska}, B. and {Hahn}, C. and {Hanif}, M.~M. S and {Herrera-Alcantar}, H.~K. and {Honscheid}, K. and {Howlett}, C. and {Huterer}, D. and {Ir{\v{s}}i{\v{c}}}, V. and {Ishak}, M. and {Joyce}, R. and {Juneau}, S. and {Kara{\c{c}}ayl{\i}}, N.~G. and {Kehoe}, R. and {Kent}, S. and {Kirkby}, D. and {Kong}, H. and {Koposov}, S.~E. and {Kremin}, A. and {Krolewski}, A. and {Lahav}, O. and {Lai}, Y. and {Lan}, T. -W. and {Landriau}, M. and {Lang}, D. and {Lasker}, J. and {Le Goff}, J.~M. and {Le Guillou}, L. and {Leauthaud}, A. and {Levi}, M.~E. and {Li}, T.~S. and {Lodha}, K. and {Magneville}, C. and {Manera}, M. and {Margala}, D. and {Martini}, P. and {Matthewson}, W. and {Maus}, M. and {McDonald}, P. and {Medina-Varela}, L. and {Meisner}, A. and {Mena-Fern{\'a}ndez}, J. and {Miquel}, R. and {Moon}, J. and {Moore}, S. and {Moustakas}, J. and {Mudur}, N. and {Mueller}, E. and {Mu{\~n}oz-Guti{\'e}rrez}, A. and {Myers}, A.~D. and {Nadathur}, S. and {Napolitano}, L. and {Neveux}, R. and {Newman}, J.~A. and {Nguyen}, N.~M. and {Nie}, J. and {Niz}, G. and {Noriega}, H.~E. and {Padmanabhan}, N. and {Paillas}, E. and {Palanque-Delabrouille}, N. and {Pan}, J. and {Penmetsa}, S. and {Percival}, W.~J. and {Pieri}, M.~M. and {Pinon}, M. and {Poppett}, C. and {Porredon}, A. and {Prada}, F. and {P{\'e}rez-Fern{\'a}ndez}, A. and {P{\'e}rez-R{\`a}fols}, I. and {Rabinowitz}, D. and {Raichoor}, A. and {Ram{\'\i}rez-P{\'e}rez}, C. and {Ramirez-Solano}, S. and {Rashkovetskyi}, M. and {Ravoux}, C. and {Rezaie}, M. and {Rich}, J. and {Rocher}, A. and {Rockosi}, C. and {Roe}, N.~A. and {Rosado-Marin}, A. and {Ross}, A.~J. and {Rossi}, G. and {Ruggeri}, R. and {Ruhlmann-Kleider}, V. and {Samushia}, L. and {Sanchez}, E. and {Saulder}, C. and {Schlafly}, E.~F. and {Schlegel}, D. and {Schubnell}, M. and {Seo}, H. and {Shafieloo}, A. and {Sharples}, R. and {Silber}, J. and {Slosar}, A. and {Smith}, A. and {Sprayberry}, D. and {Tan}, T. and {Tarl{\'e}}, G. and {Taylor}, P. and {Trusov}, S. and {Vaisakh}, R. and {Valcin}, D. and {Valdes}, F. and {Valogiannis}, G. and {Vargas-Maga{\~n}a}, M. and {Verde}, L. and {Walther}, M. and {Wang}, B. and {Wang}, M.~S. and {Weaver}, B.~A. and {Weaverdyck}, N. and {Wechsler}, R.~H. and {Weinberg}, D.~H. and {White}, M. and {Wilson}, M.~J.},
        title = "{DESI 2024 VII: Cosmological Constraints from the Full-Shape Modeling of Clustering Measurements}",
      journal = {arXiv e-prints},
         year = 2024,
        month = nov,
          eid = {arXiv:2411.12022},
        pages = {arXiv:2411.12022},
          doi = {10.48550/arXiv.2411.12022},
archivePrefix = {arXiv},
       eprint = {2411.12022},
 primaryClass = {astro-ph.CO},
       adsurl = {https://ui.adsabs.harvard.edu/abs/2024arXiv241112022D}
}

@ARTICLE{DESI_DR2_cosmology_2025,
       author = {{DESI Collaboration} and {Abdul-Karim}, M. and {Aguilar}, J. and {Ahlen}, S. and {Alam}, S. and {Allen}, L. and {Allende Prieto}, C. and {Alves}, O. and {Anand}, A. and {Andrade}, U. and {Armengaud}, E. and {Aviles}, A. and {Bailey}, S. and {Baltay}, C. and {Bansal}, P. and {Bault}, A. and {Behera}, J. and {BenZvi}, S. and {Bianchi}, D. and {Blake}, C. and {Brieden}, S. and {Brodzeller}, A. and {Brooks}, D. and {Buckley-Geer}, E. and {Burtin}, E. and {Calderon}, R. and {Canning}, R. and {Carnero Rosell}, A. and {Carrilho}, P. and {Casas}, L. and {Castander}, F.~J. and {Cereskaite}, R. and {Charles}, M. and {Chaussidon}, E. and {Chaves-Montero}, J. and {Chebat}, D. and {Chen}, X. and {Claybaugh}, T. and {Cole}, S. and {Cooper}, A.~P. and {Cuceu}, A. and {Dawson}, K.~S. and {de la Macorra}, A. and {de Mattia}, A. and {Deiosso}, N. and {Della Costa}, J. and {Demina}, R. and {Dey}, A. and {Dey}, B. and {Ding}, Z. and {Doel}, P. and {Edelstein}, J. and {Eisenstein}, D.~J. and {Elbers}, W. and {Fagrelius}, P. and {Fanning}, K. and {{Fern{\'a}ndez-Garc{\'i}a}}, E. and {Ferraro}, S. and {Font-Ribera}, A. and {Forero-Romero}, J.~E. and {Frenk}, C.~S. and {Garcia-Quintero}, C. and {Garrison}, L.~H. and {Gazta\~naga}, E. and {Gil-Mar\textbackslash'in}, H. and {Gontcho}, S. Gontcho A and {Gonzalez}, D. and {Gonzalez-Morales}, A.~X. and {Gordon}, C. and {Green}, D. and {Gutierrez}, G. and {Guy}, J. and {Hadzhiyska}, B. and {Hahn}, C. and {He}, S. and {Herbold}, M. and {Herrera-Alcantar}, H.~K. and {Ho}, M. and {Honscheid}, K. and {Howlett}, C. and {Huterer}, D. and {Ishak}, M. and {Juneau}, S. and {Kamble}, N.~V. and {Kara{\c{c}}ayl{\i}}, N.~G. and {Kehoe}, R. and {Kent}, S. and {Kim}, A.~G. and {Kirkby}, D. and {Kisner}, T. and {Koposov}, S.~E. and {Kremin}, A. and {Krolewski}, A. and {Lahav}, O. and {Lamman}, C. and {Landriau}, M. and {Lang}, D. and {Lasker}, J. and {Le Goff}, J.~M. and {Le Guillou}, L. and {Leauthaud}, A. and {Levi}, M.~E. and {Li}, Q. and {Li}, T.~S. and {Lodha}, K. and {Lokken}, M. and {Lozano-Rodr\textbackslash'iguez}, F. and {Magneville}, C. and {Manera}, M. and {Martini}, P. and {Matthewson}, W.~L. and {Meisner}, A. and {Mena-Fern\textbackslash'andez}, J. and {Menegas}, A. and {Mergulh\~ao}, T. and {Miquel}, R. and {Moustakas}, J. and {Mu\~noz-Guti\'errez}, A. and {Mu-Santos}, D. and {Myers}, A.~D. and {Nadathur}, S. and {Naidoo}, K. and {Napolitano}, L. and {Newman}, J.~A. and {Niz}, G. and {Noriega}, H.~E. and {Paillas}, E. and {Palanque-Delabrouille}, N. and {Pan}, J. and {Peacock}, J. and {Pellejero Ibanez}, Marcos and {Percival}, W.~J. and {P{\'e}rez-Fern\textbackslash'andez}, A. and {P{\'e}rez-R\textbackslash`afols}, I. and {Pieri}, M.~M. and {Poppett}, C. and {Prada}, F. and {Rabinowitz}, D. and {Raichoor}, A. and {Ram{\'i}rez-P{\'e}rez}, C. and {Rashkovetskyi}, M. and {Ravoux}, C. and {Rich}, J. and {Rocher}, A. and {Rockosi}, C. and {Rohlf}, J. and {Rom{\'a}n-Herrera}, J.~O. and {Ross}, A.~J. and {Rossi}, G. and {Ruggeri}, R. and {Ruhlmann-Kleider}, V. and {Samushia}, L. and {Sanchez}, E. and {Sanders}, N. and {Schlegel}, D. and {Schubnell}, M. and {Seo}, H. and {Shafieloo}, A. and {Sharples}, R. and {Silber}, J. and {Sinigaglia}, F. and {Sprayberry}, D. and {Tan}, T. and {Tarl\textbackslash'e}, G. and {Taylor}, P. and {Turner}, W. and {Ure'opez}, L.~A. and {Vaisakh}, R. and {Valdes}, F. and {Valogiannis}, G. and {Vargas-Maga}, M. and {Verde}, L. and {Walther}, M. and {Weaver}, B.~A. and {Weinberg}, D.~H. and {White}, M. and {Wolfson}, M. and {Y\textbackslash`eche}, C. and {Yu}, J. and {Zaborowski}, E.~A. and {Zarrouk}, P. and {Zhai}, Z. and {Zhang}, H. and {Zhao}, C. and {Zhao}, G.~B. and {Zhou}, R. and {Zou}, H.},
        title = "{DESI DR2 Results II: Measurements of Baryon Acoustic Oscillations and Cosmological Constraints}",
      journal = {arXiv e-prints},
         year = 2025,
        month = mar,
          eid = {arXiv:2503.14738},
        pages = {arXiv:2503.14738},
          doi = {10.48550/arXiv.2503.14738},
archivePrefix = {arXiv},
       eprint = {2503.14738},
 primaryClass = {astro-ph.CO},
       adsurl = {https://ui.adsabs.harvard.edu/abs/2025arXiv250314738D}
}

@ARTICLE{Jamieson2024:POP,
       author = {{Jamieson}, Drew and {Caravano}, Angelo and {Hou}, Jiamin and {Slepian}, Zachary and {Komatsu}, Eiichiro},
        title = "{Parity-odd power spectra: concise statistics for cosmological parity violation}",
      journal = {\mnras},
         year = 2024,
        month = sep,
       volume = {533},
       number = {3},
        pages = {2582-2598},
          doi = {10.1093/mnras/stae1924},
archivePrefix = {arXiv},
       eprint = {2406.15683},
 primaryClass = {astro-ph.CO},
       adsurl = {https://ui.adsabs.harvard.edu/abs/2024MNRAS.533.2582J}
}

@ARTICLE{Gao2025:parity,
       author = {{Gao}, Zucheng and {Moradinezhad Dizgah}, Azadeh and {Vlah}, Zvonimir},
        title = "{Parity in Composite-Field Galaxy Correlators}",
      journal = {arXiv e-prints},
         year = 2025,
        month = sep,
          eid = {arXiv:2509.13207},
        pages = {arXiv:2509.13207},
          doi = {10.48550/arXiv.2509.13207},
archivePrefix = {arXiv},
       eprint = {2509.13207},
 primaryClass = {astro-ph.CO},
       adsurl = {https://ui.adsabs.harvard.edu/abs/2025arXiv250913207G}
}

@ARTICLE{Kurita2025:parity,
       author = {{Kurita}, Toshiki and {Jamieson}, Drew and {Komatsu}, Eiichiro and {Schmidt}, Fabian},
        title = "{Parity Violation in Galaxy Shapes: Primordial Non-Gaussianity}",
      journal = {arXiv e-prints},
         year = 2025,
        month = sep,
          eid = {arXiv:2509.08787},
        pages = {arXiv:2509.08787},
          doi = {10.48550/arXiv.2509.08787},
archivePrefix = {arXiv},
       eprint = {2509.08787},
 primaryClass = {astro-ph.CO},
       adsurl = {https://ui.adsabs.harvard.edu/abs/2025arXiv250908787K}
}

@ARTICLE{Euclid2011:WhitePaper,
       author = {{Laureijs}, R. and {Amiaux}, J. and {Arduini}, S. and {Augu{\`e}res}, J. -L. and {Brinchmann}, J. and {Cole}, R. and {Cropper}, M. and {Dabin}, C. and {Duvet}, L. and {Ealet}, A. and {Garilli}, B. and {Gondoin}, P. and {Guzzo}, L. and {Hoar}, J. and {Hoekstra}, H. and {Holmes}, R. and {Kitching}, T. and {Maciaszek}, T. and {Mellier}, Y. and {Pasian}, F. and {Percival}, W. and {Rhodes}, J. and {Saavedra Criado}, G. and {Sauvage}, M. and {Scaramella}, R. and {Valenziano}, L. and {Warren}, S. and {Bender}, R. and {Castander}, F. and {Cimatti}, A. and {Le F{\`e}vre}, O. and {Kurki-Suonio}, H. and {Levi}, M. and {Lilje}, P. and {Meylan}, G. and {Nichol}, R. and {Pedersen}, K. and {Popa}, V. and {Rebolo Lopez}, R. and {Rix}, H. -W. and {Rottgering}, H. and {Zeilinger}, W. and {Grupp}, F. and {Hudelot}, P. and {Massey}, R. and {Meneghetti}, M. and {Miller}, L. and {Paltani}, S. and {Paulin-Henriksson}, S. and {Pires}, S. and {Saxton}, C. and {Schrabback}, T. and {Seidel}, G. and {Walsh}, J. and {Aghanim}, N. and {Amendola}, L. and {Bartlett}, J. and {Baccigalupi}, C. and {Beaulieu}, J. -P. and {Benabed}, K. and {Cuby}, J. -G. and {Elbaz}, D. and {Fosalba}, P. and {Gavazzi}, G. and {Helmi}, A. and {Hook}, I. and {Irwin}, M. and {Kneib}, J. -P. and {Kunz}, M. and {Mannucci}, F. and {Moscardini}, L. and {Tao}, C. and {Teyssier}, R. and {Weller}, J. and {Zamorani}, G. and {Zapatero Osorio}, M.~R. and {Boulade}, O. and {Foumond}, J.~J. and {Di Giorgio}, A. and {Guttridge}, P. and {James}, A. and {Kemp}, M. and {Martignac}, J. and {Spencer}, A. and {Walton}, D. and {Bl{\"u}mchen}, T. and {Bonoli}, C. and {Bortoletto}, F. and {Cerna}, C. and {Corcione}, L. and {Fabron}, C. and {Jahnke}, K. and {Ligori}, S. and {Madrid}, F. and {Martin}, L. and {Morgante}, G. and {Pamplona}, T. and {Prieto}, E. and {Riva}, M. and {Toledo}, R. and {Trifoglio}, M. and {Zerbi}, F. and {Abdalla}, F. and {Douspis}, M. and {Grenet}, C. and {Borgani}, S. and {Bouwens}, R. and {Courbin}, F. and {Delouis}, J. -M. and {Dubath}, P. and {Fontana}, A. and {Frailis}, M. and {Grazian}, A. and {Koppenh{\"o}fer}, J. and {Mansutti}, O. and {Melchior}, M. and {Mignoli}, M. and {Mohr}, J. and {Neissner}, C. and {Noddle}, K. and {Poncet}, M. and {Scodeggio}, M. and {Serrano}, S. and {Shane}, N. and {Starck}, J. -L. and {Surace}, C. and {Taylor}, A. and {Verdoes-Kleijn}, G. and {Vuerli}, C. and {Williams}, O.~R. and {Zacchei}, A. and {Altieri}, B. and {Escudero Sanz}, I. and {Kohley}, R. and {Oosterbroek}, T. and {Astier}, P. and {Bacon}, D. and {Bardelli}, S. and {Baugh}, C. and {Bellagamba}, F. and {Benoist}, C. and {Bianchi}, D. and {Biviano}, A. and {Branchini}, E. and {Carbone}, C. and {Cardone}, V. and {Clements}, D. and {Colombi}, S. and {Conselice}, C. and {Cresci}, G. and {Deacon}, N. and {Dunlop}, J. and {Fedeli}, C. and {Fontanot}, F. and {Franzetti}, P. and {Giocoli}, C. and {Garcia-Bellido}, J. and {Gow}, J. and {Heavens}, A. and {Hewett}, P. and {Heymans}, C. and {Holland}, A. and {Huang}, Z. and {Ilbert}, O. and {Joachimi}, B. and {Jennins}, E. and {Kerins}, E. and {Kiessling}, A. and {Kirk}, D. and {Kotak}, R. and {Krause}, O. and {Lahav}, O. and {van Leeuwen}, F. and {Lesgourgues}, J. and {Lombardi}, M. and {Magliocchetti}, M. and {Maguire}, K. and {Majerotto}, E. and {Maoli}, R. and {Marulli}, F. and {Maurogordato}, S. and {McCracken}, H. and {McLure}, R. and {Melchiorri}, A. and {Merson}, A. and {Moresco}, M. and {Nonino}, M. and {Norberg}, P. and {Peacock}, J. and {Pello}, R. and {Penny}, M. and {Pettorino}, V. and {Di Porto}, C. and {Pozzetti}, L. and {Quercellini}, C. and {Radovich}, M. and {Rassat}, A. and {Roche}, N. and {Ronayette}, S. and {Rossetti}, E. and {Sartoris}, B. and {Schneider}, P. and {Semboloni}, E. and {Serjeant}, S. and {Simpson}, F. and {Skordis}, C. and {Smadja}, G. and {Smartt}, S. and {Spano}, P. and {Spiro}, S. and {Sullivan}, M. and {Tilquin}, A. and {Trotta}, R. and {Verde}, L. and {Wang}, Y. and {Williger}, G. and {Zhao}, G. and {Zoubian}, J. and {Zucca}, E.},
        title = "{Euclid Definition Study Report}",
      journal = {arXiv e-prints},
         year = 2011,
        month = oct,
          eid = {arXiv:1110.3193},
        pages = {arXiv:1110.3193},
archivePrefix = {arXiv},
       eprint = {1110.3193},
 primaryClass = {astro-ph.CO},
       adsurl = {https://ui.adsabs.harvard.edu/abs/2011arXiv1110.3193L}
}

@ARTICLE{Hartlap2007:hartlap,
       author = {{Hartlap}, J. and {Simon}, P. and {Schneider}, P.},
        title = "{Why your model parameter confidences might be too optimistic. Unbiased estimation of the inverse covariance matrix}",
      journal = {\aap},
         year = 2007,
        month = mar,
       volume = {464},
       number = {1},
        pages = {399-404},
          doi = {10.1051/0004-6361:20066170},
archivePrefix = {arXiv},
       eprint = {astro-ph/0608064},
 primaryClass = {astro-ph},
       adsurl = {https://ui.adsabs.harvard.edu/abs/2007A&A...464..399H}
}

@ARTICLE{Hou2022:AnalytCov,
       author = {{Hou}, Jiamin and {Cahn}, Robert N. and {Philcox}, Oliver H.~E. and {Slepian}, Zachary},
        title = "{Analytic Gaussian covariance matrices for galaxy N -point correlation functions}",
      journal = {\prd},
         year = 2022,
        month = aug,
       volume = {106},
       number = {4},
          eid = {043515},
        pages = {043515},
          doi = {10.1103/PhysRevD.106.043515},
archivePrefix = {arXiv},
       eprint = {2108.01714},
 primaryClass = {astro-ph.CO},
       adsurl = {https://ui.adsabs.harvard.edu/abs/2022PhRvD.106d3515H}
}

@ARTICLE{hou2022:parity,
       author = {{Hou}, Jiamin and {Slepian}, Zachary and {Cahn}, Robert N.},
        title = "{Measurement of parity-odd modes in the large-scale 4-point correlation function of Sloan Digital Sky Survey Baryon Oscillation Spectroscopic Survey twelfth data release CMASS and LOWZ galaxies}",
      journal = {\mnras},
         year = 2023,
        month = may,
       volume = {522},
       number = {4},
        pages = {5701-5739},
          doi = {10.1093/mnras/stad1062},
archivePrefix = {arXiv},
       eprint = {2206.03625},
 primaryClass = {astro-ph.CO},
       adsurl = {https://ui.adsabs.harvard.edu/abs/2023MNRAS.522.5701H}
}

@ARTICLE{SE_3pt,
       author = {{Slepian}, Zachary and {Eisenstein}, Daniel J.},
        title = "{Computing the three-point correlation function of galaxies in O(N\^2) time}",
      journal = {\mnras},
         year = 2015,
        month = dec,
       volume = {454},
       number = {4},
        pages = {4142-4158},
          doi = {10.1093/mnras/stv2119},
archivePrefix = {arXiv},
       eprint = {1506.02040},
 primaryClass = {astro-ph.CO},
       adsurl = {https://ui.adsabs.harvard.edu/abs/2015MNRAS.454.4142S}
}

@ARTICLE{Jeong2012:fossil,
       author = {{Jeong}, Donghui and {Kamionkowski}, Marc},
        title = "{Clustering Fossils from the Early Universe}",
      journal = {\prl},
         year = 2012,
        month = jun,
       volume = {108},
       number = {25},
          eid = {251301},
        pages = {251301},
          doi = {10.1103/PhysRevLett.108.251301},
archivePrefix = {arXiv},
       eprint = {1203.0302},
 primaryClass = {astro-ph.CO},
       adsurl = {https://ui.adsabs.harvard.edu/abs/2012PhRvL.108y1301J}
}

@ARTICLE{Cahn2021:parity,
       author = {{Cahn}, Robert N. and {Slepian}, Zachary and {Hou}, {J}},
        title = "{Test for Cosmological Parity Violation Using the 3D Distribution of Galaxies}",
      journal = {\prl},
         year = 2023,
        month = may,
       volume = {130},
       number = {20},
          eid = {201002},
        pages = {201002},
          doi = {10.1103/PhysRevLett.130.201002},
       adsurl = {https://ui.adsabs.harvard.edu/abs/2023PhRvL.130t1002C}
}

@article{Philcox:encore,
    author = "Philcox, Oliver H. E. and Slepian, Zachary and Hou, Jiamin and Warner, Craig and Cahn, Robert N. and Eisenstein, Daniel J.X",
    title = "{ENCORE: Estimating Galaxy $N$-point Correlation Functions in $\mathcal{O}(N_{\rm g}^2)$ Time}",
    journal = "MNRAS",
    volume = "509",
    pages="2457-2481",
    year="2021",
    eprint = "2105.08722",
    archivePrefix = "arXiv",
    primaryClass = "astro-ph.IM",
    month = "5",
    year = "2022"
}

\end{document}